\documentclass[10pt,twocolumn]{article}

\usepackage[a4paper,margin=0.75in]{geometry}

\usepackage{titlesec}       
\usepackage{abstract}          
\usepackage{authblk}        
\usepackage{ulem} 
\usepackage{enumitem}
\usepackage{graphicx} 
\usepackage{nicefrac}
\usepackage{amsmath,amssymb,bm} 
\usepackage{siunitx}        
\usepackage{caption}        
\usepackage{booktabs}       
\usepackage{multirow}
\usepackage[colorlinks=true, linkcolor=blue, citecolor=blue, urlcolor=blue]{hyperref} 
\usepackage{fancyhdr}       
\usepackage{titlesec}
\usepackage{setspace}
\usepackage{orcidlink}
\usepackage{comment}
\usepackage{xspace}
\usepackage{comment}
\usepackage{makecell}

\usepackage[dvipsnames]{xcolor}
\usepackage[backend=biber, sorting=none, style=nature]{biblatex}
\renewcommand\Affilfont{\small\normalfont\raggedright}
\makeatletter
\renewcommand\AB@affilsepx{\par\protect\Affilfont}
\makeatother

\def\mlgwjax{\texttt{mlgw-jax}\xspace}
\def\mlgwbnsjax{\texttt{mlgw-bns-jax}\xspace}
\def\sharpy{\texttt{SHARPy}\xspace}
\def\blackjaxns{\texttt{BlackJax-NS}\xspace}
\def\mlgw{\texttt{mlgw}\xspace}
\def\mlgwbns{\texttt{mlgw-bns}\xspace}
\def\ripple{\texttt{ripple}\xspace}
\def\jax{\texttt{JAX}\xspace}
\def\seob{SEOBNRv5HM\xspace}
\def\teobresums{TEOBResumSPA\xspace}

\def\blackjax{\texttt{BlackJAX}\xspace}
\def\gwgpujax{\texttt{GWgpu-jax}\xspace}
\def\chieff{\mathcal{\chi}_{\text{eff}}}

\title{\textbf{Architecture Acceleration of Machine Learning Gravitational Waveform Surrogate Models}}

\author[1,2]{Adriano Frattale Mascioli\textsuperscript{$\star$}\textsuperscript{$\dagger$}\orcidlink{0000-0002-0155-3833}}
\author[1,2]{Lorenzo Piccari\textsuperscript{$\star$}\textsuperscript{$\dagger\dagger$}\orcidlink{0009-0000-0247-4339}}
\author[3,4,5]{Saulo Albuquerque\textsuperscript{$\star$}\textsuperscript{$\ddagger$}\orcidlink{0000-0003-2911-9358}}
\author[5,6]{\\ Gabriele Demasi \orcidlink{0009-0009-5320-502X}}
\author[7,8]{Giulia Capurri \orcidlink{0000-0003-0889-1015}} 
\author[5,6]{Massimo Lenti}
\author[7,8]{Angelo Ricciardone \orcidlink{0000-0002-5688-455X}}
\author[7,8]{\\ Barbara Patricelli \orcidlink{0000-0001-6709-0969}}
\author[4,5]{Gianluca M. Guidi \orcidlink{0000-0002-3061-9870}}
\author[9,10]{Giulia Stratta \orcidlink{0000-0003-1055-7980}}
\author[7,8]{Walter Del Pozzo \orcidlink{0000-0003-3978-2030}}
\author[1,2]{Francesco Pannarale \orcidlink{0000-0002-7537-3210}}

\affil[1]{Dipartimento di Fisica, ``Sapienza'' Universit\`a di Roma, Piazzale Aldo Moro 5,  00185, Roma, Italy}
\affil[2]{Dipartimento di Fisica, Sezione INFN Roma, Piazzale Aldo Moro 5,  00185, Roma, Italy}
\affil[3]{Theoretical Astrophysics, IAAT, University of T\"ubingen, Auf der Morgenstelle 10, D-72076 T\"ubingen, Germany}
\affil[4]{Universit\`a degli Studi di Urbino ``Carlo Bo'', I-61029 Urbino, Italy}
\affil[5]{INFN, Sezione di Firenze, Sesto Fiorentino (Firenze) I-50019, Italy}
\affil[6]{Dipartimento di Fisica e Astronomia, Università degli Studi di Firenze,
Via Sansone 1, Sesto Fiorentino (Firenze) I-50019, Italy}
\affil[7]{Dipartimento di Fisica “Enrico Fermi”, Università di Pisa,
Largo Bruno Pontecorvo 3, Pisa I-56127, Italy}
\affil[8]{INFN, Sezione di Pisa, Largo Bruno Pontecorvo 3, Pisa I-56127, Italy}
\affil[9]{INAF, Osservatorio di Astrofisica e Scienza dello Spazio, Via Piero Gobetti 101, 40129 Bologna, Italy}
\affil[10]{INFN, Sezione di Bologna, viale Carlo Berti Pichat 6/2, 40127 Bologna, Italy}

\date{}

\begin{document}
\normalem  
\twocolumn[
\maketitle

\begin{onecolabstract}
Accurate and computationally efficient waveform models of compact binary coalescences are a fundamental requirement for gravitational-wave data analysis. 
This work presents differentiable and hardware-accelerated implementations of the machine-learning surrogate models \mlgw~\cite{mlgw_1} and \mlgwbns~\cite{mlgw_bns} within the \jax ecosystem.
%
The proposed framework enables just-in-time compilation, vectorized paralellization, native GPU acceleration, and automatic differentiation, while preserving the original surrogate-training infrastructure and its applicability to arbitrary non-precessing waveform approximants.
Benchmarks are performed in the binary black hole case with a newly trained surrogate of the \seob approximant; on CPU, they show speed-ups in waveform evaluation time with respect to the original \mlgw implementation that exceed one order of magnitude; further, when exploiting GPU acceleration and large-batch vectorization these reach approximately two orders of magnitude.
The benchmarks performed in the case of binary neutron stars with a \teobresums surrogate model achieve similar gains on GPU.
Our \jax-based surrogate-modeling framework can be integrated into Bayesian parameter-estimation pipelines through \gwgpujax, an open-source package that connects \jax-compatible gravitational waveform generators
to \jax-native sampling algorithms. We demonstrate this capability within a nested-sampling pipeline built upon \blackjaxns~\cite{Prathaban:2025qgg},
performing parameter-estimation analyses of the GW150914 and GW170817 events with our newly built accelerated waveform surrogates. The analyses require approximately twelve and seventeen minutes, respectively, on a single GPU and yield posterior distributions consistent with those reported by the LIGO--Virgo--KAGRA Collaboration.
Beyond nested sampling, \jax automatic differentiation provides efficient gradient and Hessian evaluations of the waveform models, enabling seamless integration with gradient-based Bayesian inference methods.
\end{onecolabstract}
\vspace{1em}
 ]

\begingroup
\renewcommand\thefootnote{\fnsymbol{footnote}}
\footnotetext{%
\textsuperscript{$\star$}{These authors contributed equally to this work.}%
}
\endgroup

\begingroup
\renewcommand\thefootnote{\fnsymbol{footnote}}
\footnotetext{%
\textsuperscript{$\dagger$}E-mail: \texttt{adriano.frattalemascioli@uniroma1.it }%
}
\endgroup

\begingroup
\renewcommand\thefootnote{\fnsymbol{footnote}}
\footnotetext{%
\textsuperscript{$\dagger\dagger$}E-mail: \texttt{lorenzo.piccari@uniroma1.it }%
}
\endgroup

\begingroup
\renewcommand\thefootnote{\fnsymbol{footnote}}
\footnotetext{%
\textsuperscript{$\ddagger$}E-mail: \texttt{saulo.soaresdealbuquerquefilho@uniurb.it }%
}
\endgroup

\section{Introduction}\label{sec:intro}
The direct observation of gravitational waves (GWs) by the LIGO-Virgo Collaboration~\cite{LIGOScientific:2016aoc} opened a new era in observational astronomy. 
It confirmed Einstein's original intuition of gravitational radiation, and provided an entirely new probe of the Universe and its most compact objects, such as black holes and neutron stars. 
GW observations have enabled advances across multiple domains, ranging from astrophysics and cosmology to fundamental physics~\cite{GWTC_5_intro, GWTC_5_methods, GWTC_5_results,  LIGOScientific:2026ctl, LIGOScientific:2026uyd}.
All confirmed GW detections to date by the LIGO–Virgo–KAGRA (LVK) Collaboration~\cite{2015:LIGO,2015:VIRGO,2013:KAGRA} are consistent with transient signals produced by compact binary coalescences (CBCs)~\cite{GWTC_5_results, GWTC_5_GWOSC}. 
Currently, the most successful data analysis technique to identify and rank candidate events of this kind in interferometric data is matched filtering~\cite{GWTC_5_methods, Allen:2005fk, Usman:2015kfa, mbta, gstlal, spiir}.
This approach relies critically on both the accuracy of CBC waveform and the computational efficiency with which they can be evaluated, as large template banks must be generated and repeatedly compared to the detector data~\cite{Owen_1996, Owen_1999}. The progressive increase in detector sensitivity and event rates places growing demands on both. These requirements will become particularly stringent for third-generation detectors, such as the Einstein Telescope~\cite{ET1, ET2} and Cosmic Explorer \cite{Reitze:2019iox}, where longer-duration signals and higher signal-to-noise ratios will amplify systematic modeling errors and computational costs.

Beyond their role in searches, accurate waveform models are essential for parameter estimation, where the source properties are inferred with Bayesian statistics~\cite{GWTC_5_methods, Thrane_2019, Christensen_2022}.  
In this context, evaluating the likelihood requires repeated waveform generation. 
Moreover, gradient-based samplers, such as Hamiltonian Monte Carlo \cite{DUANE1987216, hmc_betancourt}, aim at efficiently exploring the high-dimensional CBC parameter space by exploiting the waveform derivatives with respect to the source parameters.
The ability to rapidly evaluate waveforms and their derivatives is therefore crucial throughout GW data analysis.

Over the years, a variety of waveform models have been formulated and developed,  balancing the non-trivial trade-off between computational cost and accuracy.
With time, and as we move towards more sensitive detectors, the complexity of these models is increasing, also amplifying the challenge of their differentiation. 
Also in response to these demands, surrogate models have been constructed to reproduce with high faithfulness the output of standard waveform approximants, while simultaneously reducing the evaluation cost.  In this context, the authors of \cite{mlgw_1} introduced the \mlgw (Machine Learning Gravitational Waves) Python package, implementing a machine-learning surrogate model for the rapid generation of waveforms emitted by non-precessing binary black hole (BBH) systems. The model was subsequently extended to include higher-order modes~\cite{mlgw_2}, and later adapted to binary neutron star (BNS) systems in the \mlgwbns package~\cite{mlgw_bns}. 

Recently, differentiable programming frameworks \cite{Blondel2024TheEO}, which allow for automatic differentiation, enabled advances in gravitational waveform gradient computation.
Neither \mlgw nor \mlgwbns, however, provide native support for automatic differentiation, preventing efficient evaluation of waveform gradients and higher-order derivatives.
We aim at overcoming this limitation and adopt the \jax library~\cite{jax2018github} as a differentiable programming framework.
\jax combines automatic differentiation, automatic vectorization (\texttt{vmap}), and just-in-time (\texttt{JIT}) compilation capabilities within a \texttt{NumPy}-compatible interface, while natively supporting execution on GPUs and other hardware accelerators. 
These features make \jax appealing to speed up any GW data analysis approach that requires generating waveform models, particularly those that also rely on the differentiation of waveforms. 

With this in mind, the present work lies at the intersection between machine-learning surrogate modeling techniques and hardware acceleration enabled by \jax, with the primary goal of developing fast BBH and BNS waveform generators. We therefore introduce the following software frameworks.
\begin{itemize}
    \item \mlgwjax\footnote{\url{https://github.com/adrianomascioli/MLGW-JAX}} and \mlgwbnsjax\footnote{\url{https://github.com/saulo-albuquerque-phys/MLGW-BNS-JAX}}, the \jax implementations of the \mlgw and \mlgwbns waveform-generation interfaces. 
    They provide a fully \jax-compatible surrogate modeling framework that can be trained on any non-precessing waveform approximant, provided that a representative training dataset can be generated from the target model.  This is in contrast with the approach of existing packages, such as \ripple~\cite{Edwards:2023sak}, which instead provide \jax implementations of selected waveform models obtained by manually translating each individual approximant into \jax. 
    \item \gwgpujax\footnote{\url{https://github.com/saulo-albuquerque-phys/gwgpu-jax}}, a software framework designed to integrate \jax-compatible waveform generators with \jax-native Bayesian inference algorithms, enabling end-to-end GPU-accelerated GW parameter estimation.
\end{itemize}

The remainder of this paper is organized as follows. Section~\ref{sec:waveform_modeling} briefly reviews gravitational waveform modeling, with emphasis on its role in parameter estimation and modeled searches. Section~\ref{sec:mlgw-jax} introduces \mlgwjax and \mlgwbnsjax and presents benchmarks that assess both the accuracy in reproducing the training model and the computational performance relative to the original vanilla implementation on both CPU and GPU architectures. A comparison with \ripple is also provided.
In Sec.~\ref{sec:GW_inf} we introduce \gwgpujax and illustrate the application of the proposed generators
within a Bayesian inference framework, leveraging the \jax implementation of nested sampling provided by \blackjax~\cite{cabezas2024, yallup2025nested, Prathaban:2025qgg}.
As representative applications, we carry out a parameter estimation analysis of the GW150914~\cite{LIGOScientific:2016aoc,LIGOScientific:2016vlm} and GW170817~\cite{LIGOScientific:2017vwq, LIGOScientific:2018hze} events using \mlgwjax and \mlgwbnsjax as waveform generators, respectively.
%
%
%
\section{Waveform Modeling and Generation Budget}\label{sec:waveform_modeling}
GW signal modeling is inherently characterized by a trade-off between computational efficiency and physical accuracy. Modeled searches~\cite{Usman:2015kfa, pycbc_live, mbta, gstlal, gstlal_2, spiir} prioritize rapid waveform generation and rely on template banks~\cite{template_bank1, template_bank2, template_bank3,template_bank4,template_bank5,template_bank6,template_bank7,template_bank8,template_bank9,template_bank10}. Parameter-estimation analyses generally favor completeness and accuracy, although inference pipelines can operate on different latency scales. Within the LVK Collaboration for example, \texttt{BAYESTAR}~\cite{bayestar} provides rapid sky-localization and luminosity distance estimates, whereas full analyses are typically performed with computationally more demanding frameworks such as \texttt{Bilby}~\cite{bilby}.
Ideally, inference algorithms should combine speed and accuracy, enabling complete low-latency parameter estimation. Rapid access to source properties such as component masses and inclination can provide valuable information on the likelihood of electromagnetic counterparts and support subsequent investigations in astrophysics, cosmology, and fundamental physics~\cite{LIGOScientific:2017ync, LIGOScientific:2017adf, Abbott_2017, Baker:2017hug, LIGOScientific:2018dkp, LIGOScientific:2019zcs, Burns:2019tqz}. Significant progress toward this objective was achieved in recent years~\cite{dingo, rift}.

In addition to sophisticated and efficient sampling strategies, waveform generators that are both fast and physically accurate are also crucial.\footnote{Notably, approaches based on Neural Posterior Estimation can bypass explicit waveform generation at inference time~\cite{dingo}, further reshaping the computational landscape of GW inference.}
The modeling of CBC signals relies primarily on three broad families of waveform approximants: phenomenological (Phenom) models~\cite{phenom_1, phenom_2, phenom_3, phenom_4, phenom_5, phenom_6, phenom_7, phenom_8, phenom_9,phenom_10,phenom_11},models rooted in the effective-one-body (EOB) approach~\cite{eobnr1,eobnr2,eobnr3, eobnr4, eobnr5,eobnr6,eobnr7,eobnr8,eobnr9,eobnr10, eobnr11,eobnr12,eobnr13, eobnr14, eobnr15, eobnr16, eobnr17, eobnr18}, and the surrogate models built directly from numerical relativity (NRSur) waveforms~\cite{NR_SUR_1,NR_SUR_2,NR_SUR_3,NR_SUR_4} . 

Models in the Phenom family are constructed by building closed-form, frequency-domain representations of the waveform, calibrated to numerical-relativity simulations and post-Newtonian/EOB information across the inspiral, merger, and ringdown regimes. 
By directly modeling the amplitude and phase with an analytic ansatz, 
Phenom models achieve high computational efficiency while retaining faithfulness to numerical-relativity results within their calibration region. 
In contrast, EOB models map the relativistic two-body problem onto an effective test particle evolving in a deformed background spacetime. 
Despite the significantly higher computational cost due to the time-domain evolution of the dynamical system, they remain a benchmark in terms of physical accuracy and robustness, the state of the art models being \texttt{SEOBNRv5HM}~\cite{eobnr14, eobnr15} and \texttt{TEOBResumS-Dalì}~\cite{eobnr16,eobnr17}.

The waveform generation time can approximately be decomposed as $t_{\text{overhead}}+N_{\text{points}}t_{\text{point}}$~\cite{mlgw_bns}, where $N_{\text{points}}$ denotes the number of grid points at which the waveform is evaluated, $t_{\text{overhead}}$ denotes the fixed initialization cost per model call, independent of the number of evaluated points, while $t_{\text{point}}$ represents the marginal computational cost per waveform sample, scaling linearly with the total number of points. The generation time of both families of approximants has significantly improved over time, primarily via a reduction of $t_{\text{overhead}}$. 

More recently, data-driven approaches based on Reduced Order Modeling — also known as dimensional or complexity reduction — alongside the broader class of surrogate models have been actively developed~\cite{roq_1, roq_2, Purrer_2014, Purrer_2016, Tiglio_2022}. 
These techniques aim at eliminating redundancies in the data by exploiting methods from spectral theory, data science, machine learning and artificial intelligence, thereby achieving drastic reductions in waveform generation time while preserving high fidelity, acting also on $N_{\text{points}}$~\cite{multibanding, mlgw_bns} besides $t_{\text{overhead}}$.\footnote{$t_{\text{point}}$ is typically of the order of a few hundred nanoseconds on standard hardware, and it is ultimately bounded by the underlying clock speed and the number of floating-point operations required per evaluation. 
In practice, this contribution varies far less across waveform models than $t_{\text{overhead}}$~\cite{mlgw_bns}.} 

A complementary class of techniques, such as Reduced Order Quadrature~\cite{roq_1,roq_2,roq_3,roq_4,roq_5,roq_6, mlgw_bns} and Relative binning~\cite{relative_binning}, focuses on accelerating likelihood evaluations without modifying the underlying waveform model. 
Other approaches aim at constructing genuine surrogate models~\cite{Purrer_2014, surrogate} by approximating the map from the intrinsic binary parameters to the gravitational waveform, drastically reducing the overhead. NR Surrogates ~\cite{NR_SUR_1,NR_SUR_2,NR_SUR_3,NR_SUR_4}, such as \texttt{NRSur7dq4}, typically rely on directly using waveform templates produced by numerical relativity expensive simulations, and then building an orthonormal reduced basis through greedy algorithms, and subsequently fitting the parametric dependence of the projection coefficients onto this basis. The parametric dependence of the projection coefficients can be modeled using interpolation schemes. Recently, deep learning algorithms have also been trained from numerical relativity templates, leading to a neural-network surrogate model for \texttt{NRSur7dq4}~\cite{Purrer:2026wgp}.

Finally, beyond improvements to the internal software structure of waveform models, additional performance gains can be achieved through hardware acceleration. 
Algorithms exhibiting intrinsic parallelism and amenable to vectorization can efficiently leverage modern GPU architectures, which are particularly well-suited for large-scale batched computations. 
In this context, the Python framework \jax~\cite{jax2018github} provides a powerful infrastructure for writing hardware-accelerated software. 
\section{Machine Learning Surrogate Waveform Models in \jax}\label{sec:mlgw-jax}
This section introduces our \jax-based surrogate waveform framework.
We begin by outlining the main features of the baseline models \mlgw and \mlgwbns~\cite{mlgw_1,mlgw_2,mlgw_bns} and then we describe their translation into the \jax ecosystem, highlighting the modifications required to exploit automatic differentiation, just-in-time compilation, and hardware acceleration, among others. 
Finally, we present a systematic evaluation of the computational performance and accuracy of the resulting models.

\subsection{Binary Black Holes}
%
In this subsection we briefly review the main technical features of the \mlgw model, referring to the most up-to-date version of the package described in~\cite{mlgw_2}.
The complex time-domain waveform $h(t)$ describing the emission from a non-precessing BBH system can be decomposed into spin-2 spherical harmonics~\cite{Estelles:2021gvs}:
\begin{equation}\label{eq_waveform}
\begin{split}
&h(t; D_L, \iota, \varphi_c, M, \bm{\vartheta}) = \\
= \, &h_{+}(t; D_L, \iota, \varphi_c, M, \bm{\vartheta}) - i h_{\times}(t; D_L, \iota, \varphi_c, M, \bm{\vartheta}) = \\
= \, &\frac{G M}{c^{2} D_L}
\sum_{\ell=2}^{\infty}
\sum_{m=-\ell}^{\ell}
{}^{-2}Y_{\ell m}(\iota, \varphi_c)\,
h_{\ell m}(t/M; \bm{\vartheta}) \ ,
\end{split}
\end{equation}
where $G$ and $c$ are the gravitational constant and the speed of light, $D_L$ is the luminosity distance, $\iota$ is the inclination angle between the orbital angular momentum and the line of sight, $\varphi_c$ is the coalescence phase, and $M$ is the total mass of the binary,
while $\bm{\vartheta} = \left (q, \chi_{1z}, \chi_{2z} \right )$ denotes the relevant intrinsic physical parameters of the source, namely, the mass ratio $q = m_2/m_1\leq 1$, 
and the dimensionless spins, $\chi_{1z}$ and $\chi_{2z}$, which are aligned with the orbital angular momentum of the binary.
\mlgw learns the mapping
\begin{equation}
\bm{\vartheta} \longrightarrow h_{\ell m}(t,\bm{\vartheta}) \ ,
\end{equation} 
for each $(\ell, m)$ mode. 
This is achieved through an Artificial Neural Network (ANN) regression task.
Specifically, the model employs multiple ANNs, that perform the regression separately on the amplitude and phase of each mode, according to the decomposition
\begin{equation}
h_{\ell m}(t,\bm{\vartheta})=A_{\ell m}(t,\bm{\vartheta})e^{i\phi_{\ell m}(t,\bm{\vartheta})} \ .
\end{equation}

The ANNs are trained on a dataset of waveforms generated by a time-domain approximant and evaluated on a fixed time grid, conveniently spaced to provide a faithful yet compact representation of the waveform.
To improve the representation power, each ANN is provided with an augmented version $\tilde{\bm{\vartheta}}$ of the input vector $\bm{\vartheta}$ that includes additional mass and spin combinations, such as the chirp mass $\mathcal{M}=Mq^{3/5}/(1+q)^{6/5}$ and the effective spin $\chieff=(\chi_{1z}+q \, \chi_{2z})/(1+q)$, as well as polynomial combinations of $ \left (q, \chi_{1z}, \chi_{2z} \right )$ up to a chosen order. 
The selection of the augmented features to be included in $\tilde{\bm{\vartheta}}$, along with other ANNs hyperparameters, such as the number of layers and the activation functions, needs to be determined through a validation procedure. 

Given the intrinsic redundancy present in GW signals, particularly during the early inspiral phase, a principal component analysis (PCA) is employed to approximate the waveforms and reduce the dimensionality of the regression targets.
Let $\mathbf{f} \in \mathbb{R}^D$ denote a generic regression target vector, its approximation $\hat{\mathbf{f}}$ as provided by the PCA is: 
\begin{equation}\label{eq:PCA}
\hat{\mathbf{f}} = \sum_{i=0}^{K-1} g_i \mathbf{H}_i = \sum_{i=0}^{K-1} \langle \mathbf{f} | \mathbf{H}_i \rangle \mathbf{H}_i \ ,
\end{equation}
where the notation $\langle \cdot , \cdot \rangle$ denotes the Euclidean scalar product in $\mathbb{R}^D$, and the vectors $\mathbf{H}_i$ are the principal components.
The principal components are given by the first $K$ eigenvectors of the $D \times D$ covariance matrix of the training dataset.
In our case, $\mathbf{f}$ represents either the amplitude or the phase of a given waveform mode, while the dimensionality $D$ corresponds to the number of points in the fixed time grid.
From Eq.~\eqref{eq:PCA}, it follows that the choice of $K$ controls the quality of the approximation. 
After PCA dimensionality reduction, the ANNs regression task consists in learning the mapping between $\tilde{\bm{\vartheta}}$ and the reduced representation of the waveform amplitude and phase. 
For a more detailed discussion of the technical aspects of the model, including the selection of ANNs hyperparameters, we refer the reader to \cite{mlgw_2}.

Once the training process is complete, \mlgw reconstructs each waveform mode $h_{\ell m}(t,\bm{\vartheta})$ in its reduced PCA representation by loading the trained ANNs and evaluating them at the input physical parameters.
The predicted reduced amplitude and phase are then interpolated onto a generic time grid provided by the user.
Subsequently, the inverse PCA transformation is applied to recover their full-dimensional representation.
Finally, the complete waveform $h(t; D_L, \iota, \varphi_c, M, \bm{\vartheta})$ is assembled according to Eq.~\eqref{eq_waveform}.
We refer to this reconstruction procedure as the \textit{waveform generation interface} of the model. 
%
\subsubsection{\jax implementation}
\label{sec:mlgw_jax_implementation}
The framework we present here, \mlgwjax, consists of a translation of the waveform generation interface of the original \mlgw model into \jax.
The training infrastructure is preserved, while we gain the advantages of a full \jax implementation, the main ones of which are are summarized below.
\begin{itemize}
\item \texttt{JIT} compilation --- Compilation via the \texttt{XLA} (Accelerated Linear Algebra)~\cite{XLA} machine-learning compiler allows computational graphs to be optimized and compiled ahead of execution, often resulting in substantial speed-ups for numerically intensive workloads.
\item Hardware acceleration --- Since \jax is agnostic to the underlying hardware architecture, it can seamlessly leverage hardware accelerators such as GPUs. Moreover, its functional transformations, including automatic vectorization (\texttt{vmap}) and parallel mapping (\texttt{pmap}), facilitate efficient parallel execution, making it particularly suitable for large-scale waveform generation.
\item \jax automatic differentiation --- This enables direct interoperability with gradient-based samplers (cfr. Sec.~\ref{sec:GW_inf}).
\end{itemize}


The original \mlgw code relied on the \texttt{TensorFlow}~\cite{tensorflow2015-whitepaper} 
deep-learning framework and the package \texttt{KERAS}~\cite{chollet2015keras} 
built on it.
The main development effort focused on translating the \mlgw \texttt{GW\_generator} module to \jax.
We extensively modified the nested methods of the classes responsible for generating the individual waveform modes and the main class that assembles them into the full waveform returned to the user.
All the dynamic and runtime-dependent definitions were removed, and \jax's vectorization utilities and its \texttt{NumPy}-compatible application programming interface \texttt{JAX.numpy} (\texttt{jnp}) were employed. 
A particularly delicate aspect was the conversion of the trained ANNs into functions compatible with \jax. 
For this purpose, the \texttt{tf2jax}~\cite{tf2jax} 
library developed by Google DeepMind proved to be a valuable tool.
Modifications to the methods responsible for calculating the augmented input $\tilde{\bm{\vartheta}}$ and to the inverse-PCA mapping were also necessary. 
As an additional feature, \mlgwjax also supports native frequency-domain waveform generation.
This is achieved via a \texttt{GW\_FD\_generator} module which applies \jax built-in Fast Fourier Transform functions to the output of the standard time-domain generator.

The implementation of \mlgwjax is publicly available at the GitHub repository: \href{https://github.com/adrianomascioli/MLGW}{\texttt{github.com/adrianomascioli/MLGW}}. 

\subsubsection{Speed and accuracy tests}
\label{sec:BBH-speed-accuracy}
%
%
To quantitatively assess \mlgwjax in terms of computational performance and accuracy, we construct and use an \mlgw surrogate model trained on \seob~\cite{eobnr14}, adopting the same training procedure and same hyperparameter configuration of~\cite{mlgw_2}.\footnote{In~\cite{mlgw_2}, the authors constructed an \mlgw surrogate model trained on \texttt{SEOBNRv4HM}~\cite{Cotesta:2018fcv} with a training dataset of $6.8 \times 10^4$ waveforms. For each mode, they retained $K = 4$ principal components for the amplitude and $K = 6$ for the phase.}
To simplify the notation, throughout the remainder of this paper any results presented for \mlgw and \mlgwjax are understood to refer to this specific surrogate model.

To assess the accuracy of the new surrogate model and validate our \jax implementation of the \mlgw waveform-generation interface, we compare two sets of waveforms generated at $10^4$ parameter space points that differ from the ones used to train the surrogate model: one set is generated with the target model \seob, while the other is generated with \mlgwjax.
The total mass $M$ is fixed to $20 M_{\odot}$, while the other waveform parameters are drawn from the following distributions: q $\in \mathcal{U}[0.1, 1.0]$, $\{\chi_{1z}, \chi_{2z}\} \in \mathcal{U}[-0.9, 0.9]$, $\phi_c \in \mathcal{U}[0, 2\pi]$, and $\cos \ \iota \in \mathcal{U}[-1, 1]$, where $\mathcal{U}[a, b]$ denotes a uniform distribution between $a$ and $b$. 
To account for different waveform lengths, we draw the starting frequency $f_{\mathrm{start}}$ from $\mathcal{U}[15, 75\ \mathrm{Hz}]$.

We adopt the \textit{symphony mismatch} $\mathcal{F}_{\mathrm{sym}}$~\cite{mlgw_2} as figure of merit for the quantitative comparison of the two waveform sets, as this has been shown to be the most suitable quantity to compare waveforms that include higher order modes~\cite{Harry:2017weg, Harry:2016ijz}. $\mathcal{F}_{\mathrm{sym}}$ ranges from 0 to 1, with values closer to zero indicating a better agreement between the two waveforms being compared.
It is defined as $\mathcal{F}_{\mathrm{sym}} = 1 - \mathcal{M}_{\mathrm{sym}}$, where the match $\mathcal{M}_{\mathrm{sym}}$ is given by:
\begin{align}\label{eq:Msym}
&\mathcal{M}_{\mathrm{sym}} = \\
&= \max_{t_c}
\frac{
\langle\hat{s}|\hat{h}_{+}\rangle^{2}
+\langle\hat{s}|\hat{h}_{\times}\rangle^{2}
-2\langle\hat{h}_{+}|\hat{h}_{\times}\rangle
\langle\hat{s}|\hat{h}_{+}\rangle
\langle\hat{s}|\hat{h}_{\times}\rangle
}{
1-\langle\hat{h}_{+}|\hat{h}_{\times}\rangle^{2}
}\ .\nonumber
\end{align}
In the definition above, $t_c$ is the time of coalescence
and
\begin{equation}\label{eq:overlap}
    \langle a|b \rangle= 4 \mathcal{R}\int_{f_{\text{min}}}^{f_{\text{max}}}\frac{\tilde{a}^*(f)\tilde{b}(f)}{S_n(f)}df
\end{equation}
denotes the inner product weighted by the noise power spectral density $S_n(f)$, which characterizes the frequency-dependent noise level of the instrument.  Further, the hat symbol in Eq.~\eqref{eq:Msym} indicates waveforms normalized with respect to the inner product defined in Eq.~\eqref{eq:overlap}.
The fraction on the right hand side of Eq.~\eqref{eq:Msym} is therefore a normalized signal-to-noise ratio for the signal $s(t)$ and a given template with polarizations $h_+$ and $h_\times$, maximized over sky-location parameters $\alpha$ and $\delta$, and polarization angle $\psi$, as discussed in~\cite{Harry:2016ijz} and provided by the \texttt{PyCBC} library~\cite{alex_nitz_2024_10473621}.
In our specific comparison, $s$ is taken to be the waveform generated with \seob, while $h_+$ and $h_\times$ are the polarizations of the \mlgwjax approximant with the same parameters; finally, a flat power spectral density is adopted in order to obtain detector-agnostic results.

\begin{figure}[t!]
    \centering
    \includegraphics[width=1.0\linewidth]{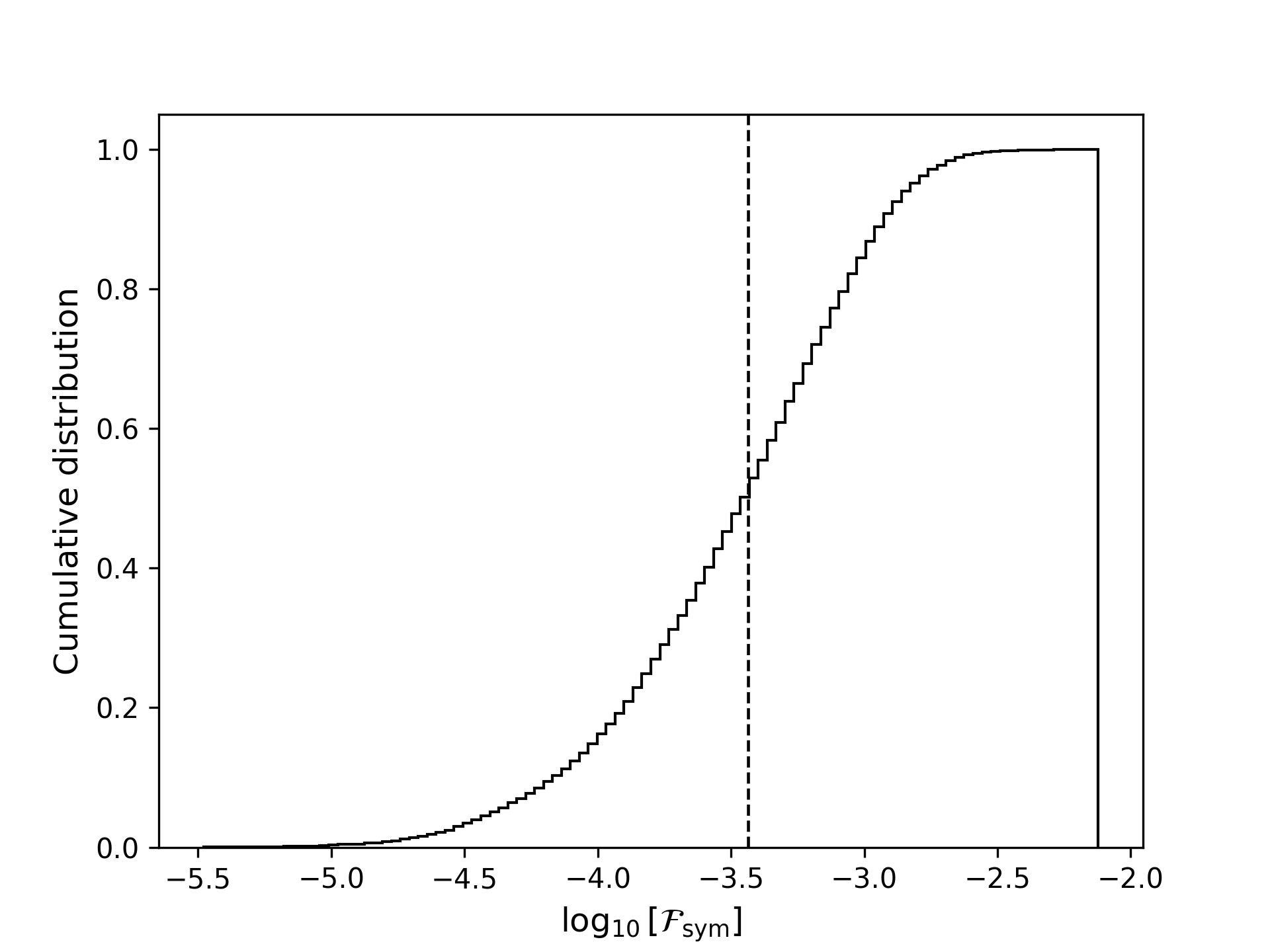}
    \caption{Cumulative distribution of the base-10 logarithm of $\mathcal{F}_{\mathrm{sym}}$ between $10^4$ waveforms generated with the \mlgwjax model and the training model \seob. The vertical dashed line represents the median of the distribution.}
    \label{fig:mlgw_jax_vs_seobnrv4_mismatch_histo}
\end{figure}
\begin{figure}[t!]
    \centering
    \includegraphics[width=1.0\linewidth]{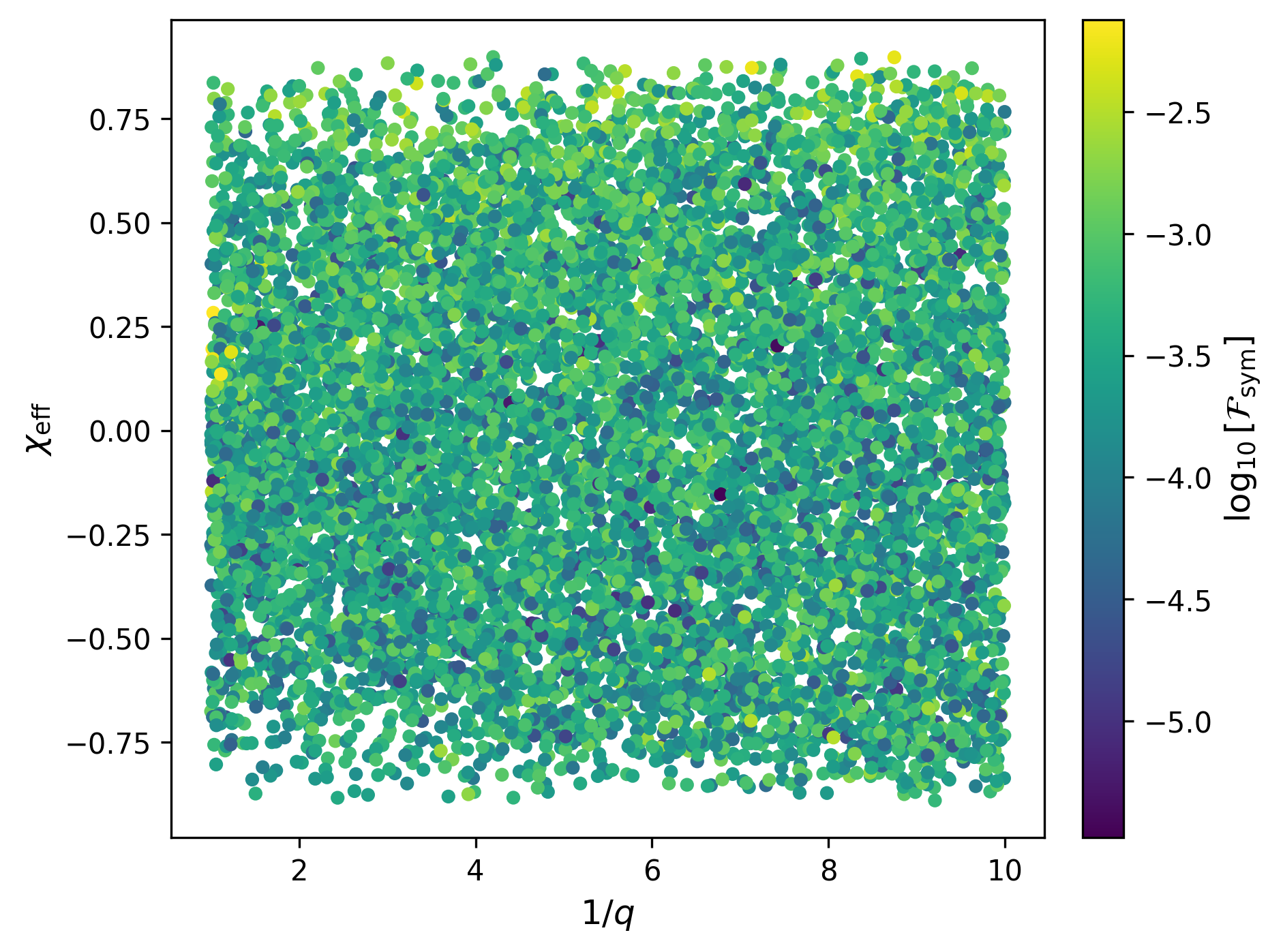}
    \hspace*{0.02\linewidth}
    \includegraphics[width=1.0\linewidth]{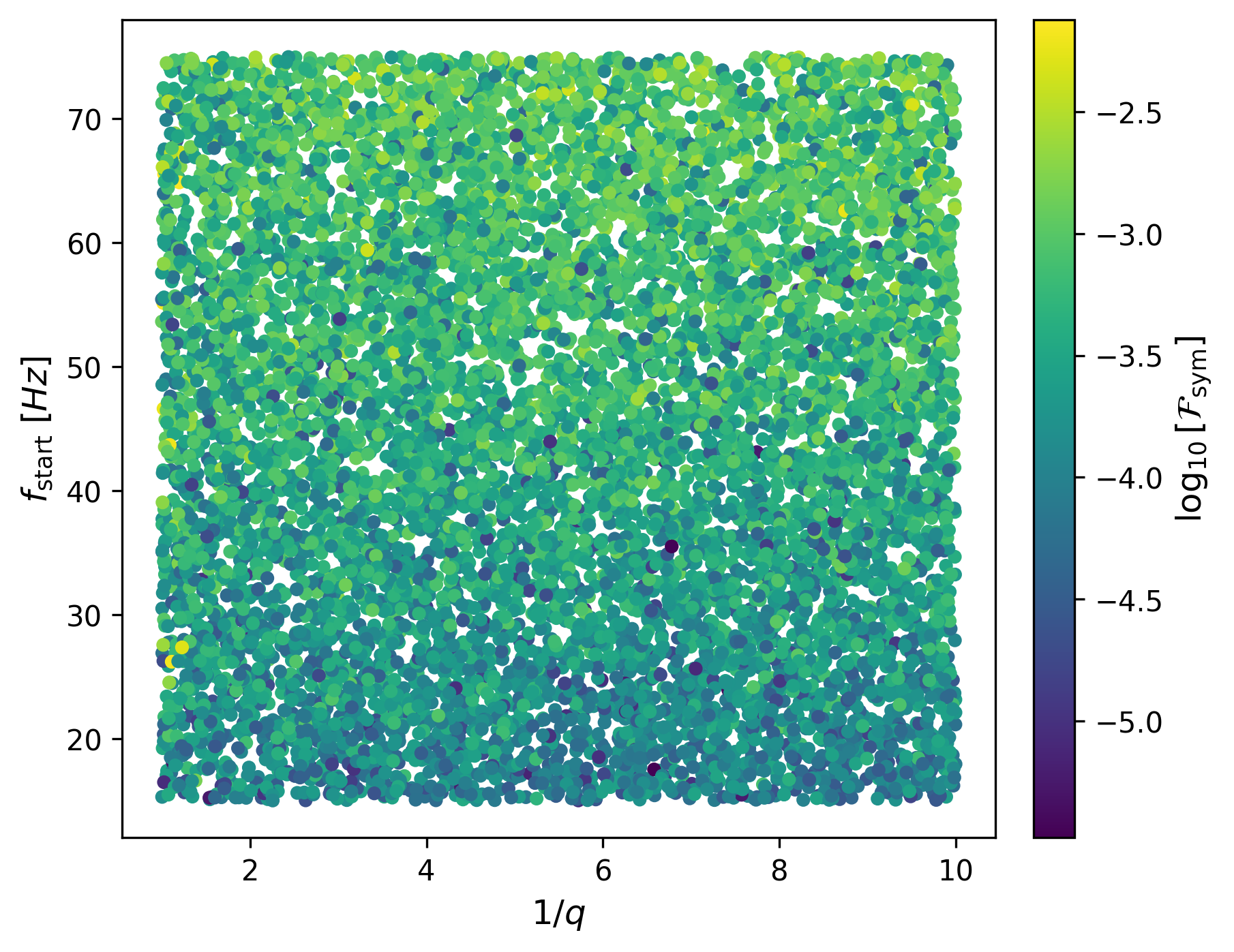}
    \par\vspace{2ex}
    \caption{Scatter plots showing the distribution of $\mathcal{F}_{\mathrm{sym}}$ (color scale) in the $(1/q, \chieff)$ (top panel) and $(1/q, f_{\mathrm{start}})$ (bottom panel) planes. $\mathcal{F}_{\mathrm{sym}}$ is calculated between $10^4$ waveforms generated with \mlgwjax model and the training model \seob.
    }
    \label{fig:mlgw_jax_vs_seobnrv4_mismatch_scatter}
\end{figure}

Figure ~\ref{fig:mlgw_jax_vs_seobnrv4_mismatch_histo} shows the cumulative distribution of $\log_{10}\mathcal{F}_{\mathrm{sym}}$. 
All signals exhibit $\mathcal{F}_{\mathrm{sym}} < 10^{-2.1}$, with a median value of $10^{-\mathrm{3.4}}$. 
In Fig.~\ref{fig:mlgw_jax_vs_seobnrv4_mismatch_scatter}, we display two scatter plots showing the distribution of $\log_{10}\mathcal{F}_{\mathrm{sym}}$ in the $(1/q, \chieff)$ and $(1/q, f_{\mathrm{start}})$ planes, where $\log_{10}\mathcal{F}_{\mathrm{sym}}$ is represented by the color scale.
$\mathcal{F}_{\mathrm{sym}}$ is evenly distributed across the parameter space. The accuracy slightly decreases for nearly equal-mass binaries with large positive $\chieff$ and for shorter signals, i.e., at higher values of $f_{\mathrm{start}}$. 
However, even in these regions, the median value remains below $10^{-3.3}$. 
A similar behavior was already reported in~\cite{mlgw_2}, indicating that it is independent of our \jax implementation.

\begin{figure}[t!]
    \centering
    \includegraphics[width=1.0\linewidth]{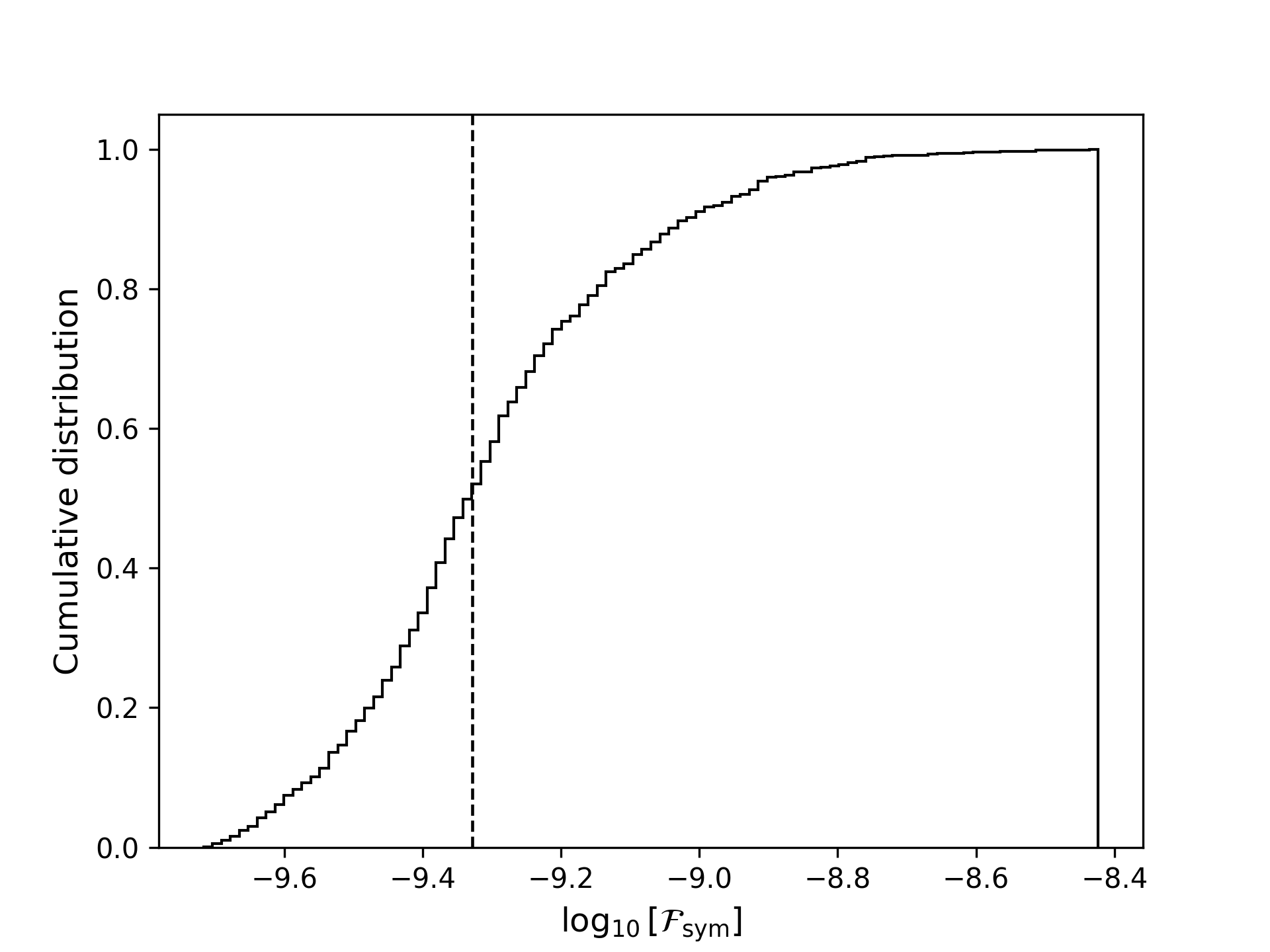}
    \caption{Cumulative distribution of the base-10 logarithm of $\mathcal{F}_{\mathrm{sym}}$ between $10^3$ waveforms generated with the \mlgwjax model and \mlgw. The vertical dashed line represents the median of the distribution.}
    \label{fig:mlgw_jax_vs_mlgw_mismatch_histo}
\end{figure}

The accuracy of \mlgwjax in reproducing the training model is therefore consistent with that obtained in~\cite{mlgw_2} with the original implementation of \mlgw. 
To further validate this point, we perform an additional test in which the two compared sets of waveforms are generated with \mlgwjax and \mlgw, respectively, rather than with \mlgwjax and the target model. 
The waveforms are generated at $10^3$ parameter space points, drawn from the same distributions as in the previous test.
The cumulative distribution of the $\log_{10}\mathcal{F}_{\mathrm{sym}}$ for this second benchmark is shown in Figure ~\ref{fig:mlgw_jax_vs_mlgw_mismatch_histo}.
The base-10 logarithm values of the minimum, median and maximum of the mismatch distribution are $-9.72$, $-9.33$, and $-8.42$, respectively.
This demonstrates that our \jax implementation of the waveform generation interface does not degrade the quality of the model. The two tests therefore probe different quantities. The comparison against \seob in Figs. ~\ref{fig:mlgw_jax_vs_seobnrv4_mismatch_scatter} and ~\ref{fig:mlgw_jax_vs_mlgw_mismatch_histo}  measures the total error of the surrogate pipeline, which combines the approximation error of the trained model with any difference introduced by the re-implementation. The comparison in Fig. ~\ref{fig:mlgw_jax_vs_mlgw_mismatch_histo} isolates the latter, and yields mismatches about six orders of magnitude smaller than the former (median $10^{-9.33}$ against $10^{-3.4}$). The error budget is thus entirely dominated by the surrogate approximation, and the translation to \jax contributes negligibly to it.
%

%
We also perform speed benchmarks to quantify the computational gain achieved by \mlgwjax in terms of waveform evaluation time with respect to the original implementation. 

\begin{figure}[t!]
    \centering
    \includegraphics[width=1.0\linewidth]{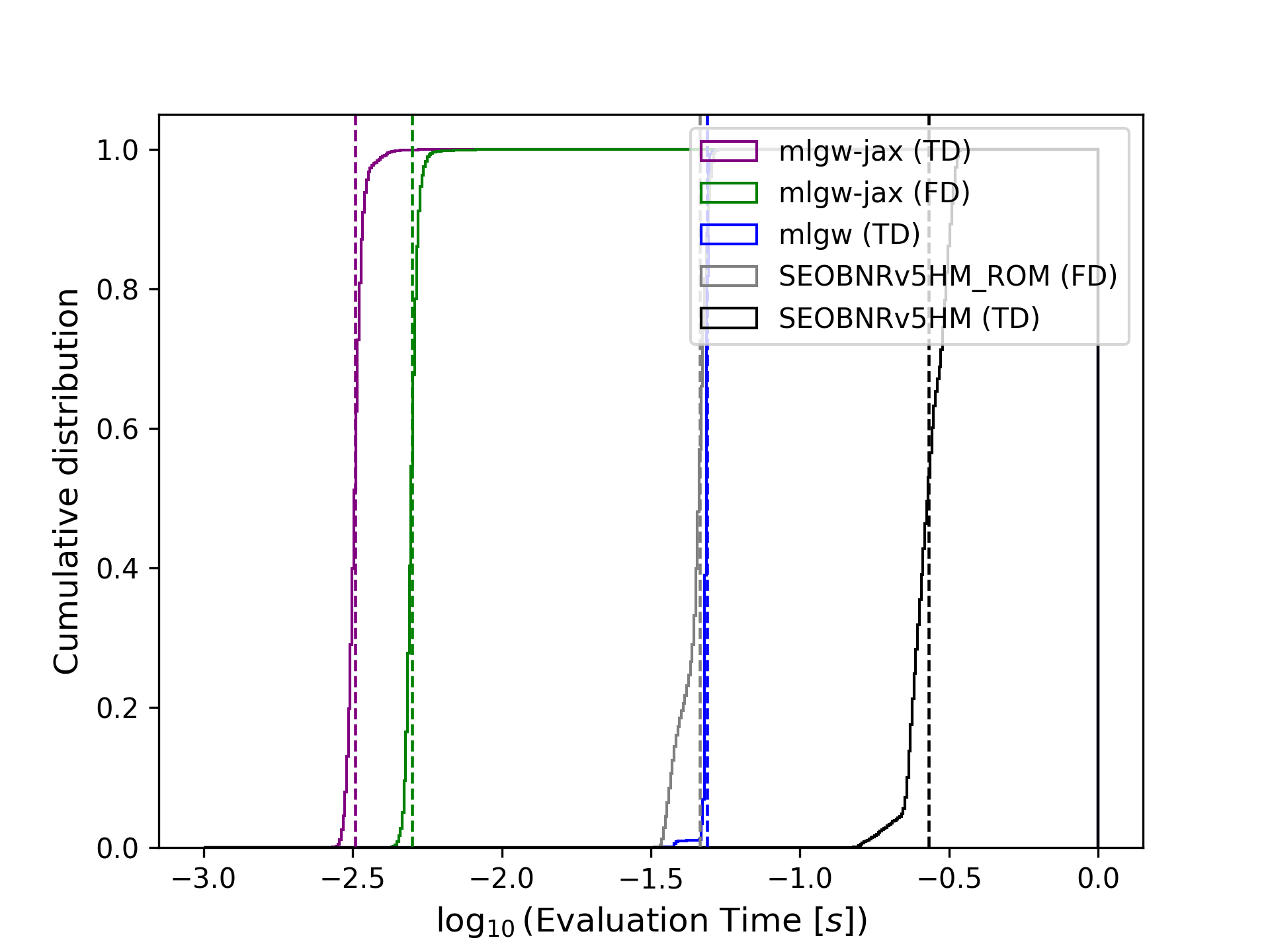}
    \caption{Cumulative distribution of the base-10 logarithm of the evaluation time for the serial generation of waveforms on CPU for five configurations: time-domain (TD) waveform generation with \mlgw (blue lines) and \mlgwjax (purple lines), frequency-domain (FD) waveform generation with \mlgwjax (green lines), all using our surrogate model trained on \seob; frequency-domain waveform generation with the \texttt{SEOBNRv5HM\_ROM} surrogate (grey lines); and time-domain waveform generation with the full model \seob (black lines). For each histogram, the vertical dashed line represents the median of the distribution. Each set is composed of $10^4$ waveforms, generated with a signal duration of ten seconds and a sampling frequency $f_s = 8192 \ \mathrm{Hz}$.
    }
    \label{fig:exec_time_comparison_cumulative_8192}
\end{figure}

As a first benchmark we consider the serial generation on a CPU of a set of $10^4$ waveforms over the same parameter space adopted for the accuracy study.
We compare three configurations: 1) time-domain generation with \mlgw, 2) time-domain generation with \mlgwjax, and 3) frequency-domain generation with \mlgwjax exploiting the new frequency-domain generator.
The waveforms are generated with a signal duration of ten seconds and a sampling frequency $f_s = 8192 \ \mathrm{Hz}$. 
The cumulative distributions of the base-10 logarithm of the evaluation times are shown in Fig.~\ref{fig:exec_time_comparison_cumulative_8192}.
The latter also includes the distributions relative to two additional sets of $10^4$ waveforms: one generated using the widely used frequency-domain surrogate \texttt{SEOBNRv5HM\_ROM} and evaluated with \texttt{LALSuite}~\cite{lalsuite}, the other adopting the full \seob model and evaluated with the Python package \texttt{pySEOBNR}~\cite{Mihaylov:2023bkc}.
This result, as all the other CPU-based benchmarks presented for \mlgwjax, was obtained on an Intel Core i7-10700 CPU%
~\footnote{8 cores, 16 threads at 2.90GHz, with 16 GB of RAM.}. 
The \jax ability of \texttt{JIT} compiling was exploited, with the compilation overhead contribution excluded from the evaluation times reported.
The same has been done for the all the evaluation-time benchmarks in this work.
It is evident that \mlgwjax is significantly faster than \mlgw. 
The average speed-up factor is $\bar{t}_\mlgw/\bar{t}_\mlgwjax \sim 17$, corresponding to a $\sim 1.2$ orders of magnitude improvement.
The speed-up factor is smaller for the frequency-domain generation: $\bar{t}_\mlgw/\bar{t}^{FD}_\mlgwjax \sim 10$. 
This is expected, since the \mlgwjax frequency-domain generator applies a \jax built-in Fast Fourier Transform function to the time-domain output, thereby introducing an additional computational step. 
Nevertheless, the frequency-domain generation still exhibits a speed-up of one order of magnitude.
We also note that \mlgwjax exhibits a comparable speed-up relative to \texttt{SEOBNRv5HM\_ROM}, and is therefore nearly two orders of magnitude faster than the full \seob model.

As a second benchmark, we repeat the test by comparing the \mlgwjax model with the \jax implementation of the IMRPhenomXAS waveform model provided by the \ripple package~\cite{Edwards:2023sak}. 
This time, since IMRPhenomXAS models only the dominant $(\ell, m) = (2, 2)$ mode, we restrict \mlgwjax to the same mode in order to ensure a fair comparison. Accordingly, we reduce the sampling frequency to $f_s = 2048 \ \mathrm{Hz}$. 

\begin{figure}[t!]
    \centering
    \includegraphics[width=1.0\linewidth]{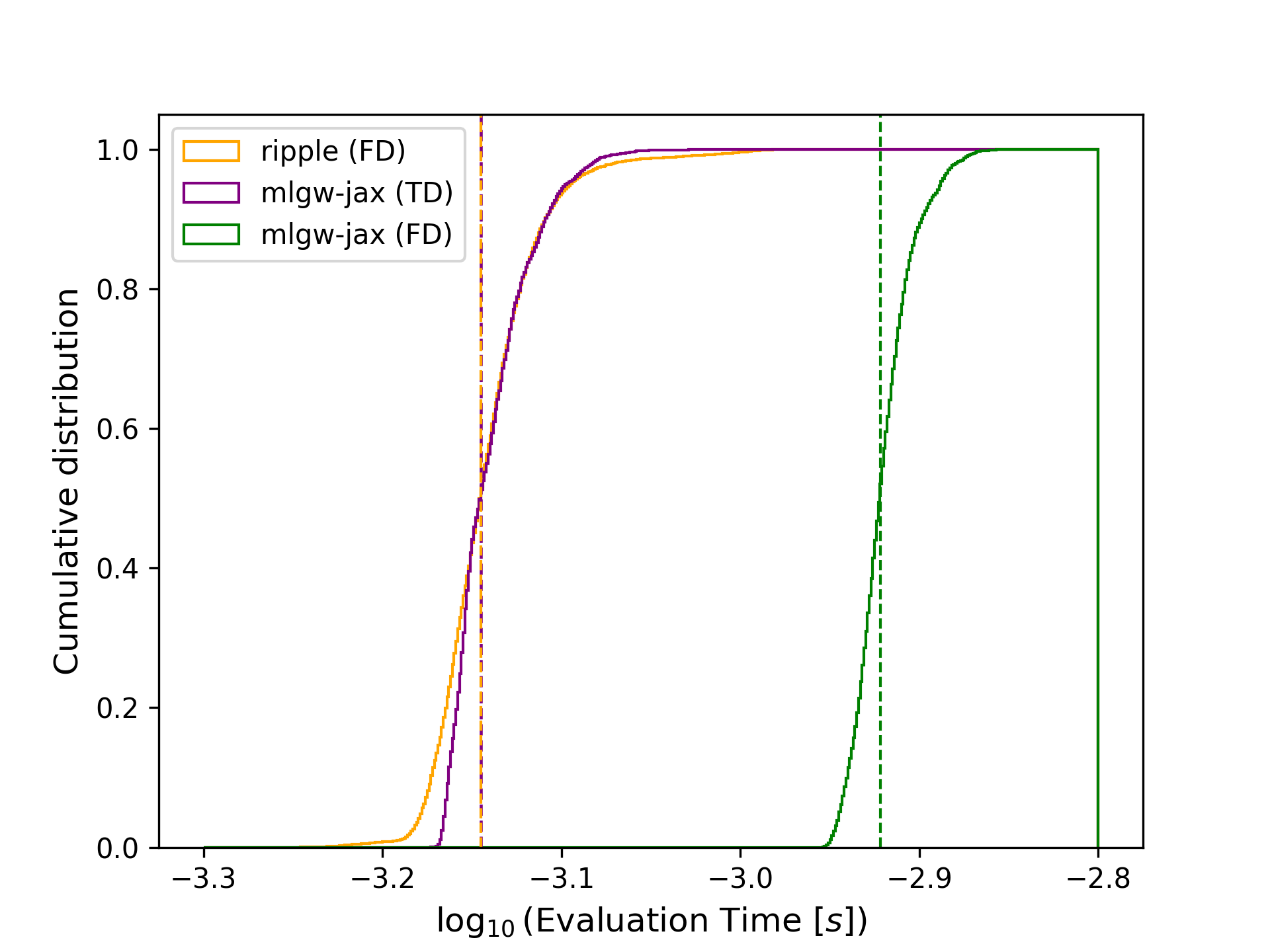}
    \caption{Cumulative distribution of the base-10 logarithm of the evaluation time for the serial generation of waveforms on CPU for three configurations: time-domain (TD) generation with \mlgwjax (purple line), frequency-domain waveform generation with \mlgwjax (green lines), both using our surrogate model trained on \seob; frequency-domain (FD) generation with the implementation of the IMRPhenomXAS model provided by \ripple (orange lines). For each histogram, the vertical dashed line represents the median of the distribution. Each set is composed of $10^4$ waveforms including only the dominant $(\ell, m) = (2, 2)$ mode, generated with a signal duration of ten seconds, and a sampling frequency $f_s = 2048 \ \mathrm{Hz}$.
    }
    \label{fig:exec_time_comparison_cumulative_22}
\end{figure}

The results are shown in Fig.~\ref{fig:exec_time_comparison_cumulative_22}. 
We find that the CPU performance of \mlgwjax is comparable to that of \ripple. The comparison is performed using the native implementations of the two frameworks, namely the frequency-domain generator provided by \texttt{ripple} and the time-domain implementation of \texttt{mlgw-jax}. For completeness and as a reference, we also report the frequency-domain version of \texttt{mlgw-jax}, provided through the \texttt{GW\_FD\_generator} module. Since a time-domain implementation of \texttt{ripple} is not available, a direct comparison in the time domain cannot be performed on equal footing.

%
One of the key advantages of the \jax implementation is the ability to exploit hardware accelerators such as GPUs.
To this end, we repeat the previous benchmarks on a single NVIDIA L40S GPU.
Since the original \mlgw implementation cannot benefit from GPU acceleration, it is not included in this benchmark.
In this case, we exploit \jax automatic parallelization capabilities by applying the \texttt{vmap} transformation to both the \mlgwjax and \ripple waveform generators. 
To fully highlight the benefits of GPU acceleration, we increase the workload by generating a batch of $10^5$ waveforms with each model. 

The average evaluation time per waveform is reported in Tab.~\ref{tab:gpu_comparison}. 
First, we note that \mlgwjax and \ripple exhibit comparable performance also with regards to this benchmark.
Most importantly, GPU acceleration provides an additional order of magnitude of speed-up with respect to the previously reported CPU evaluation time for both \mlgwjax configurations.
Remarkably, generating six additional higher-order modes with \mlgwjax require less than three times the computational time needed for the dominant mode alone.
\begin{table}[t!]
    \centering
    \caption{Average evaluation time per waveform obtained on GPU when generating a batch of $10^5$ waveforms in the frequency-domain with three different configurations. Two use a sampling frequency $f_s = 2048 \ \mathrm{Hz}$ and only include the dominant $(\ell, m) = (2, 2)$ mode (the first two rows, with \ripple and \mlgwjax), while the third one uses a sampling frequency $f_s = 8192 \ \mathrm{Hz}$ and \mlgwjax including all seven available modes, i.e., $(l,m)\in\{(2,2), (2,1), (3,2),(3,3), (4,3), (4,4), (5,5)\}$.
    }
    \renewcommand{\arraystretch}{1.4}
    \begin{tabular}{@{\hspace{0.2cm}}c@{\hspace{0.2cm}}c@{\hspace{0.2cm}}}
        \addlinespace[0.5em]
        \toprule[1.0pt]
        \toprule[1.0pt]
        \multirow{ 2}{*}{Model} & Average \\
         & evaluation time\\ 
        \midrule[1.0pt]
        \ripple (IMRPhenomXAS)             & $ 30\ \mu$s\\
        \mlgwjax $(2,2)$-mode  & $ 28\ \mu$s\\
        \mlgwjax     & $ 86\ \mu$s\\
        \bottomrule[1.0pt]
        \bottomrule[1.0pt]
    \end{tabular}\label{tab:gpu_comparison}
\end{table}
All benchmarks were performed using double-precision arithmetic. We used \jax \texttt{v0.9.0.1} and \texttt{jaxlib} \texttt{v0.9.0.1}. For GPU benchmarks, we used \texttt{CUDA} \texttt{v13.0} and \texttt{NVIDIA} driver \texttt{v580.95.05}. The BBH and BNS benchmarks were carried out independently, on different machines and at different times, and therefore rely on different versions of the \jax stack and of the GPU drivers; the comparisons reported in each case are internally consistent, as each model and its \jax counterpart were always benchmarked within the same software environment.
%
%
\subsection{Binary Neutron Stars}\label{sec:mlgwbns}
%
\mlgwbns is a neural-network-based surrogate model that, through a regression task, learns the mapping between the intrinsic parameters of a BNS source and the $(2,2)$-mode of the corresponding gravitational waveform.
Introduced in Ref.~\cite{mlgw_bns}, the code retains the general training structure of \mlgw, while incorporating several modifications tailored to the specific features of BNS signals.
We briefly summarize these differences in the following, and refer the reader to the original paper for a complete discussion of the technical details.

Firstly, \mlgwbns works directly in the frequency domain to optimize waveform generation in a Bayesian parameter estimation context. Furthermore, unlike \mlgw, the training dataset consists of the residual amplitude and phase between the target waveform approximant and the analytical post-Newtonian (PN) baseline provided by the TaylorF2 model. 
This allows for a simplification of the neural networks involved.
Specifically, the maps targeted by the regression are to the following quantities:
\begin{align}
    \Delta A(f;\bm{\theta}) &= \log\!\frac{A_{\rm target}(f;\bm{\theta})}{A_{\rm PN}(f;\bm{\theta})} \label{eq:resA}\\
    \Delta\phi(f;\bm{\theta}) &= \phi_{\rm target}(f;\bm{\theta}) - \phi_{\rm PN}(f;\bm{\theta})\,, \label{eq:resphi}
\end{align}
where $\bm{\theta} = \{q, \chi_{1z}, \chi_{2z}, \Lambda_1, \Lambda_2\}$ is the intrinsic parameters vector, with $\Lambda_1, \Lambda_2$ the dimensionless tidal deformability parameters of the two neutron stars.
Once a surrogate is trained, the predicted residuals are recombined with the PN baseline to reconstruct the full waveforms. 

As for \mlgwbns, the dimensionality of the residual representation is reduced to facilitate the regression task. This is achieved in three stages. 
1) A reduction of the number of points in the frequency grid on which the training dataset of residuals are evaluated, achieved by applying a multibanding approach; 2) a dataset-dependent greedy downsampling; and 3) a PCA applied to the downsampled residuals. Those final principal components are the outputs of the neural network regression model.

\subsubsection{\jax implementation}
The \mlgwbns code relies on the \texttt{scikit-learn} library~\cite{scikit} for the neural network implementation, and on \texttt{scipy}~\cite{scipy} for the interpolation of the waveform from the downsampled frequency grid to the one specified by the user.
Following the same philosophy adopted for \mlgwjax, we translated the waveform generation interface of the model into \jax.
Similarly to Sec.~\ref{sec:mlgw_jax_implementation}, we describe here the key technical steps of our re-implementation.

%
%
%
All the neural-network feed-forward operations, PCA reconstruction, and TaylorF2 baseline evaluation were re-written using \jax built-in functions, while preserving the double-precision accuracy required for GW applications.

Since standard \texttt{scipy} interpolation routines are not \texttt{JIT}-traceable, we implemented a fully differentiable cubic-spline interpolator using \jax-native operations instead. Particular care was devoted to optimizing this operation for GPU execution, removing the main computational bottleneck of the original implementation. Further details about the implementation of the cubic-spline interpolator are presented in the appendix \ref{app:binning_marginalization}. The resulting code can be found in the following repository: \href{https://github.com/saulo-albuquerque-phys/mlgw_bns_jax}{\texttt{github.com/saulo-albuquerque-phys/mlgw-bns-jax}}. 
\subsubsection{Speed and accuracy tests}\label{sec:bns_timing_mismat}
We benchmark also \mlgwbnsjax in terms of both waveform accuracy and computational performance.
For all tests, we employ the \mlgwbns surrogate model trained on \teobresums and introduced in Ref.~\cite{mlgw_bns}.
To simplify the notation, similarly to the BBH case, throughout the remainder of this paper any results presented for \mlgwbns and \mlgwbnsjax are understood to refer to this specific surrogate model.

\begin{figure}[t!]
    \centering
    \includegraphics[width=1.0\linewidth]{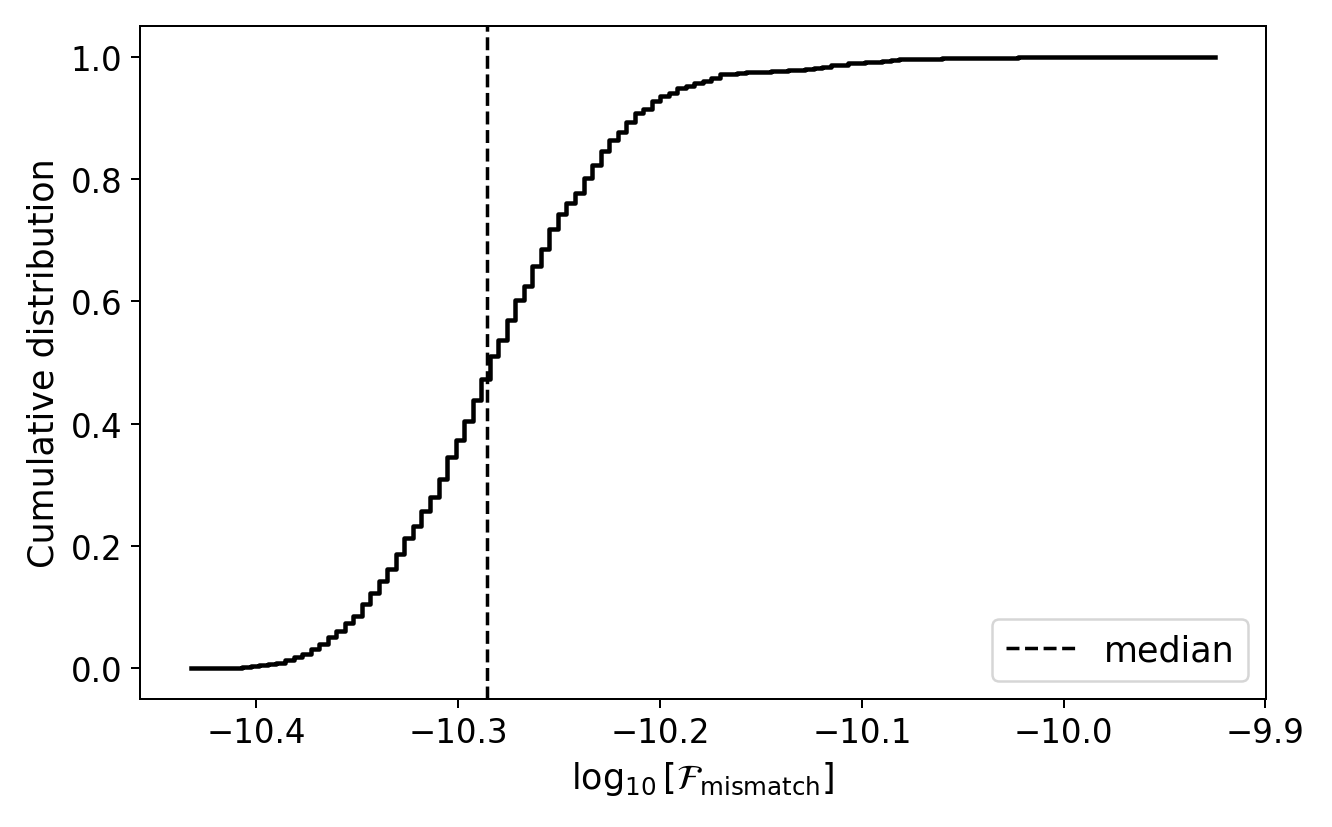}
    \caption{Cumulative distribution of the base-10 logarithm of the mismatch between two sets of $10^3$ waveforms generated with \mlgwbnsjax and \mlgwbns. The vertical dashed line represents the median of the distribution. Since waveforms are calculated directly in the frequency domain, the mismatch is reported here, rather than the \textit{symphony} mismatch introduced in Sec.~\ref{sec:BBH-speed-accuracy}.  }
    \label{fig:mismatch_mlgwbnsjax}
\end{figure}

To verify that the \jax implementation does not introduce additional numerical inaccuracies with respect to the original \mlgwbns code, we compute the mismatch between a set of $10^3$ waveforms generated by \mlgwbnsjax and an equivalent set generated with \mlgwbns. 
The source parameters are randomly drawn within the ranges used for the training, namely, $M\in \mathcal{U}[2.4, 3.1] \ M_\odot$, $q \in \mathcal{U}[0.5, 1.0]$, $\{\chi_{1z}, \chi_{2z}\} \in \mathcal{U}[-0.5, 0.5]$, $\{\Lambda_1, \Lambda_2\} \in [5, 5000]$. For the extrinsic parameters, we draw $D_L\in\mathcal{U}[20, 120]\ \textrm{Mpc}$ and $\cos \iota \in \mathcal{U}[-1, 1]$. 
Since \mlgwbnsjax is directly calculated in the frequency domain, and only the quadrupole mode is considered here, the \textit{symphony mismatch} formula in Eq.~\eqref{eq:Msym} can be reduced to the ordinary mismatch formula $\mathcal{F}\equiv 1- \big|\langle {\hat h}_{\rm jax}, {\hat h}_{\rm orig}\rangle\big|$, where ${\hat h}_{\rm jax}$ and ${\hat h}_{\rm orig}$ are the normalized \mlgwbnsjax and \mlgwbns frequency domain templates, respectively. 
Figure \ref{fig:mismatch_mlgwbnsjax} reports the results of this comparison.  The base-10 logarithm values of the minimum, median and maximum of the mismatch distribution are $-10.43$, $-10.28$ and $-9.92$, respectively.
This confirms that the waveforms generated by \mlgwbnsjax are numerically identical to those produced by \mlgwbns.
Consequently, the surrogate approximation error with respect to the target model \teobresums is preserved and remains consistent with the one reported in Ref.~\cite{mlgw_bns}.\footnote{The authors of Ref.~\cite{mlgw_bns} reported a median mismatch of $\sim\!10^{-5}$ between the surrogate and the target model.}

\begin{figure}[t!]
    \centering
    \includegraphics[width=1.0\linewidth]{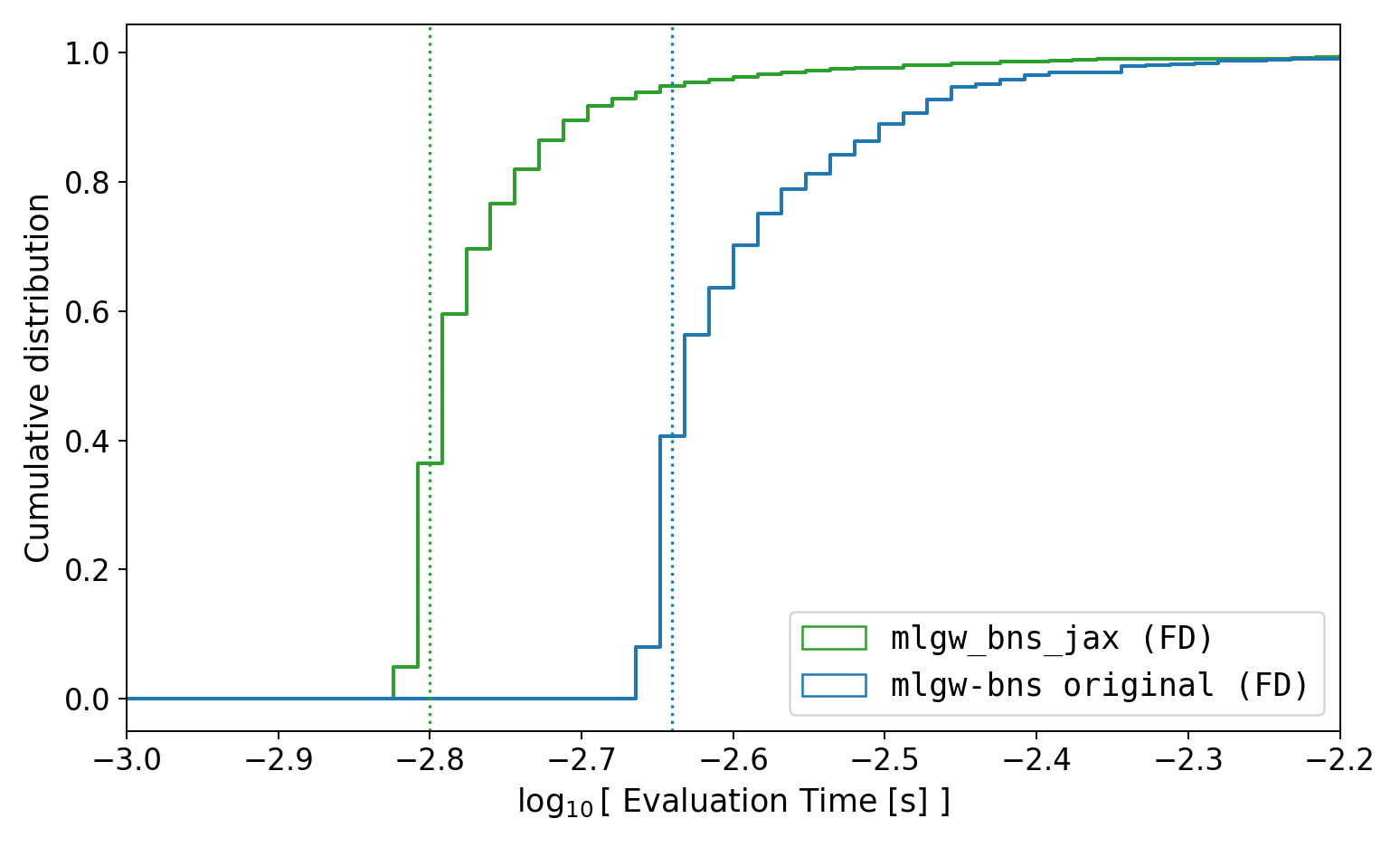}
    \caption{Cumulative distribution of the base-10 logarithm of the evaluation time for the serial generation of waveforms on CPU for two configurations: frequency-domain generation with \mlgwbns (blue line) and \mlgwbnsjax (green line). Each set is composed of $2 \times 10^3$ waveforms, generated with a signal duration of four seconds, and a sampling frequency $f_s = 2048 \ \mathrm{Hz}$.
\label{fig:exec_time_comparison_cumulative_mlgwbns}}
\end{figure} 

All benchmarks were performed using double-precision arithmetic. CPU benchmarks used \texttt{jax} \texttt{v0.4.30} and \texttt{jaxlib} \texttt{v0.4.30}. GPU benchmarks used \texttt{jax} \texttt{v0.4.31}, \texttt{jaxlib} \texttt{v0.4.31}, \texttt{CUDA} \texttt{v13.0}, and \texttt{NVIDIA} driver \texttt{v580.82.07}.

%




%

To assess computational performance, we consider the serial CPU generation of $2\times10^3$ waveforms over the same parameter space used for the accuracy study.
Two waveform sets are generated, one with \mlgwbns, one with \mlgwbnsjax, with a signal duration of four seconds and a sampling frequency $f_s = 2048 \ \mathrm{Hz}$.
The benchmark was performed on an Intel Core i5-8259U CPU~\footnote{4 cores at 2.3 GHz, with 8 GB of RAM.}, exploiting \texttt{JIT} compilation whenever applicable.
Figure \ref{fig:exec_time_comparison_cumulative_mlgwbns} provides the comparison between the cumulative distribution of the base-10 logarithm of the evaluation time with the two configurations.

\mlgwbnsjax is faster than \mlgwbns, with a median speed-up factor of $\bar{t}_\mlgwbns/\bar{t}_\mlgwbnsjax \sim 1.45$, corresponding to a gain of approximately $0.2$ orders of magnitude.
The improvement in serial CPU performance is therefore more modest than that observed for the BBH case in Sec.~\ref{sec:BBH-speed-accuracy}, likely reflecting the higher degree of optimization already present in the original \mlgwbns implementation.
However, substantially larger gains are achieved when exploiting batched generation on GPU architectures. In fact, on a single NVIDIA A100 device, we measure an average evaluation time per waveform of approximately $3\ \mu$s for a batch size of $3 \times 10^5$ waveforms with sampling frequency $f_s = 2048 \ \mathrm{Hz}$ and a duration of $T=4\ $s.
This corresponds to a speed-up of approximately $2.7$ orders of magnitude relative to the serial CPU evaluation. This batch size is the largest that fits in the $80$ GB of VRAM of the device; beyond it, the memory required to hold the full frequency-domain strain of every template in the batch exceeds the available memory.

\section{Example Application: Boosting Gravitational-Wave Inference}\label{sec:GW_inf}
%
We now consider our \jax-based surrogate waveform framework in the broader context of achieving  rapid and fully reliable GW inference. 
A long-term goal of advancing GW inference infrastructures is to enable and enhance multimessenger observations and follow-up campaigns~\cite{LIGOScientific:2017ync} by determining in low-latency not only the sky location of the source~\cite{bayestar} but also additional source parameters of astrophysical relevance, such as the component masses.

GW parameter estimation is carried out by sampling the \textit{posterior} distribution $p(\bm{\theta}|s)$, namely the probability of the parameters $\bm{\theta}$ given that the data $s$ is observed, according to Bayes' Theorem: 
\begin{equation}\label{bayes}
    p(\bm{\theta}|s)=\frac{\mathcal{L}(s|\bm{\theta})\pi(\bm{\theta})}{\mathcal{Z}}\,,
\end{equation}
where $\pi(\bm{\theta})$ denotes the \textit{prior} distribution, representing the knowledge we have about the parameters $\bm{\theta}$ before the data $s$ is collected,  $\mathcal{L}(s|\bm{\theta})$ is the \textit{likelihood} function quantifying the probability of recording the data $s$ if it contains a GW signal with parameters $\bm{\theta}$,  and the denominator $\mathcal{Z} = \int d\bm{\theta} \mathcal{L}(s|\bm{\theta}) \pi(\bm{\theta})$ is a normalization constant called \textit{evidence} that guarantees that the posterior is a properly normalized probability distribution. 

To apply Eq.~\eqref{bayes} to GW data analysis, the likelihood needs to be evaluated repeatedly over the parameter space, for a large number of vectors $\bm{\theta}$.
Assuming the noise of the detectors is Gaussian and stationary,
\begin{equation} \label{eq_likelihood}
    \mathcal{L}(s|\bm{\theta}) \propto \exp\left(-\frac{1}{2}\langle s-h(\bm{\theta}) |  s-h(\bm{\theta})\rangle\right) \ ,
\end{equation}
where the inner product $\langle a,b\rangle$ is defined in Eq.~\eqref{eq:overlap}.

It is evident from Eqs.~\eqref{bayes} and \eqref{eq_likelihood} that sampling the posterior requires multiple evaluations of the likelihood and therefore of waveform templates. Hence, the waveform evaluation time can easily become a bottleneck of this process. 
The other major contributor to the overall computational cost of GW inference is the sampling algorithm itself, but this is beyond the scope of this paper.

As an example of the benefits of the waveform acceleration features explored in this work, and in order to further validate them, we perform parameter-estimation analyses of the GW150914~\cite{LIGOScientific:2016aoc, LIGOScientific:2016vlm} and GW170817~\cite{LIGOScientific:2017vwq, LIGOScientific:2018hze} events generating waveform templates with \mlgwjax and  \mlgwbnsjax, respectively. In order to do so, we provide a general \jax interface between \jax-based waveform generators, including ours, and \blackjaxns, a \jax sampling algorithm that implements Nested Sampling~\cite{2006:Skilling} within the \blackjax~\cite{cabezas2024} modular library of Bayesian inference algorithms implemented in \jax.  
Specifically, we employ the acceptance-walk nested sampler implemented in the \texttt{blackjax\_ns\_gw} package~\cite{yallup2025nested, Prathaban:2025qgg}.
This interface goes under the name \gwgpujax and is available at \href{https://github.com/saulo-albuquerque-phys/gwgpu_jax}{\texttt{github.com/saulo-albuquerque-phys/gwgpu\_jax}}. Its main features are the following.
\begin{itemize}
    \item Users can generate waveforms with any surrogate model currently available in our \jax surrogate modeling framework, as well as with waveform models provided by \ripple. 
    \item It contains a built-in likelihood function in \jax that is \texttt{JIT} compatible and makes use of \texttt{vmap}.
    \item Interferometer data, such as any strain data from the Gravitational-Wave Open Science Center (GWOSC), may be imported along with the geometrical configuration of the detectors.
    \item Our \jax GW likelihood is connected to the \blackjaxns application programming interface, allowing for fast and accurate parameter estimation with GPU-accelerated performance. 
\end{itemize}


\subsection[GW150914]{Binary Black Hole Application: GW150914}
%
%
For the GW150914 analysis, we download 32 seconds of inteferometer data around the event time for the LIGO Hanford (H1) and LIGO Livingston (L1) detectors. This full segment length is used to estimate the Welch power spectral density. For the parameter estimation, we use a smaller segment length of 4 seconds, with the merger time exactly $3.5$ seconds after the initial time.  The sampling rate is fixed to $4096 \, \mathrm{Hz}$. 
The priors and sampler settings adopted in the analysis are summarized in Tab.~\ref{tab:pe_params}.
The analysis is performed on a single NVIDIA A100 GPU and \emph{takes approximately 12 minutes}. 

\begin{figure}[t!]
    \centering
    \includegraphics[width=1.0\linewidth]{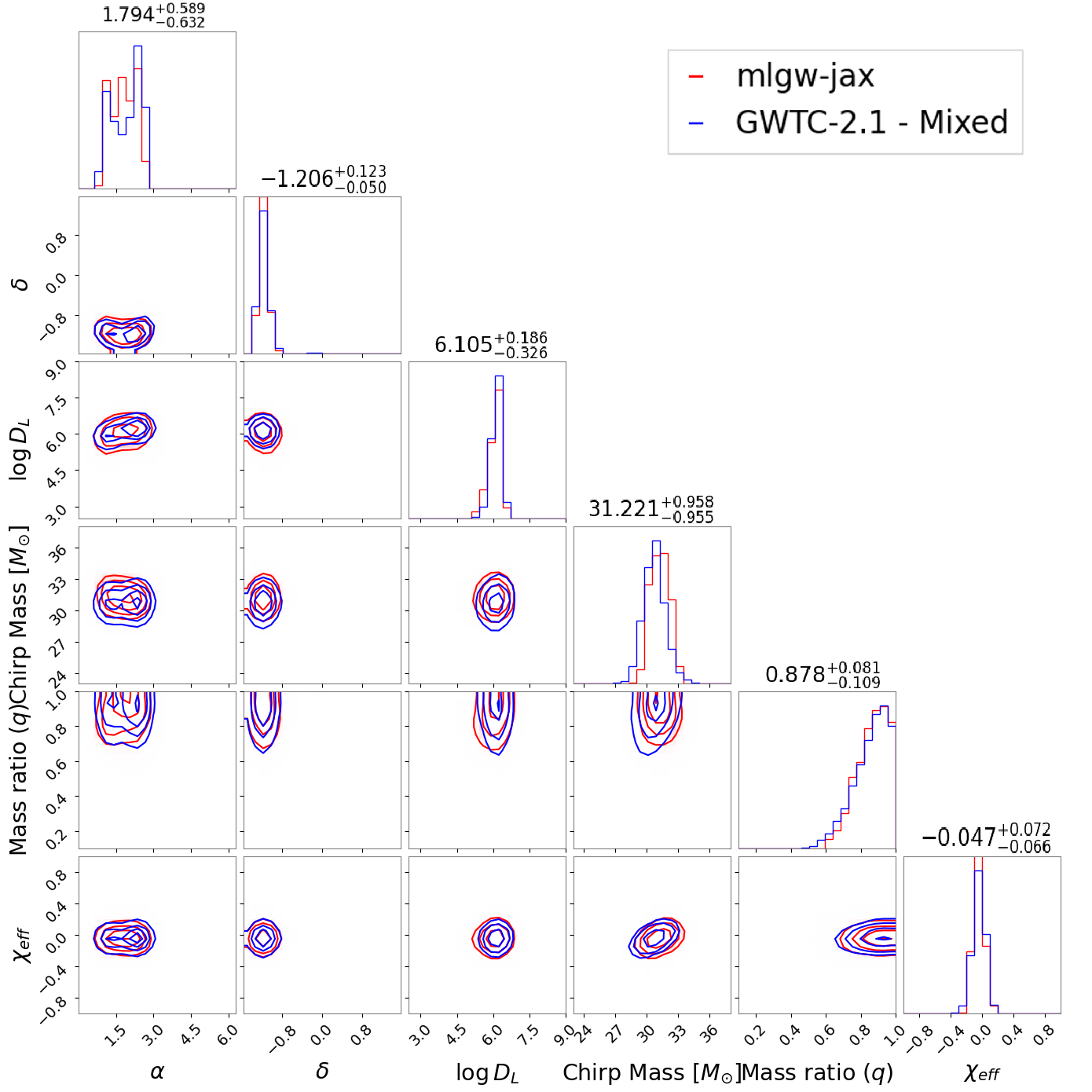}\label{corner}
    \caption{Corner plot comparison between the posterior samples for GW150914 that we obtained with \blackjaxns and \mlgwjax (in red), and those publicly available from the GWTC-2.1 catalog (in black), produced with \texttt{Bilby}~\cite{bilby} and the precessing waveform models IMRPhenomXPHM and SEOBNRv5PHM (details in~\cite{gwtc-2.1}).
    The selected set of parameters shown includes  the right ascension, the declination, the logarithm of the luminosity distance, chirp mass, mass ratio, and effective spin parameter. 
    }
    \label{fig:corner}
\end{figure}

Figure \ref{fig:corner} compares the posterior samples obtained from our analysis of GW150914 with those from GWTC-2.1~\cite{gwtc-2.1} 
and publicly available on GWOSC\footnote{\url{https://gwosc.org/eventapi/html/GWTC-2.1-confident/GW150914/v4/}.}.
We find very good agreement between the two posterior distributions, bearing in mind that, unlike the LVK analysis, our analysis with the \mlgwjax surrogate trained on SEOBNRv5HM is restricted to aligned spins.  When precession effects are small, as in this case, neglecting them typically produces a small bias in $\theta_{JN}$, the angle between the total angular momentum and the line of sight. Since the amplitude of the signal depends on $\theta_{JN}$ and on the luminosity distance through a nearly degenerate combination, and since the relative amplitudes of the subdominant multipoles depend on both $\theta_{JN}$ and the mass ratio, such a bias does not remain confined to the viewing angle but propagates primarily into $D_L$ and, to a lesser extent, into $q$ and the aligned spins \cite{degeneracy_1,degeneracy_2,degeneracy_6}.

\begin{table}[t!]
    \centering
    \renewcommand{\arraystretch}{1.2}
    \begin{tabular}{l@{\hspace{0.51cm}}c@{\hspace{0.6cm}}c@{\hspace{0.21cm}}}
        \toprule[2.0pt]
        \toprule[1.0pt]
        Parameter & Prior Classification & Prior Range \\
        \midrule[1.0pt]
        $\alpha$ & uniform &$[0, 2\pi]$  \\
        $\delta$ & $\cos$ & $[-\pi/2, \pi/2]$ \\
        $\ln D_{L}$ & volumetric & $[5, 6.55]$   \\
        $\theta_{JN}$ & $\sin$ &$[0, \pi]$\\
        $\phi_c$ & uniform & $[0, 2\pi]$ \\
        $\psi$ & uniform & $[0, \pi]$  \\
        $\mathcal{M}$ & uniform  & $[26, 33] \ M_\odot$  \\
        $q$ & uniform & $[0.6, 1.0]$  \\
        $t_{c}$ & uniform & $[-0.1, 0.1]\ $s  \\
        $\chi_{1}$ & uniform & $[-0.4, 0.4]$\\
        $\chi_{2}$ & uniform & $[-0.4, 0.4]$  \\
        \bottomrule[1.0pt]
    \end{tabular}
    \begin{tabular}
{@{\hspace{0.2cm}}l@{\hspace{0.8cm}}l@{\hspace{0.2cm}}}
    \toprule[1.0pt]
    \blackjaxns configuration parameter & Value \\
    \midrule[1.0pt]
        Number of live particles & 2500 \\
        Batch deleted particles & 125 \\
        Duration & 4 s \\
        Sampling Rate & 4096 Hz \\
        \bottomrule[1.0pt]
        \bottomrule[2.0pt]
    \end{tabular}
    \caption{Prior distributions (upper panel), and sampler data settings (lower panel) of the parameter-estimation analysis performed on GW150914 with \blackjaxns using the \mlgwjax waveform surrogate model. From top to bottom, the parameters reported in the upper table are: the right ascension angle $\alpha$; the declination angle $\delta$; the logarithm of the luminosity distance $\ln{D_{L}}$; the angle between the total angular momentum and the line of sight $\theta_{JN}$ (in the absence of precession $\theta_{JN}$ coincides with $\iota$); the phase at coalescence $\phi_c$; the polarization angle $\psi$; the chirp mass $\mathcal{M}$; the mass ratio $q \equiv m_2/m_1$; the coalescence time $t_{c}$; and the aligned spins $\chi_{1}$ and $\chi_{2}$.  The priors denoted as ``$\cos$'' and ``$\sin$'' are uniform in sine and cosine, respectively.}
    \label{tab:pe_params}
\end{table}

%
%
\subsection[GW170817]{Binary Neutron Star Application: GW170817}\label{sec:mlgwbnspe}
%
We analyze the BayesWave-deglitched data for GW170817. We download 1024\,s of strain data sampled at 4096\,Hz for each of the three detectors: LIGO Hanford (H1), LIGO Livingston (L1), and Virgo (V1). From these data, we extract an analysis segment of duration $T = 128$\,s  with merger time at approximately $t_{c} = 112$\,s.  

The priors and sampler settings of this analysis are summarized in Tab.~\ref{tab:pe_params_2}.
Unlike several previous analyses of GW170817, as in~\cite{mlgw_bns}, we treat the right ascension and declination as free parameters and do not fix the sky-position to that of the known electromagnetic counterpart.
To reduce the computational cost, we employ relative binning and an analytical marginalization over the coalescence time and phase; Appendix \ref{app:binning_marginalization} provides additional technical details on these technicques within our implementation.
\emph{The analysis completed in approximately 17 minutes} on a single NVIDIA A100 GPU.

\begin{figure}
\centering
\includegraphics[width=\columnwidth]{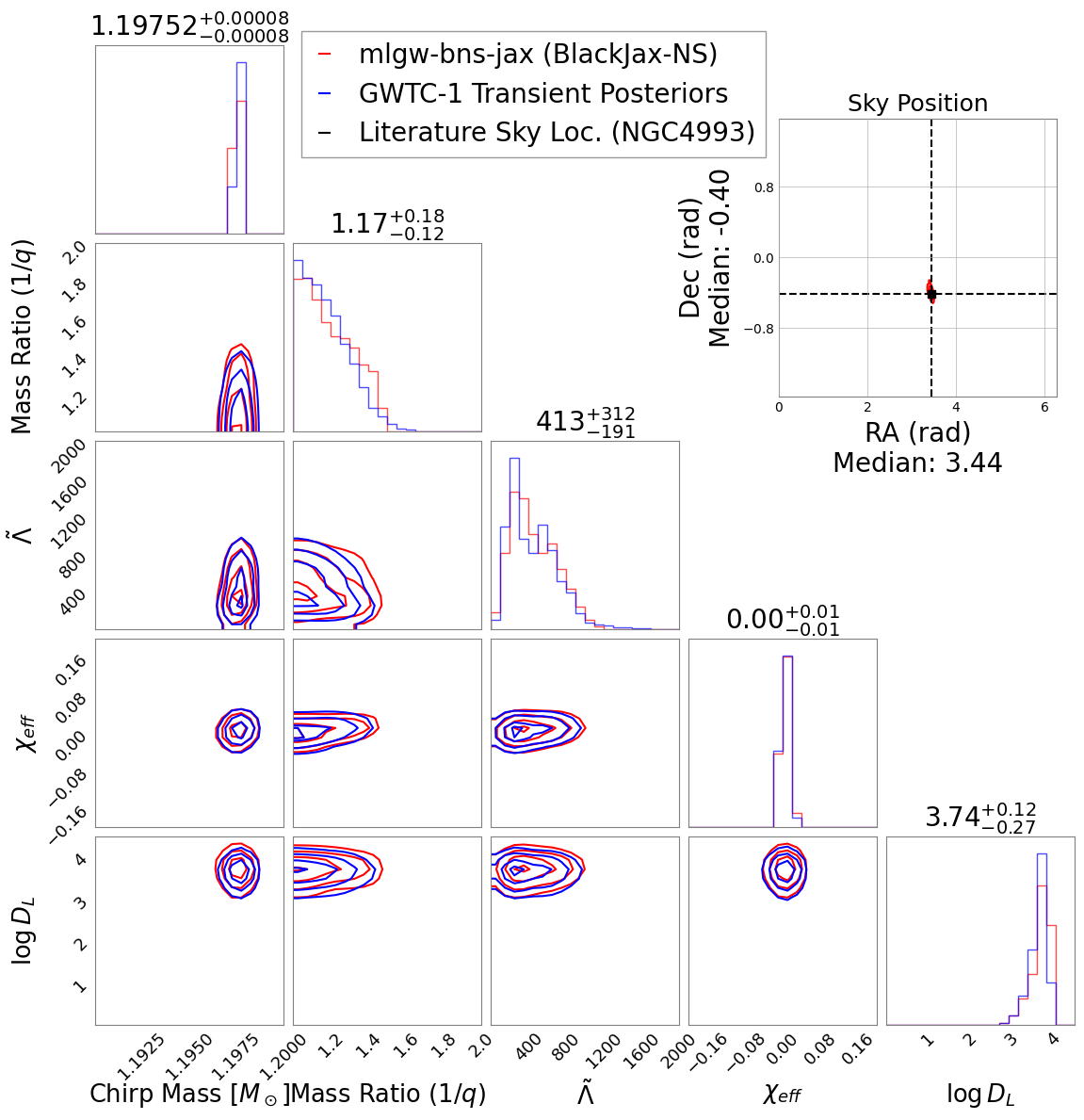}
\caption{Corner plot comparison between the posterior samples for GW170817 that we obtained with \blackjaxns and \mlgwbnsjax (in red), and those from the GWTC-1 catalog(in black), produced with \texttt{Bilby}~\cite{bilby} and the precessing waveform models IMRPhenomXPHM and SEOBNRv5PHM (details in~\cite{gwtc-1}. The selected set of parameters shown includes the chirp mass, the inverse of the mass ratio, the viewing angle comprised between the total angular momentum and the line of sight, the mass-weighted binary tidal deformability parameter which can be seen in Ref.~\cite{mlgw_bns}, and the effective spin parameter. The inset in the upper right shows the 2D histograms for the sky-location angles inferred by our parameter estimation.} 
\label{fig:BNS_PE}
\end{figure}

The corner plot in Fig.~\ref{fig:BNS_PE} compares the posterior distributions obtained in our analysis with those reported in the GWTC-1 catalog~\cite{gwtc-1} and publicly available through the GWOSC.\footnote{\url{https://gwosc.org/eventapi/html/GWTC-1-confident/GW170817/v3/}} 
The inset in the upper-right corner shows the joint posterior distribution of the sky-location angles.
Overall, we find good agreement.
Importantly, despite adopting agnostic priors on the sky-location parameters, the recovered posterior yields a 90\% credible region of approximately 23 square degrees fully consistent with the position of the host galaxy NGC~4993.
The posterior median values for the right ascension and declination ($3.45$ and $-0.41$, respectively) are in full agreement with the ones reported in literature~\cite{Coulter:2017wya}.

%
%
\begin{table}[t!]
    \centering
    \renewcommand{\arraystretch}{1.2}
    \begin{tabular}{l@{\hspace{0.8cm}}c@{\hspace{0.6cm}}c@{\hspace{0.19cm}}}
        \toprule[2.0pt]
        \toprule[1.0pt]
        Parameter & Prior Classification & Prior Range \\
        \midrule[1.0pt]
        $\alpha$ & uniform & $[0, 2\pi]$  \\
        $\delta$ & $\cos$ & $[-\pi/2, \pi/2]$ \\
        $\ln D_{L}$ & volumetric & $[1.6, 4.17] $  \\
        $\theta_{JN}$ & $\sin $ & $[0, \pi]$\\
        $\psi$ & uniform & $[0, \pi]$  \\
        $\mathcal{M}$ & uniform & $[1.9, 2.0] \ M_\odot$  \\
        $q$ & uniform &$[0.68, 1.0]$  \\
        $\chi_{1}$ & uniform & $[-0.2, 0.2]$\\
        $\chi_{2}$ & uniform & $[-0.2, 0.2]$  \\
        $\Lambda_{1}$ & uniform & $[5, 1200]$\\
        $\Lambda_{2}$ & uniform & $[5, 1200]$  \\
        \bottomrule[1.0pt]
    \end{tabular}
    \begin{tabular}
        {@{\hspace{0.2cm}}l@{\hspace{0.8cm}}l@{\hspace{0.2cm}}}
        \toprule[1.0pt]
        \blackjaxns configuration parameter & Value \\
        \midrule[1.0pt]
        Number of live particles & 2500 \\
        Batch deleted particles & 125 \\
        Duration & 128 s \\
        Sampling Rate & 4096 Hz \\
        \bottomrule[1.0pt]
        \bottomrule[2.0pt]
    \end{tabular}
    \caption{Prior distributions (upper panel), and sampler data settings (lower panel) of the parameter-estimation analysis performed on GW170817 with \blackjaxns using the \mlgwbnsjax waveform surrogate model. The notation follows the one in Table \ref{tab:pe_params}, with the addition of the dimensionless tidal parameters $\Lambda_1$ and $\Lambda_2$.}
\label{tab:pe_params_2}
\end{table}

%
%

\section{Conclusions}\label{sec:Conclusions}
%
In this work, we presented a surrogate-modeling framework obtained by translating into the \jax ecosystem the waveform-generation interfaces of the machine-learning-based software \mlgw~\cite{mlgw_1,mlgw_2} and \mlgwbns~\cite{mlgw_bns}, with the names \mlgwjax and \mlgwbnsjax, respectively.
By retaining the original training infrastructure while replacing the waveform-generation backend, the proposed framework enables the construction of fully \jax-compatible surrogate models for non-precessing BBH time-domain and BNS frequency-domain waveform approximants, provided that a representative train set of waveforms can be generated from the target model.
The resulting surrogates benefit from \jax capabilities including just-in-time (\texttt{JIT}) compilation, vectorized execution, hardware acceleration, and automatic differentiation.

By means of dedicated accuracy tests, we demonstrated that our waveform-generation interfaces \mlgwjax and \mlgwbnsjax reproduce the predictions of the trained surrogate models with the same accuracy level of the original \mlgw and \mlgwbns implementations. This confirms that the translation to \jax does not degrade waveform generation through the introduction of additional numerical inaccuracies.
We performed all benchmarks involving \mlgwjax using a newly trained surrogate of the \seob model, whereas for \mlgwbnsjax we adopted the \teobresums surrogate already introduced in the seminal \mlgwbns paper~\cite{mlgw_bns}.

Computational performance benchmarks on both CPU and GPU architectures demonstrated the benefits of the proposed framework.
\mlgwjax reduces the waveform evaluation time by approximately $1.2$ orders of magnitude with respect to the original \mlgw implementation on CPU. 
When leveraging GPU acceleration and large-batch vectorization, the speed-up increases to nearly two orders of magnitudes, reaching generation times of order $10^{-5}\ $s per waveform for batches of $10^5$. 
For \mlgwbnsjax, the gain on CPU is more modest, corresponding to approximately $0.2$ orders of magnitude relative to \mlgwbns.
On GPU, however, we reach an average evaluation time per waveform of order $10^{-6}\ $s for batches of $3 \times 10^5$, corresponding to a speed-up of almost three orders of magnitude relative to the serial CPU evaluation.
Overall, these results highlighted both the scalability of the implementation and its suitability for high-throughput inference applications.

Finally, we demonstrated the use of the proposed framework in combination with the \jax-based implementation of the nested sampling algorithm provided by \blackjaxns, in the context of achieving rapid and accurate GW inference algorithms.
As proof-of-concept applications, we performed a parameter-estimation analysis of the GW150914 and GW170817 events using, respectively, \mlgwjax and \mlgwbnsjax to generate waveform templates.
The resulting posterior distributions are in good agreement with those reported in the GWTC catalogs released by the LVK Collaboration, while requiring a total run time of approximately 12 minutes and 17 minutes for the BBH and BNS analysis, respectively.
Remarkably, despite adopting agnostic priors on the sky-location parameters, our analysis successfully recovers a 23 square degrees portion of the sky that comprises the host galaxy NGC~4993.

These results show that, when coupled with \jax-based sampling algorithms, the proposed framework enables efficient Bayesian inference while retaining the physical fidelity of the surrogate models within their domain of validity, namely non-precessing BBH systems and dominant-mode BNS signals, producing results in a timely fashion that would be beneficial for low-latency multimessenger follow-up campaigns during upcoming LVK observing runs, as well as for meeting the high computational demands posed by next-generation GW-detectors science~\cite{ET3}.

Moreover, we emphasize that, in addition to pure fast waveform evaluation, our framework's capability of providing efficient gradient and Hessian evaluations of the surrogates via \jax automatic differentiation makes it particularly suited for gradient-based Bayesian inference algorithms.
In this direction, we report that we already successfully ran end-to-end analyses with \mlgwjax in combination with \sharpy~\cite{sharpy}, a \jax-compatible gradient-based sampler previously introduced by the authors of this work. 

More broadly, the combination of surrogate modeling with differentiable, and hardware-accelerated computation, as provided by \mlgwjax and \mlgwbnsjax, enables the use of waveform models that are simultaneously computationally efficient and physically detailed. By covering subdominant multipoles and additional dynamical effects, these can fairly rapidly provide more accurate estimation of strongly correlated parameters, such as luminosity distance and orbital inclination, helping to alleviate, or even break, residual degeneracies that persist in simplified waveform descriptions~\cite{degeneracy_1, degeneracy_2, degeneracy_3, degeneracy_4, degeneracy_5, degeneracy_6, degeneracy_7, LIGOScientific:2017adf}.

A natural next step for this work is the extension of the \mlgw surrogate-training framework, which already allows for dominant higher-order modes, to waveform approximants modeling binaries with misaligned spins and non-negligible orbital eccentricity.
Beyond improving parameter estimation, incorporating precession and eccentricity induced effects is essential for robust source classification and population studies~\cite{population_1,population_2, LIGOScientific:2026ctl}, enabling a clearer discrimination between formation channels and strengthening the astrophysical interpretation of detected events~\cite{Mapelli:2021taw, Samsing_2018, gerosa_2018}.

\section*{Software}
This work used the following software: 
\blackjax~\cite{cabezas2024},
\blackjaxns~\cite{Prathaban:2025qgg},
\texttt{JAX}~\cite{jax2018github},
\texttt{LALSuite}~\cite{lalsuite},
\texttt{matplotlib}~\cite{matplotlib},
\mlgw~\cite{mlgw_1, mlgw_2},
\mlgwbns~\cite{mlgw_bns},
\texttt{numpy}~\cite{numpy},
\texttt{PySEOBNR}~\cite{Mihaylov:2023bkc},
\ripple~\cite{Edwards:2023sak},
\texttt{scipy}~\cite{scipy},
\texttt{tf2jax}~\cite{tf2jax}.

\section*{Acknowledgments}
This work was supported by the project BIGA -- “Boosting Inference for Gravitational-wave Astrophysics” funded by the MUR Progetti di Ricerca di Rilevante Interesse Nazionale (PRIN) Bando 2022 - grant 20228TLHPE - CUP I53D23000630006.
LP acknowledges funding support from the project GWExtend -- “Extending Gravitational-Wave Search and Characterization Capabilities for Compact Binary Coalescences” by the Bando per la Ricerca Scientifica 2025 - Progetti Avvio alla Ricerca - CUP B83C25004300005.
S. A acknowledges support from the Alexander von Humboldt Foundation. This study was financed in part by the Coordenação de Aperfeiçoamento de Pessoal de Nível Superior – Brasil (CAPES) – Finance Code 001.
GD acknowledges financial support from the National Recovery and Resilience Plan (PNRR), Mission 4 Component 2 Investment 1.4 - National Center for HPC, Big Data and Quantum Computing, funded by the European Union - NextGenerationEU - CUP B83C22002830001.
FP acknowledges support from the ICSC—Centro Nazionale di Ricerca in High Performance Computing, Big Data and Quantum Computing, funded by the European Union--NextGenerationEU.
This research has made use of data or software obtained from the Gravitational Wave Open Science Center (gwosc.org), a service of the LIGO Scientific Collaboration, the Virgo Collaboration, and KAGRA. The software development in this framework was assisted by Anthropic's Claude (Claude Opus 4.8, via Claude Code), which helped with code development, debugging, and cross-checking throughout this project.
This material is based upon work supported by NSF's LIGO Laboratory which is a major facility fully funded by the National Science Foundation, as well as the Science and Technology Facilities Council (STFC) of the United Kingdom, the Max-Planck-Society (MPS), and the State of Niedersachsen/Germany for support of the construction of Advanced LIGO and construction and operation of the GEO600 detector. 
Additional support for Advanced LIGO was provided by the Australian Research Council. 
Virgo is funded, through the European Gravitational Observatory (EGO), by the French Centre National de Recherche Scientifique (CNRS), the Italian Istituto Nazionale di Fisica Nucleare (INFN) and the Dutch Nikhef, with contributions by institutions from Belgium, Germany, Greece, Hungary, Ireland, Japan, Monaco, Poland, Portugal, Spain. 
KAGRA is supported by Ministry of Education, Culture, Sports, Science and Technology (MEXT), Japan Society for the Promotion of Science (JSPS) in Japan; National Research Foundation (NRF) and Ministry of Science and ICT (MSIT) in Korea; Academia Sinica (AS) and National Science and Technology Council (NSTC) in Taiwan. 

\begin{appendix}
\section{Further details on \mlgwbnsjax implementation}
\label{app:binning_marginalization}

\subsection{Cubic-Splines Interpolation in \jax}

As originally proposed in Ref.~\cite{mlgw_bns}, in \mlgwbns vanilla implementation, the reconstructed principal components obtained from the neural network regression predict waveform amplitude and phase only at a fixed set of downsampled frequency nodes. Accordingly, each template evaluation requires resampling these quantities onto the user-given frequency grid, even when that grid is further reduced with relative binning, as we shall explain. Equivalently to the original framework, we perform this resampling here with a cubic spline. In this interpolation scheme, the coefficients are obtained from solving a tridiagonal linear system for the spline moments. 
Since this calculation is run inside the \texttt{JIT}-compiled, \texttt{vmap}-batched waveform call (\texttt{jax\_import\_n\_predict.py}), its implementation has a direct impact on performance, and we therefore provide two interchangeable back-ends. 

The first is a \emph{serial} solver that carries out the forward-elimination and back-substitution sweeps of the Thomas algorithm as a \texttt{jax.lax.scan}; it performs the minimal amount of arithmetic, $\mathcal{O}(N_{\rm nodes})$ operations, but is intrinsically sequential, with a computational depth that also scales as $\mathcal{O}(N_{\rm nodes})$. 
The second approach, on the other hand, recasts the two substitution sweeps as first-order linear recurrences and evaluates them with a parallel associative scan, namely (\texttt{jax.lax.associative\_scan}). This parallel scan reduces the depth to $\mathcal{O}(\log N_{\rm nodes})$ at the cost of a larger total work count, $\mathcal{O}(N_{\rm nodes}\log N_{\rm nodes})$. 
The two back-ends return identical waveforms (agreeing at the $\sim10^{-33}$ level in strain), so the choice between them is purely one of hardware efficiency. 
On a CPU, where instructions are executed essentially serially, the smaller operation count of the \texttt{lax.scan} solver makes it the faster option---by a factor of a few in our single-template benchmarks, where it also outperforms the original \texttt{mlgw-bns} evaluation, as we see in Fig.~\ref{fig:exec_time_comparison_cumulative_mlgwbns}. 
On a GPU, by contrast, the abundant parallelism hides the additional arithmetic of the prefix scan and rewards its shallow depth, so the \texttt{associative\_scan} solver attains the higher batched throughput. 

\begin{figure}[t!]
    \centering
    \includegraphics[width=1.0\linewidth]{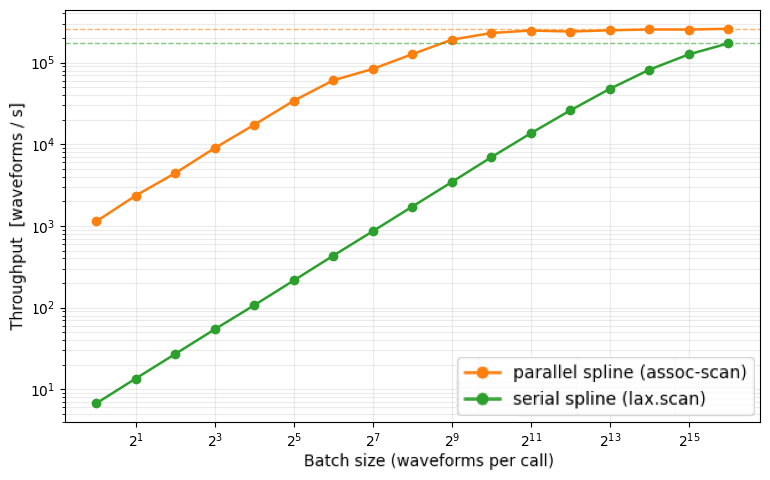}
    \caption{Performance comparison between serial and parallel cubic splines interpolation for waveform generation in batches. Notice that the parallel splines provides superior execution speed for all batch sizes. It is also possible to notice that both approaches are tending to plateau somewhere along $300.000$ waveforms per second, which is the natural limit provided by this GPU device.
\label{fig:GPU_time_mlgwbns}}
\end{figure}

To demonstrate the enhanced performance of parallel splines, in Fig.~\ref{fig:GPU_time_mlgwbns} we plot the template generation speed (number of waveform templates per second) for different corresponding batch sizes in a GPU device (NVIDIA A100). The batch size  scale ranges from $1$ to $3\times 10^5$, where this superior bound was set by the VRAM memory overhead limit. We consider template generation from the same region of parameter space discussed in Sec.~\ref{sec:bns_timing_mismat}. The template duration and sampling rate were equivalently set to four seconds and $2048\,$Hz, respectively.

Therefore, we default to the serial solver for CPU execution and to the parallel solver for the GPU runs that drive the nested-sampling exploration described in Sec.~\ref{sec:GW_inf}. In Sec.~\ref{sec:mlgwbnspe}, we show that combining the enhanced GPU performance provided by this parallel interpolator, and two additional key implementations, namely \textit{relative binning} and \textit{time and phase analytical marginalization}, we could further enhance BNS parameter estimation with \mlgwbnsjax to $17$ minutes. In the following, we summarize the key concepts of relative binning, time and phase analytical marginalizations.

\subsection{Relative binning and analytical marginalization}

For a BNS parameter estimation, two extra dimmensions are added to the usual 15-dimmensional BBH parameter space, namely the two tidal deformability parameters $\Lambda_{1,2}$.
In such a high-dimensional parameter space, the proper exploration of the posterior distribution is severely obstructed by the large memory demanded for the calculation of each waveform individually. 
For each template, for instance, we evaluate its strain in a frequency grid generated by a $T = 128\,$s duration  with a sampling rate of $4096\,$Hz, leading to a frequency grid of approximately $253 \times 10^3$ bins for each one of the three interferometers analyzed simultaneously. 
Even the most powerful GPU devices such as the NVIDIA A100 ($80$ GB VRAM) cannot properly sample a significant large batch with more than a few dozen of parallelized templates without causing a memory overhead problem. 
To overcome this problem, two naive solutions could possibly be considered in a first moment: reducing the number of particles; or decreasing the number of bins in the given frequency grid. 
While, a reduction of the particle number would fatally lead to poorly-mixed parameter estimations with a scarce exploration of the posterior distribution, reducing the number of bins uniformly can bias the parameter estimation by introducing artificial features in the results. 
To overcome this problem, replacing the uniform reduction of the grid size by a calibrated reduction becomes then the only alternative to chase GPU performance on this parameter estimation. 

A few approaches are available in literature for the goal of reducing the frequency grid for template evaluation while retaining a significant part of its accuracy, among them we can highlight the Reduced-Order-Quadrature scheme~\cite{roq_1,roq_2,roq_3,roq_4,roq_5,roq_6}, which has been used in the original \mlgwbns paper~\cite{mlgw_bns}. 
Through that approach, the authors have successufly achieved a reduction of three order of magnitudes for the grid size, dropping the number of bins from 253k to 263 grid points. 
This reduction came at a small cost for the accuracy of the parameter estimation, as they have shown in their results. 
They managed to recover approximately the same posterior distribution for GW170817 parameter estimation. 
In our framework here, however, we have decided to use a different approach for that same task, namely, we replace the Reduced-Order-Quadrature by the relative-binning method of Zackay et al.~\cite{relative_binning}.

The relative-binning method reduces the cost of each likelihood evaluation from $\mathcal{O}(N_{\rm freq})$
to $\mathcal{O}(N_{\rm bins})$, where $N_{\rm bins} \ll N_{\rm freq}$.  
The key observation is that for a candidate template $h(\bm{\theta})$ close to a fiducial template $h(\bm{\theta}_0)$, the ratio $h(f_j;\bm{\theta})/h(f_j;\bm{\theta}_0)$ is slowly varying across each frequency bin $j$, allowing it to be approximated as constant within that bin.

We partition the analysis band $[f_{\rm min}, f_{\rm max}]$ into approximately $400-1600$ logarithmically spaced bins, compared to $N_{\rm freq} \sim 253000$ for the full likelihood. 
We implement the relative binning likelihood as a \jax function in \texttt{relative\_binning.py}, fully compatible with \texttt{jax.jit} and \texttt{jax.vmap} so that it can be batched across nested-sampling live points on GPU.

Finally, additionally to the relative binning, we also implement analytical marginalization of the coalescence time and phase, just like in the original framework \cite{mlgw_bns}. By marginalizing over both $t_c$ and $\phi_c$ analytically, the sampled parameter space is reduced by two dimensions, removing the two sharpest and most difficult-to-sample posteriors. We use $N_{t_c} = 3000$ uniformly spaced points over $[\Delta t_{c,\min}, \Delta t_{c,\max}] = [-10, +10]$\,ms. All $N_{t_c}$ evaluations are performed simultaneously by \jax as a single batched matrix multiplication, costing $\mathcal{O}(N_{t_c} N_{\rm bins})$ per likelihood call rather than $N_{t_c}$ independent full-grid evaluations. Finally, after the analytical marginalization over $t_c$ and $\phi_c$, the sampler operates in an 11-dimensional parameter space:
$\{\ln D_L,\, \theta_{JN},\, \psi,\, \alpha,\, \delta,\,
\mathcal{M}_c,\, q,\, \chi_1,\, \chi_2,\, \Lambda_1,\, \Lambda_2\}$.

\end{appendix}

\printbibliography

@article{GWTC_5_intro,
    author = "Abac, A. G. and others",
    collaboration = "LIGO Scientific, VIRGO, KAGRA",
    title = "{GWTC-5.0: An Introduction to Version 5.0 of the Gravitational-Wave Transient Catalog}",
    eprint = "2605.27223",
    archivePrefix = "arXiv",
    primaryClass = "gr-qc",
    reportNumber = "LIGO-P2500701",
    month = "5",
    year = "2026"
}

@article{GWTC_5_results,
      title={GWTC-5.0: Observations from the Second Part of the Fourth LIGO-Virgo-KAGRA Observing Run and Updates to the Gravitational-Wave Transient Catalog},
      author="Abac, A. G. and others",
      collaboration = "LIGO Scientific, VIRGO, KAGRA",
      year={2026},
      eprint={2605.27225},
      archivePrefix={arXiv},
      primaryClass={gr-qc},
      url={https://arxiv.org/abs/2605.27225}, 
}

@article{GWTC_5_methods,
    author = "Abac, A. G. and others",
    collaboration = "LIGO Scientific, VIRGO, KAGRA",
    title = "{GWTC-5.0: Methods for Identifying and Characterizing Gravitational-wave Transients}",
    eprint = "2605.27224",
    archivePrefix = "arXiv",
    primaryClass = "gr-qc",
    reportNumber = "LIGO-P2600166",
    month = "5",
    year = "2026"
}

@article{GWTC_5_GWOSC,
    author = "Abac, A. G. and others",
    collaboration = "LIGO Scientific, VIRGO, KAGRA",
    title = "{Open Data from LIGO, Virgo, and KAGRA through the Second Part of the Fourth Observing Run}",
    eprint = "2605.27090",
    archivePrefix = "arXiv",
    primaryClass = "gr-qc",
    reportNumber = "LIGO-P2600085",
    month = "5",
    year = "2026"
}

@ARTICLE{2015:LIGO,
       author = {{LIGO Scientific Collaboration} and {Aasi}, J. and {Abbott}, B.~P. and {Abbott}, R. and {Abbott}, T. and {Abernathy}, M.~R. and {Ackley}, K. and {Adams}, C. and {Adams}, T. and {Addesso}, P. and {Adhikari}, R.~X. and {Adya}, V. and {Affeldt}, C. and {Aggarwal}, N. and {Aguiar}, O.~D. and {Ain}, A. and {Ajith}, P. and {Alemic}, A. and {Allen}, B. and {Amariutei}, D. and {Anderson}, S.~B. and {Anderson}, W.~G. and {Arai}, K. and {Araya}, M.~C. and {Arceneaux}, C. and {Areeda}, J.~S. and {Ashton}, G. and {Ast}, S. and {Aston}, S.~M. and {Aufmuth}, P. and {Aulbert}, C. and {Aylott}, B.~E. and {Babak}, S. and {Baker}, P.~T. and {Ballmer}, S.~W. and {Barayoga}, J.~C. and {Barbet}, M. and {Barclay}, S. and {Barish}, B.~C. and {Barker}, D. and {Barr}, B. and {Barsotti}, L. and {Bartlett}, J. and {Barton}, M.~A. and {Bartos}, I. and {Bassiri}, R. and {Batch}, J.~C. and {Baune}, C. and {Behnke}, B. and {Bell}, A.~S. and {Bell}, C. and {Benacquista}, M. and {Bergman}, J. and {Bergmann}, G. and {Berry}, C.~P.~L. and {Betzwieser}, J. and {Bhagwat}, S. and {Bhandare}, R. and {Bilenko}, I.~A. and {Billingsley}, G. and {Birch}, J. and {Biscans}, S. and {Biwer}, C. and {Blackburn}, J.~K. and {Blackburn}, L. and {Blair}, C.~D. and {Blair}, D. and {Bock}, O. and {Bodiya}, T.~P. and {Bojtos}, P. and {Bond}, C. and {Bork}, R. and {Born}, M. and {Bose}, Sukanta and {Brady}, P.~R. and {Braginsky}, V.~B. and {Brau}, J.~E. and {Bridges}, D.~O. and {Brinkmann}, M. and {Brooks}, A.~F. and {Brown}, D.~A. and {Brown}, D.~D. and {Brown}, N.~M. and {Buchman}, S. and {Buikema}, A. and {Buonanno}, A. and {Cadonati}, L. and {Calder{\'o}n Bustillo}, J. and {Camp}, J.~B. and {Cannon}, K.~C. and {Cao}, J. and {Capano}, C.~D. and {Caride}, S. and {Caudill}, S. and {Cavagli{\`a}}, M. and {Cepeda}, C. and {Chakraborty}, R. and {Chalermsongsak}, T. and {Chamberlin}, S.~J. and {Chao}, S. and {Charlton}, P. and {Chen}, Y. and {Cho}, H.~S. and {Cho}, M. and {Chow}, J.~H. and {Christensen}, N. and {Chu}, Q. and {Chung}, S. and {Ciani}, G. and {Clara}, F. and {Clark}, J.~A. and {Collette}, C. and {Cominsky}, L. and {Constancio}, Jr., M. and {Cook}, D. and {Corbitt}, T.~R. and {Cornish}, N. and {Corsi}, A. and {Costa}, C.~A. and {Coughlin}, M.~W. and {Countryman}, S. and {Couvares}, P. and {Coward}, D.~M. and {Cowart}, M.~J. and {Coyne}, D.~C. and {Coyne}, R. and {Craig}, K. and {Creighton}, J.~D.~E. and {Creighton}, T.~D. and {Cripe}, J. and {Crowder}, S.~G. and {Cumming}, A. and {Cunningham}, L. and {Cutler}, C. and {Dahl}, K. and {Dal Canton}, T. and {Damjanic}, M. and {Danilishin}, S.~L. and {Danzmann}, K. and {Dartez}, L. and {Dave}, I. and {Daveloza}, H. and {Davies}, G.~S. and {Daw}, E.~J. and {DeBra}, D. and {Del Pozzo}, W. and {Denker}, T. and {Dent}, T. and {Dergachev}, V. and {DeRosa}, R.~T. and {DeSalvo}, R. and {Dhurandhar}, S. and {D{\textasciiacute}{\i}az}, M. and {Di Palma}, I. and {Dojcinoski}, G. and {Dominguez}, E. and {Donovan}, F. and {Dooley}, K.~L. and {Doravari}, S. and {Douglas}, R. and {Downes}, T.~P. and {Driggers}, J.~C. and {Du}, Z. and {Dwyer}, S. and {Eberle}, T. and {Edo}, T. and {Edwards}, M. and {Edwards}, M. and {Effler}, A. and {Eggenstein}, H.-B. and {Ehrens}, P. and {Eichholz}, J. and {Eikenberry}, S.~S. and {Essick}, R. and {Etzel}, T. and {Evans}, M. and {Evans}, T. and {Factourovich}, M. and {Fairhurst}, S. and {Fan}, X. and {Fang}, Q. and {Farr}, B. and {Farr}, W.~M. and {Favata}, M. and {Fays}, M. and {Fehrmann}, H. and {Fejer}, M.~M. and {Feldbaum}, D. and {Ferreira}, E.~C. and {Fisher}, R.~P. and {Frei}, Z. and {Freise}, A. and {Frey}, R. and {Fricke}, T.~T. and {Fritschel}, P. and {Frolov}, V.~V. and {Fuentes-Tapia}, S. and {Fulda}, P. and {Fyffe}, M. and {Gair}, J.~R.},
        title = "{Advanced LIGO}",
      journal = {Classical and Quantum Gravity},
         year = 2015,
        month = apr,
       volume = {32},
       number = {7},
          eid = {074001},
        pages = {074001},
          doi = {10.1088/0264-9381/32/7/074001},
archivePrefix = {arXiv},
       eprint = {1411.4547},
 primaryClass = {gr-qc},
       adsurl = {https://ui.adsabs.harvard.edu/abs/2015CQGra..32g4001L}
}

@ARTICLE{2015:VIRGO,
       author = {{Acernese}, F. and {Agathos}, M. and {Agatsuma}, K. and {Aisa}, D. and {Allemandou}, N. and {Allocca}, A. and {Amarni}, J. and {Astone}, P. and {Balestri}, G. and {Ballardin}, G. and {Barone}, F. and {Baronick}, J.-P. and {Barsuglia}, M. and {Basti}, A. and {Basti}, F. and {Bauer}, Th S. and {Bavigadda}, V. and {Bejger}, M. and {Beker}, M.~G. and {Belczynski}, C. and {Bersanetti}, D. and {Bertolini}, A. and {Bitossi}, M. and {Bizouard}, M.~A. and {Bloemen}, S. and {Blom}, M. and {Boer}, M. and {Bogaert}, G. and {Bondi}, D. and {Bondu}, F. and {Bonelli}, L. and {Bonnand}, R. and {Boschi}, V. and {Bosi}, L. and {Bouedo}, T. and {Bradaschia}, C. and {Branchesi}, M. and {Briant}, T. and {Brillet}, A. and {Brisson}, V. and {Bulik}, T. and {Bulten}, H.~J. and {Buskulic}, D. and {Buy}, C. and {Cagnoli}, G. and {Calloni}, E. and {Campeggi}, C. and {Canuel}, B. and {Carbognani}, F. and {Cavalier}, F. and {Cavalieri}, R. and {Cella}, G. and {Cesarini}, E. and {Mottin}, E. Chassande- and {Chincarini}, A. and {Chiummo}, A. and {Chua}, S. and {Cleva}, F. and {Coccia}, E. and {Cohadon}, P.-F. and {Colla}, A. and {Colombini}, M. and {Conte}, A. and {Coulon}, J.-P. and {Cuoco}, E. and {Dalmaz}, A. and {D'Antonio}, S. and {Dattilo}, V. and {Davier}, M. and {Day}, R. and {Debreczeni}, G. and {Degallaix}, J. and {Del{\'e}glise}, S. and {Pozzo}, W. Del and {Dereli}, H. and {Rosa}, R. De and {Fiore}, L. Di and {Lieto}, A. Di and {Virgilio}, A. Di and {Doets}, M. and {Dolique}, V. and {Drago}, M. and {Ducrot}, M. and {Endr{\H{o}}czi}, G. and {Fafone}, V. and {Farinon}, S. and {Ferrante}, I. and {Ferrini}, F. and {Fidecaro}, F. and {Fiori}, I. and {Flaminio}, R. and {Fournier}, J.-D. and {Franco}, S. and {Frasca}, S. and {Frasconi}, F. and {Gammaitoni}, L. and {Garufi}, F. and {Gaspard}, M. and {Gatto}, A. and {Gemme}, G. and {Gendre}, B. and {Genin}, E. and {Gennai}, A. and {Ghosh}, S. and {Giacobone}, L. and {Giazotto}, A. and {Gouaty}, R. and {Granata}, M. and {Greco}, G. and {Groot}, P. and {Guidi}, G.~M. and {Harms}, J. and {Heidmann}, A. and {Heitmann}, H. and {Hello}, P. and {Hemming}, G. and {Hennes}, E. and {Hofman}, D. and {Jaranowski}, P. and {Jonker}, R.~J.~G. and {Kasprzack}, M. and {K{\'e}f{\'e}lian}, F. and {Kowalska}, I. and {Kraan}, M. and {Kr{\'o}lak}, A. and {Kutynia}, A. and {Lazzaro}, C. and {Leonardi}, M. and {Leroy}, N. and {Letendre}, N. and {Li}, T.~G.~F. and {Lieunard}, B. and {Lorenzini}, M. and {Loriette}, V. and {Losurdo}, G. and {Magazz{\`u}}, C. and {Majorana}, E. and {Maksimovic}, I. and {Malvezzi}, V. and {Man}, N. and {Mangano}, V. and {Mantovani}, M. and {Marchesoni}, F. and {Marion}, F. and {Marque}, J. and {Martelli}, F. and {Martellini}, L. and {Masserot}, A. and {Meacher}, D. and {Meidam}, J. and {Mezzani}, F. and {Michel}, C. and {Milano}, L. and {Minenkov}, Y. and {Moggi}, A. and {Mohan}, M. and {Montani}, M. and {Morgado}, N. and {Mours}, B. and {Mul}, F. and {Nagy}, M.~F. and {Nardecchia}, I. and {Naticchioni}, L. and {Nelemans}, G. and {Neri}, I. and {Neri}, M. and {Nocera}, F. and {Pacaud}, E. and {Palomba}, C. and {Paoletti}, F. and {Paoli}, A. and {Pasqualetti}, A. and {Passaquieti}, R. and {Passuello}, D. and {Perciballi}, M. and {Petit}, S. and {Pichot}, M. and {Piergiovanni}, F. and {Pillant}, G. and {Piluso}, A. and {Pinard}, L. and {Poggiani}, R. and {Prijatelj}, M. and {Prodi}, G.~A. and {Punturo}, M. and {Puppo}, P. and {Rabeling}, D.~S. and {R{\'a}cz}, I. and {Rapagnani}, P. and {Razzano}, M. and {Re}, V. and {Regimbau}, T. and {Ricci}, F. and {Robinet}, F. and {Rocchi}, A. and {Rolland}, L. and {Romano}, R. and {Rosi{\'n}ska}, D. and {Ruggi}, P. and {Saracco}, E.},
        title = "{Advanced Virgo: a second-generation interferometric gravitational wave detector}",
      journal = {Classical and Quantum Gravity},
         year = 2015,
        month = jan,
       volume = {32},
       number = {2},
          eid = {024001},
        pages = {024001},
          doi = {10.1088/0264-9381/32/2/024001},
archivePrefix = {arXiv},
       eprint = {1408.3978},
 primaryClass = {gr-qc},
       adsurl = {https://ui.adsabs.harvard.edu/abs/2015CQGra..32b4001A}
}

@ARTICLE{2013:KAGRA,
       author = {{Aso}, Yoichi and {Michimura}, Yuta and {Somiya}, Kentaro and {Ando}, Masaki and {Miyakawa}, Osamu and {Sekiguchi}, Takanori and {Tatsumi}, Daisuke and {Yamamoto}, Hiroaki},
        title = "{Interferometer design of the KAGRA gravitational wave detector}",
      journal = {Phys. Rev. D},
         year = 2013,
        month = aug,
       volume = {88},
       number = {4},
          eid = {043007},
        pages = {043007},
          doi = {10.1103/PhysRevD.88.043007},
archivePrefix = {arXiv},
       eprint = {1306.6747},
 primaryClass = {gr-qc},
       adsurl = {https://ui.adsabs.harvard.edu/abs/2013PhRvD..88d3007A}
}

@article{LIGOScientific:2016aoc,
	author = "Abbott, B. P. and others",
	collaboration = "LIGO Scientific, Virgo",
	title = "{Observation of Gravitational Waves from a Binary Black Hole Merger}",
	eprint = "1602.03837",
	archivePrefix = "arXiv",
	primaryClass = "gr-qc",
	reportNumber = "LIGO-P150914",
	doi = "10.1103/PhysRevLett.116.061102",
	journal = "Phys. Rev. Lett.",
	volume = "116",
	number = "6",
	pages = "061102",
	year = "2016"
}

@article{LIGOScientific:2016vlm,
	author = "Abbott, B. P. and others",
	collaboration = "LIGO Scientific, Virgo",
	title = "{Properties of the Binary Black Hole Merger GW150914}",
	eprint = "1602.03840",
	archivePrefix = "arXiv",
	primaryClass = "gr-qc",
	reportNumber = "LIGO-P1500218",
	doi = "10.1103/PhysRevLett.116.241102",
	journal = "Phys. Rev. Lett.",
	volume = "116",
	number = "24",
	pages = "241102",
	year = "2016"
}

@article{LIGOScientific:2017vwq,
    author = "Abbott, B. P. and others",
    collaboration = "LIGO Scientific, Virgo",
    title = "{GW170817: Observation of Gravitational Waves from a Binary Neutron Star Inspiral}",
    eprint = "1710.05832",
    archivePrefix = "arXiv",
    primaryClass = "gr-qc",
    reportNumber = "LIGO-P170817",
    doi = "10.1103/PhysRevLett.119.161101",
    journal = "Phys. Rev. Lett.",
    volume = "119",
    number = "16",
    pages = "161101",
    year = "2017"
}

@article{LIGOScientific:2017ync,
	author = "Abbott, B. P. and others",
	collaboration = "LIGO Scientific, Virgo, Fermi GBM, INTEGRAL, IceCube, AstroSat Cadmium Zinc Telluride Imager Team, IPN, Insight-Hxmt, ANTARES, Swift, AGILE Team, 1M2H Team, Dark Energy Camera GW-EM, DES, DLT40, GRAWITA, Fermi-LAT, ATCA, ASKAP, Las Cumbres Observatory Group, OzGrav, DWF (Deeper Wider Faster Program), AST3, CAASTRO, VINROUGE, MASTER, J-GEM, GROWTH, JAGWAR, CaltechNRAO, TTU-NRAO, NuSTAR, Pan-STARRS, MAXI Team, TZAC Consortium, KU, Nordic Optical Telescope, ePESSTO, GROND, Texas Tech University, SALT Group, TOROS, BOOTES, MWA, CALET, IKI-GW Follow-up, H.E.S.S., LOFAR, LWA, HAWC, Pierre Auger, ALMA, Euro VLBI Team, Pi of Sky, Chandra Team at McGill University, DFN, ATLAS Telescopes, High Time Resolution Universe Survey, RIMAS, RATIR, SKA South Africa/MeerKAT",
	title = "{Multi-messenger Observations of a Binary Neutron Star Merger}",
	eprint = "1710.05833",
	archivePrefix = "arXiv",
	primaryClass = "astro-ph.HE",
	reportNumber = "LIGO-P1700294, VIR-0802A-17, FERMILAB-PUB-17-478-A-AE-CD",
	doi = "10.3847/2041-8213/aa91c9",
	journal = "Astrophys. J. Lett.",
	volume = "848",
	number = "2",
	pages = "L12",
	year = "2017"
}

@article{Usman:2015kfa,
	author = "Usman, Samantha A. and others",
	title = "{The PyCBC search for gravitational waves from compact binary coalescence}",
	eprint = "1508.02357",
	archivePrefix = "arXiv",
	primaryClass = "gr-qc",
	reportNumber = "LIGO-P1500086",
	doi = "10.1088/0264-9381/33/21/215004",
	journal = "Class. Quant. Grav.",
	volume = "33",
	number = "21",
	pages = "215004",
	year = "2016"
}

@article{Allen:2005fk,
	author = "Allen, B. and Anderson, Warren G. and Brady, Patrick R. and Brown, Duncan A. and Creighton, Jolien D. E.",
	title = "{FINDCHIRP: An Algorithm for detection of gravitational waves from inspiraling compact binaries}",
	eprint = "gr-qc/0509116",
	archivePrefix = "arXiv",
	doi = "10.1103/PhysRevD.85.122006",
	journal = "Phys. Rev. D",
	volume = "85",
	pages = "122006",
	year = "2012"
}

@article{Thrane_2019,
   title={An introduction to Bayesian inference in gravitational-wave astronomy: Parameter estimation, model selection, and hierarchical models},
   volume={36},
   ISSN={1448-6083},
   url={http://dx.doi.org/10.1017/pasa.2019.2},
   DOI={10.1017/pasa.2019.2},
   journal={Publications of the Astronomical Society of Australia},
   publisher={Cambridge University Press (CUP)},
   author={Thrane, Eric and Talbot, Colm},
   year={2019} }

@article{Christensen_2022,
   title={Parameter estimation with gravitational waves},
   volume={94},
   ISSN={1539-0756},
   url={http://dx.doi.org/10.1103/RevModPhys.94.025001},
   DOI={10.1103/revmodphys.94.025001},
   number={2},
   journal={Reviews of Modern Physics},
   publisher={American Physical Society (APS)},
   author={Christensen, Nelson and Meyer, Renate},
   year={2022},
   month=apr }

@article{eobnr1,
  title = {Effective one-body approach to general relativistic two-body dynamics},
  author = {Buonanno, A. and Damour, T.},
  journal = {Phys. Rev. D},
  volume = {59},
  issue = {8},
  pages = {084006},
  numpages = {24},
  year = {1999},
  month = Mar,
  publisher = {American Physical Society},
  doi = {10.1103/PhysRevD.59.084006},
  url = {https://link.aps.org/doi/10.1103/PhysRevD.59.084006}
}

@article{eobnr2,
  title = {Transition from inspiral to plunge in binary black hole coalescences},
  author = {Buonanno, Alessandra and Damour, Thibault},
  journal = {Phys. Rev. D},
  volume = {62},
  issue = {6},
  pages = {064015},
  numpages = {24},
  year = {2000},
  month = Aug,
  publisher = {American Physical Society},
  doi = {10.1103/PhysRevD.62.064015},
  url = {https://link.aps.org/doi/10.1103/PhysRevD.62.064015}
}

@article{eobnr3,
  title = {Determination of the last stable orbit for circular general relativistic binaries at the third post-Newtonian approximation},
  author = {Damour, Thibault and Jaranowski, Piotr and Sch\"afer, Gerhard},
  journal = {Phys. Rev. D},
  volume = {62},
  issue = {8},
  pages = {084011},
  numpages = {21},
  year = {2000},
  month = Sep,
  publisher = {American Physical Society},
  doi = {10.1103/PhysRevD.62.084011},
  url = {https://link.aps.org/doi/10.1103/PhysRevD.62.084011}
}

@article{eobnr4,
  title = {Coalescence of two spinning black holes: An effective one-body approach},
  author = {Damour, Thibault},
  journal = {Phys. Rev. D},
  volume = {64},
  issue = {12},
  pages = {124013},
  numpages = {22},
  year = {2001},
  month = Nov,
  publisher = {American Physical Society},
  doi = {10.1103/PhysRevD.64.124013},
  url = {https://link.aps.org/doi/10.1103/PhysRevD.64.124013}
}

@article{eobnr5,
  title = {Effective one body approach to the dynamics of two spinning black holes with next-to-leading order spin-orbit coupling},
  author = {Damour, Thibault and Jaranowski, Piotr and Sch\"afer, Gerhard},
  journal = {Phys. Rev. D},
  volume = {78},
  issue = {2},
  pages = {024009},
  numpages = {18},
  year = {2008},
  month = Jul,
  publisher = {American Physical Society},
  doi = {10.1103/PhysRevD.78.024009},
  url = {https://link.aps.org/doi/10.1103/PhysRevD.78.024009}
}

@article{eobnr6,
  title = {Improved analytical description of inspiralling and coalescing black-hole binaries},
  author = {Damour, Thibault and Nagar, Alessandro},
  journal = {Phys. Rev. D},
  volume = {79},
  issue = {8},
  pages = {081503},
  numpages = {5},
  year = {2009},
  month = Apr,
  publisher = {American Physical Society},
  doi = {10.1103/PhysRevD.79.081503},
  url = {https://link.aps.org/doi/10.1103/PhysRevD.79.081503}
}

@article{eobnr7,
  title = {New effective-one-body description of coalescing nonprecessing spinning black-hole binaries},
  author = {Damour, Thibault and Nagar, Alessandro},
  journal = {Phys. Rev. D},
  volume = {90},
  issue = {4},
  pages = {044018},
  numpages = {13},
  year = {2014},
  month = Aug,
  publisher = {American Physical Society},
  doi = {10.1103/PhysRevD.90.044018},
  url = {https://link.aps.org/doi/10.1103/PhysRevD.90.044018}
}

@article{eobnr8,
  title = {Improved effective-one-body model of spinning, nonprecessing binary black holes for the era of gravitational-wave astrophysics with advanced detectors},
  author = {Boh\'e, Alejandro and Shao, Lijing and Taracchini, Andrea and Buonanno, Alessandra and Babak, Stanislav and Harry, Ian W. and Hinder, Ian and Ossokine, Serguei and P\"urrer, Michael and Raymond, Vivien and Chu, Tony and Fong, Heather and Kumar, Prayush and Pfeiffer, Harald P. and Boyle, Michael and Hemberger, Daniel A. and Kidder, Lawrence E. and Lovelace, Geoffrey and Scheel, Mark A. and Szil\'agyi, B\'ela},
  journal = {Phys. Rev. D},
  volume = {95},
  issue = {4},
  pages = {044028},
  numpages = {29},
  year = {2017},
  month = Feb,
  publisher = {American Physical Society},
  doi = {10.1103/PhysRevD.95.044028},
  url = {https://link.aps.org/doi/10.1103/PhysRevD.95.044028}
}

@article{eobnr9,
  title = {Time-domain effective-one-body gravitational waveforms for coalescing compact binaries with nonprecessing spins, tides, and self-spin effects},
  author = {Nagar, Alessandro and Bernuzzi, Sebastiano and Del Pozzo, Walter and Riemenschneider, Gunnar and Akcay, Sarp and Carullo, Gregorio and Fleig, Philipp and Babak, Stanislav and Tsang, Ka Wa and Colleoni, Marta and Messina, Francesco and Pratten, Geraint and Radice, David and Rettegno, Piero and Agathos, Michalis and Fauchon-Jones, Edward and Hannam, Mark and Husa, Sascha and Dietrich, Tim and Cerd\'a-Duran, Pablo and Font, Jos\'e A. and Pannarale, Francesco and Schmidt, Patricia and Damour, Thibault},
  journal = {Phys. Rev. D},
  volume = {98},
  issue = {10},
  pages = {104052},
  numpages = {40},
  year = {2018},
  month = Nov,
  publisher = {American Physical Society},
  doi = {10.1103/PhysRevD.98.104052},
  url = {https://link.aps.org/doi/10.1103/PhysRevD.98.104052}
}

@article{eobnr10,
  title = {Improved resummation of post-Newtonian multipolar waveforms from circularized compact binaries},
  author = {Damour, Thibault and Iyer, Bala R. and Nagar, Alessandro},
  journal = {Phys. Rev. D},
  volume = {79},
  issue = {6},
  pages = {064004},
  numpages = {29},
  year = {2009},
  month = Mar,
  publisher = {American Physical Society},
  doi = {10.1103/PhysRevD.79.064004},
  url = {https://link.aps.org/doi/10.1103/PhysRevD.79.064004}
}

@article{phenom_4,
   title={Frequency-domain gravitational waves from nonprecessing black-hole binaries. II. A phenomenological model for the advanced detector era},
   volume={93},
   ISSN={2470-0029},
   url={http://dx.doi.org/10.1103/PhysRevD.93.044007},
   DOI={10.1103/physrevd.93.044007},
   number={4},
   journal={Physical Review D},
   publisher={American Physical Society (APS)},
   author={Khan, Sebastian and Husa, Sascha and Hannam, Mark and Ohme, Frank and Pürrer, Michael and Forteza, Xisco Jiménez and Bohé, Alejandro},
   year={2016},
   month=feb }

@article{phenom_3,
  title = {Frequency-domain gravitational waves from nonprecessing black-hole binaries. I. New numerical waveforms and anatomy of the signal},
  author = {Husa, Sascha and Khan, Sebastian and Hannam, Mark and P\"urrer, Michael and Ohme, Frank and Forteza, Xisco Jim\'enez and Boh\'e, Alejandro},
  journal = {Phys. Rev. D},
  volume = {93},
  issue = {4},
  pages = {044006},
  numpages = {19},
  year = {2016},
  month = "2",
  publisher = {American Physical Society},
  doi = {10.1103/PhysRevD.93.044006},
  url = {https://link.aps.org/doi/10.1103/PhysRevD.93.044006}
}

@article{phenom_1,
  title = {Inspiral-Merger-Ringdown Waveforms for Black-Hole Binaries with Nonprecessing Spins},
  author = {Ajith, P. and Hannam, M. and Husa, S. and Chen, Y. and Br\"ugmann, B. and Dorband, N. and M\"uller, D. and Ohme, F. and Pollney, D. and Reisswig, C. and Santamar\'{\i}a, L. and Seiler, J.},
  journal = {Phys. Rev. Lett.},
  volume = {106},
  issue = {24},
  pages = {241101},
  numpages = {4},
  year = {2011},
  month = Jun,
  publisher = {American Physical Society},
  doi = {10.1103/PhysRevLett.106.241101},
  url = {https://link.aps.org/doi/10.1103/PhysRevLett.106.241101}
}

@ARTICLE{phenom_7,
       author = {{Pratten}, Geraint and {Garc{\'\i}a-Quir{\'o}s}, Cecilio and {Colleoni}, Marta and {Ramos-Buades}, Antoni and {Estell{\'e}s}, H{\'e}ctor and {Mateu-Lucena}, Maite and {Jaume}, Rafel and {Haney}, Maria and {Keitel}, David and {Thompson}, Jonathan E. and {Husa}, Sascha},
        title = "{Computationally efficient models for the dominant and subdominant harmonic modes of precessing binary black holes}",
      journal = "Phys. Rev. D",
         year = 2021,
        month = "5",
       volume = {103},
       number = {10},
          eid = {104056},
        pages = {104056},
          doi = {10.1103/PhysRevD.103.104056},
archivePrefix = {arXiv},
       eprint = {2004.06503},
 primaryClass = {gr-qc},
       adsurl = {https://ui.adsabs.harvard.edu/abs/2021PhRvD.103j4056P}
}

@article{phenom_5,
  title = {Multimode frequency-domain model for the gravitational wave signal from nonprecessing black-hole binaries},
  author = {Garc\'{\i}a-Quir\'os, Cecilio and Colleoni, Marta and Husa, Sascha and Estell\'es, H\'ector and Pratten, Geraint and Ramos-Buades, Antoni and Mateu-Lucena, Maite and Jaume, Rafel},
  journal = {Phys. Rev. D},
  volume = {102},
  issue = {6},
  pages = {064002},
  numpages = {33},
  year = {2020},
  month = "9",
  publisher = {American Physical Society},
  doi = {10.1103/PhysRevD.102.064002},
  url = {https://link.aps.org/doi/10.1103/PhysRevD.102.064002}
}

@article{phenom_2,
   title={Simple Model of Complete Precessing Black-Hole-Binary Gravitational Waveforms},
   volume={113},
   ISSN={1079-7114},
   url={http://dx.doi.org/10.1103/PhysRevLett.113.151101},
   DOI={10.1103/physrevlett.113.151101},
   number={15},
   journal={Physical Review Letters},
   publisher={American Physical Society (APS)},
   author={Hannam, Mark and Schmidt, Patricia and Bohé, Alejandro and Haegel, Leïla and Husa, Sascha and Ohme, Frank and Pratten, Geraint and Pürrer, Michael},
   year={2014},
   month="10" }

@article{phenom_8,
   title={Phenomenological time domain model for dominant quadrupole gravitational wave signal of coalescing binary black holes},
   volume={103},
   ISSN={2470-0029},
   url={http://dx.doi.org/10.1103/PhysRevD.103.124060},
   DOI={10.1103/physrevd.103.124060},
   number={12},
   journal={Physical Review D},
   publisher={American Physical Society (APS)},
   author={Estellés, Héctor and Ramos-Buades, Antoni and Husa, Sascha and García-Quirós, Cecilio and Colleoni, Marta and Haegel, Leïla and Jaume, Rafel},
   year={2021},
   month="6" }

@article{phenom_9,
  title = {Time-domain phenomenological model of gravitational-wave subdominant harmonics for quasicircular nonprecessing binary black hole coalescences},
  author = {Estell\'es, H\'ector and Husa, Sascha and Colleoni, Marta and Keitel, David and Mateu-Lucena, Maite and Garc\'{\i}a-Quir\'os, Cecilio and Ramos-Buades, Antoni and Borchers, Angela},
  journal = {Phys. Rev. D},
  volume = {105},
  issue = {8},
  pages = {084039},
  numpages = {24},
  year = {2022},
  month = "4",
  publisher = {American Physical Society},
  doi = {10.1103/PhysRevD.105.084039},
  url = {https://link.aps.org/doi/10.1103/PhysRevD.105.084039}
}

@article{phenom_10,
  title = {New twists in compact binary waveform modeling: A fast time-domain model for precession},
  author = {Estell\'es, H\'ector and Colleoni, Marta and Garc\'{\i}a-Quir\'os, Cecilio and Husa, Sascha and Keitel, David and Mateu-Lucena, Maite and Planas, Maria de Lluc and Ramos-Buades, Antoni},
  journal = {Phys. Rev. D},
  volume = {105},
  issue = {8},
  pages = {084040},
  numpages = {19},
  year = {2022},
  month = "4",
  publisher = {American Physical Society},
  doi = {10.1103/PhysRevD.105.084040},
  url = {https://link.aps.org/doi/10.1103/PhysRevD.105.084040}
}

@article{phenom_6,
   title={Setting the cornerstone for a family of models for gravitational waves from compact binaries: The dominant harmonic for nonprecessing quasicircular black holes},
   volume={102},
   ISSN={2470-0029},
   url={http://dx.doi.org/10.1103/PhysRevD.102.064001},
   DOI={10.1103/physrevd.102.064001},
   number={6},
   journal={Physical Review D},
   publisher={American Physical Society (APS)},
   author={Pratten, Geraint and Husa, Sascha and García-Quirós, Cecilio and Colleoni, Marta and Ramos-Buades, Antoni and Estellés, Héctor and Jaume, Rafel},
   year={2020},
   month=sep }

@article{phenom_11,
    author = "Ramos-Buades, Antoni and Henry, Quentin and Haney, Maria",
    title = "{Fast frequency-domain phenomenological modeling of eccentric aligned-spin binary black holes}",
    eprint = "2601.03340",
    archivePrefix = "arXiv",
    primaryClass = "gr-qc",
    month = "1",
    year = "2026"
}

@article{eobnr12,
  title = {Effective-one-body model for black-hole binaries with generic mass ratios and spins},
  author = {Taracchini, Andrea and Buonanno, Alessandra and Pan, Yi and Hinderer, Tanja and Boyle, Michael and Hemberger, Daniel A. and Kidder, Lawrence E. and Lovelace, Geoffrey and Mrou\'e, Abdul H. and Pfeiffer, Harald P. and Scheel, Mark A. and Szil\'agyi, B\'ela and Taylor, Nicholas W. and Zenginoglu, Anil},
  journal = {Phys. Rev. D},
  volume = {89},
  issue = {6},
  pages = {061502},
  numpages = {6},
  year = {2014},
  month = Mar,
  publisher = {American Physical Society},
  doi = {10.1103/PhysRevD.89.061502},
  url = {https://link.aps.org/doi/10.1103/PhysRevD.89.061502}
}

@article{eobnr11,
   title={Inspiral-merger-ringdown multipolar waveforms of nonspinning black-hole binaries using the effective-one-body formalism},
   volume={84},
   ISSN={1550-2368},
   url={http://dx.doi.org/10.1103/PhysRevD.84.124052},
   DOI={10.1103/physrevd.84.124052},
   number={12},
   journal={Physical Review D},
   publisher={American Physical Society (APS)},
   author={Pan, Yi and Buonanno, Alessandra and Boyle, Michael and Buchman, Luisa T. and Kidder, Lawrence E. and Pfeiffer, Harald P. and Scheel, Mark A.},
   year={2011},
   month=dec }

@article{eobnr13,
   title={Waveform model for an eccentric binary black hole based on the effective-one-body-numerical-relativity formalism},
   volume={96},
   ISSN={2470-0029},
   url={http://dx.doi.org/10.1103/PhysRevD.96.044028},
   DOI={10.1103/physrevd.96.044028},
   number={4},
   journal={Physical Review D},
   publisher={American Physical Society (APS)},
   author={Cao, Zhoujian and Han, Wen-Biao},
   year={2017},
   month=aug }

@article{eobnr14,
   title={Laying the foundation of the effective-one-body waveform models SEOBNRv5: Improved accuracy and efficiency for spinning nonprecessing binary black holes},
   volume={108},
   ISSN={2470-0029},
   url={http://dx.doi.org/10.1103/PhysRevD.108.124035},
   DOI={10.1103/physrevd.108.124035},
   number={12},
   journal={Physical Review D},
   publisher={American Physical Society (APS)},
   author={Pompili, Lorenzo and Buonanno, Alessandra and Estellés, Héctor and Khalil, Mohammed and van de Meent, Maarten and Mihaylov, Deyan P. and Ossokine, Serguei and Pürrer, Michael and Ramos-Buades, Antoni and Mehta, Ajit Kumar and Cotesta, Roberto and Marsat, Sylvain and Boyle, Michael and Kidder, Lawrence E. and Pfeiffer, Harald P. and Scheel, Mark A. and Rüter, Hannes R. and Vu, Nils and Dudi, Reetika and Ma, Sizheng and Mitman, Keefe and Melchor, Denyz and Thomas, Sierra and Sanchez, Jennifer},
   year={2023},
   month=dec }

@article{eobnr15,
  title = {Theoretical groundwork supporting the precessing-spin two-body dynamics of the effective-one-body waveform models SEOBNRv5},
  author = {Khalil, Mohammed and Buonanno, Alessandra and Estell\'es, H\'ector and Mihaylov, Deyan P. and Ossokine, Serguei and Pompili, Lorenzo and Ramos-Buades, Antoni},
  journal = {Phys. Rev. D},
  volume = {108},
  issue = {12},
  pages = {124036},
  numpages = {36},
  year = {2023},
  month = "12",
  publisher = {American Physical Society},
  doi = {10.1103/PhysRevD.108.124036},
  url = {https://link.aps.org/doi/10.1103/PhysRevD.108.124036}
}

@article{eobnr16,
  title = {Analytic systematics in next generation of effective-one-body gravitational waveform models for future observations},
  author = {Nagar, Alessandro and Rettegno, Piero and Gamba, Rossella and Albanesi, Simone and Albertini, Angelica and Bernuzzi, Sebastiano},
  journal = {Phys. Rev. D},
  volume = {108},
  issue = {12},
  pages = {124018},
  numpages = {28},
  year = {2023},
  month = {12},
  publisher = {American Physical Society},
  doi = {10.1103/PhysRevD.108.124018},
  url = {https://link.aps.org/doi/10.1103/PhysRevD.108.124018}
}

@article{eobnr18,
   title={Third post-Newtonian dynamics for eccentric orbits and aligned spins in the effective-one-body waveform model seobnrv5ehm},
   volume={112},
   ISSN={2470-0029},
   url={http://dx.doi.org/10.1103/rb1c-nx5f},
   DOI={10.1103/rb1c-nx5f},
   number={4},
   journal={Physical Review D},
   publisher={American Physical Society (APS)},
   author={Gamboa, Aldo and Khalil, Mohammed and Buonanno, Alessandra},
   year={2025},
   month=aug }

@article{eobnr17,
    author = "Nagar, Alessandro and Gamba, Rossella and Rettegno, Piero and Fantini, Veronica and Bernuzzi, Sebastiano",
    title = "{Effective-one-body waveform model for noncircularized, planar, coalescing black hole binaries: The importance of radiation reaction}",
    eprint = "2404.05288",
    archivePrefix = "arXiv",
    primaryClass = "gr-qc",
    doi = "10.1103/PhysRevD.110.084001",
    journal = "Phys. Rev. D",
    volume = "110",
    number = "8",
    pages = "084001",
    year = "2024"
}

@article{Purrer_2014,
   title={Frequency-domain reduced order models for gravitational waves from aligned-spin compact binaries},
   volume={31},
   ISSN={1361-6382},
   url={http://dx.doi.org/10.1088/0264-9381/31/19/195010},
   DOI={10.1088/0264-9381/31/19/195010},
   number={19},
   journal={Classical and Quantum Gravity},
   publisher={IOP Publishing},
   author={Pürrer, Michael},
   year={2014},
   month=sep, pages={195010} }

@article{Purrer_2016,
   title={Frequency domain reduced order model of aligned-spin effective-one-body waveforms with generic mass ratios and spins},
   volume={93},
   ISSN={2470-0029},
   url={http://dx.doi.org/10.1103/PhysRevD.93.064041},
   DOI={10.1103/physrevd.93.064041},
   number={6},
   journal={Physical Review D},
   publisher={American Physical Society (APS)},
   author={Pürrer, Michael},
   year={2016},
   month=mar }

@article{Tiglio_2022,
   title={Reduced order and surrogate models for gravitational waves},
   volume={25},
   ISSN={1433-8351},
   url={http://dx.doi.org/10.1007/s41114-022-00035-w},
   DOI={10.1007/s41114-022-00035-w},
   number={1},
   journal={Living Reviews in Relativity},
   publisher={Springer Science and Business Media LLC},
   author={Tiglio, Manuel and Villanueva, Aarón},
   year={2022},
   month=apr }

@article{ripple,
    author = "Edwards, Thomas D. P. and Wong, Kaze W. K. and Lam, Kelvin K. H. and Coogan, Adam and Foreman-Mackey, Daniel and Isi, Maximiliano and Zimmerman, Aaron",
    title = "{Differentiable and hardware-accelerated waveforms for gravitational wave data analysis}",
    eprint = "2302.05329",
    archivePrefix = "arXiv",
    primaryClass = "astro-ph.IM",
    doi = "10.1103/PhysRevD.110.064028",
    journal = "Phys. Rev. D",
    volume = "110",
    number = "6",
    pages = "064028",
    year = "2024"
}

@article{mlgw_1,
    author = "Schmidt, Stefano and Breschi, Matteo and Gamba, Rossella and Pagano, Giulia and Rettegno, Piero and Riemenschneider, Gunnar and Bernuzzi, Sebastiano and Nagar, Alessandro and Del Pozzo, Walter",
    title = "{Machine Learning Gravitational Waves from Binary Black Hole Mergers}",
    eprint = "2011.01958",
    archivePrefix = "arXiv",
    primaryClass = "gr-qc",
    doi = "10.1103/PhysRevD.103.043020",
    journal = "Phys. Rev. D",
    volume = "103",
    number = "4",
    pages = "043020",
    year = "2021"
}

@article{mlgw_2,
    author = "Grimbergen, Tim and Schmidt, Stefano and Kalaghatgi, Chinmay and van den Broeck, Chris",
    title = "{Generating higher order modes from binary black hole mergers with machine learning}",
    eprint = "2402.06587",
    archivePrefix = "arXiv",
    primaryClass = "gr-qc",
    doi = "10.1103/PhysRevD.109.104065",
    journal = "Phys. Rev. D",
    volume = "109",
    number = "10",
    pages = "104065",
    year = "2024"
}

@software{jax2018github,
  author = {Bradbury, James and others},
  title = {{JAX}: composable transformations of {P}ython+{N}um{P}y programs},
  url = {http://github.com/jax-ml/jax},
  year = {2018},
}

@misc{tensorflow2015-whitepaper,
title={ {TensorFlow}: Large-Scale Machine Learning on Heterogeneous Systems},
url={https://www.tensorflow.org/},
note={Software available from tensorflow.org},
author={
    Mart\'{i}n~Abadi and
    Ashish~Agarwal and
    Paul~Barham and
    Eugene~Brevdo and
    Zhifeng~Chen and
    Craig~Citro and
    Greg~S.~Corrado and
    Andy~Davis and
    Jeffrey~Dean and
    Matthieu~Devin and
    Sanjay~Ghemawat and
    Ian~Goodfellow and
    Andrew~Harp and
    Geoffrey~Irving and
    Michael~Isard and
    Yangqing Jia and
    Rafal~Jozefowicz and
    Lukasz~Kaiser and
    Manjunath~Kudlur and
    Josh~Levenberg and
    Dandelion~Man\'{e} and
    Rajat~Monga and
    Sherry~Moore and
    Derek~Murray and
    Chris~Olah and
    Mike~Schuster and
    Jonathon~Shlens and
    Benoit~Steiner and
    Ilya~Sutskever and
    Kunal~Talwar and
    Paul~Tucker and
    Vincent~Vanhoucke and
    Vijay~Vasudevan and
    Fernanda~Vi\'{e}gas and
    Oriol~Vinyals and
    Pete~Warden and
    Martin~Wattenberg and
    Martin~Wicke and
    Yuan~Yu and
    Xiaoqiang~Zheng},
  year={2015},
}

@misc{tf2jax,
  author       = {{Google DeepMind}},
  title        = {tf2jax},
  howpublished = {\url{https://github.com/deepmind/tf2jax}},
  year         = {2023}
}

@ARTICLE{bilby,
       author = {{Ashton}, Gregory and {H{\"u}bner}, Moritz and {Lasky}, Paul D. and {Talbot}, Colm and {Ackley}, Kendall and {Biscoveanu}, Sylvia and {Chu}, Qi and {Divakarla}, Atul and {Easter}, Paul J. and {Goncharov}, Boris and {Hernandez Vivanco}, Francisco and {Harms}, Jan and {Lower}, Marcus E. and {Meadors}, Grant D. and {Melchor}, Denyz and {Payne}, Ethan and {Pitkin}, Matthew D. and {Powell}, Jade and {Sarin}, Nikhil and {Smith}, Rory J.~E. and {Thrane}, Eric},
        title = "{BILBY: A User-friendly Bayesian Inference Library for Gravitational-wave Astronomy}",
      journal = {Astrophys. J. Suppl. Ser.},
         year = 2019,
        month = apr,
       volume = {241},
       number = {2},
          eid = {27},
        pages = {27},
          doi = {10.3847/1538-4365/ab06fc},
archivePrefix = {arXiv},
       eprint = {1811.02042},
 primaryClass = {astro-ph.IM},
       adsurl = {https://ui.adsabs.harvard.edu/abs/2019ApJS..241...27A}
}

@article{dingo,
  title = {Real-Time Gravitational Wave Science with Neural Posterior Estimation},
  author = {Dax, Maximilian and Green, Stephen R. and Gair, Jonathan and Macke, Jakob H. and Buonanno, Alessandra and Sch\"olkopf, Bernhard},
  journal = {Phys. Rev. Lett.},
  volume = {127},
  issue = {24},
  pages = {241103},
  numpages = {7},
  year = {2021},
  month = "12",
  publisher = {American Physical Society},
  doi = {10.1103/PhysRevLett.127.241103},
  url = {https://link.aps.org/doi/10.1103/PhysRevLett.127.241103}
}

@article{rift,
    author = "Wagner, Katelyn J. and O'Shaughnessy, R. and Yelikar, A. and Manning, N. and Fernando, D. and Lange, J. and Tiwari, V. and Fernando, A. and Williams, D.",
    title = "{Narrowing RIFT: Focused simulation-based-inference for interpreting exceptional GW sources}",
    eprint = "2505.11655",
    archivePrefix = "arXiv",
    primaryClass = "astro-ph.IM",
    month = "5",
    year = "2025"
}

@article{bayestar,
    author = "Singer, Leo P. and Price, Larry R.",
    title = "{Rapid Bayesian position reconstruction for gravitational-wave transients}",
    eprint = "1508.03634",
    archivePrefix = "arXiv",
    primaryClass = "gr-qc",
    reportNumber = "LIGO-P1500009-V3, LIGO-P1500009-V4, LIGO-P1500009-V5, LIGO-P1500009-V6, LIGO-P1500009-V7, LIGO-P1500009-V8",
    doi = "10.1103/PhysRevD.93.024013",
    journal = "Phys. Rev. D",
    volume = "93",
    number = "2",
    pages = "024013",
    year = "2016"
}

@article{template_bank1,
  title = {Choice of filters for the detection of gravitational waves from coalescing binaries},
  author = {Sathyaprakash, B. S. and Dhurandhar, S. V.},
  journal = {Phys. Rev. D},
  volume = {44},
  issue = {12},
  pages = {3819--3834},
  numpages = {0},
  year = {1991},
  month = {12},
  publisher = {American Physical Society},
  doi = {10.1103/PhysRevD.44.3819},
  url = {https://link.aps.org/doi/10.1103/PhysRevD.44.3819}
}

@article{template_bank2,
   title={Template-based searches for gravitational waves: efficient lattice covering of flat parameter spaces},
   volume={24},
   ISSN={1361-6382},
   url={http://dx.doi.org/10.1088/0264-9381/24/19/S11},
   DOI={10.1088/0264-9381/24/19/s11},
   number={19},
   journal={Classical and Quantum Gravity},
   publisher={IOP Publishing},
   author={Prix, Reinhard},
   year={2007},
   month=sep, pages={S481–S490} }

@article{template_bank3,
  title = {Template bank for gravitational waveforms from coalescing binary black holes: Nonspinning binaries},
  author = {Ajith, P. and Babak, S. and Chen, Y. and Hewitson, M. and Krishnan, B. and Sintes, A. M. and Whelan, J. T. and Br\"ugmann, B. and Diener, P. and Dorband, N. and Gonzalez, J. and Hannam, M. and Husa, S. and Pollney, D. and Rezzolla, L. and Santamar\'{\i}a, L. and Sperhake, U. and Thornburg, J.},
  journal = {Phys. Rev. D},
  volume = {77},
  issue = {10},
  pages = {104017},
  numpages = {22},
  year = {2008},
  month = "5",
  publisher = {American Physical Society},
  doi = {10.1103/PhysRevD.77.104017},
  url = {https://link.aps.org/doi/10.1103/PhysRevD.77.104017}
}

@article{template_bank4,
   title={Stochastic template placement algorithm for gravitational wave data analysis},
   volume={80},
   ISSN={1550-2368},
   url={http://dx.doi.org/10.1103/PhysRevD.80.104014},
   DOI={10.1103/physrevd.80.104014},
   number={10},
   journal={Physical Review D},
   publisher={American Physical Society (APS)},
   author={Harry, I. W. and Allen, B. and Sathyaprakash, B. S.},
   year={2009},
   month=nov }

@article{template_bank5,
    author = "Roy, Soumen and Sengupta, Anand S. and Thakor, Nilay",
    title = "{Hybrid geometric-random template-placement algorithm for gravitational wave searches from compact binary coalescences}",
    eprint = "1702.06771",
    archivePrefix = "arXiv",
    primaryClass = "gr-qc",
    reportNumber = "LIGO-P1700021",
    doi = "10.1103/PhysRevD.95.104045",
    journal = "Phys. Rev. D",
    volume = "95",
    number = "10",
    pages = "104045",
    year = "2017"
}

@article{template_bank6,
   title={Stochastic template bank for gravitational wave searches for precessing neutron-star–black-hole coalescence events},
   volume={95},
   ISSN={2470-0029},
   url={http://dx.doi.org/10.1103/PhysRevD.95.064056},
   DOI={10.1103/physrevd.95.064056},
   number={6},
   journal={Physical Review D},
   publisher={American Physical Society (APS)},
   author={Indik, Nathaniel and Haris, K. and Dal Canton, Tito and Fehrmann, Henning and Krishnan, Badri and Lundgren, Andrew and Nielsen, Alex B. and Pai, Archana},
   year={2017},
   month=mar }

@article{template_bank7,
   title={Efficient gravitational wave template bank generation with differentiable waveforms},
   volume={106},
   ISSN={2470-0029},
   url={http://dx.doi.org/10.1103/PhysRevD.106.122001},
   DOI={10.1103/physrevd.106.122001},
   number={12},
   journal={Physical Review D},
   publisher={American Physical Society (APS)},
   author={Coogan, Adam and Edwards, Thomas D. P. and Chia, Horng Sheng and George, Richard N. and Freese, Katherine and Messick, Cody and Setzer, Christian N. and Weniger, Christoph and Zimmerman, Aaron},
   year={2022},
   month="12" }

@article{template_bank8,
    author = "Hanna, Chad and others",
    title = "{Binary tree approach to template placement for searches for gravitational waves from compact binary mergers}",
    eprint = "2209.11298",
    archivePrefix = "arXiv",
    primaryClass = "gr-qc",
    doi = "10.1103/PhysRevD.108.042003",
    journal = "Phys. Rev. D",
    volume = "108",
    number = "4",
    pages = "042003",
    year = "2023"
}

@article{template_bank9,
   title={Efficient Stochastic Template Bank Using Inner Product Inequalities},
   volume={975},
   ISSN={1538-4357},
   url={http://dx.doi.org/10.3847/1538-4357/ad7d87},
   DOI={10.3847/1538-4357/ad7d87},
   number={2},
   journal={The Astrophysical Journal},
   publisher={American Astronomical Society},
   author={Kacanja, Keisi and Nitz, Alexander H. and Wu, Shichao and Cusinato, Marco and Dhurkunde, Rahul and Harry, Ian and Dal Canton, Tito and Pannarale, Francesco},
   year={2024},
   month=nov, pages={212} }

@article{template_bank10,
    author = "Piccari, Lorenzo and Pannarale, Francesco",
    title = "{Accounting for Tidal Deformability in Binary Neutron Star Template Banks}",
    eprint = "2509.24461",
    archivePrefix = "arXiv",
    primaryClass = "gr-qc",
    reportNumber = "LIGO-P2500579",
    month = "9",
    year = "2025"
}

@article{pycbc_live,
   title={Rapid detection of gravitational waves from compact binary mergers with PyCBC Live},
   volume={98},
   ISSN={2470-0029},
   url={http://dx.doi.org/10.1103/PhysRevD.98.024050},
   DOI={10.1103/physrevd.98.024050},
   number={2},
   journal={Physical Review D},
   publisher={American Physical Society (APS)},
   author={Nitz, Alexander H. and Dal Canton, Tito and Davis, Derek and Reyes, Steven},
   year={2018},
   month="7" }

@article{gwtc-1,
    author = "Abbott, B. P. and others",
    collaboration = "LIGO Scientific, Virgo",
    title = "{GWTC-1: A Gravitational-Wave Transient Catalog of Compact Binary Mergers Observed by LIGO and Virgo during the First and Second Observing Runs}",
    eprint = "1811.12907",
    archivePrefix = "arXiv",
    primaryClass = "astro-ph.HE",
    reportNumber = "LIGO-P1800307",
    doi = "10.1103/PhysRevX.9.031040",
    journal = "Phys. Rev. X",
    volume = "9",
    number = "3",
    pages = "031040",
    year = "2019"
}

@article{gwtc-2.1,
    author = "Abbott, R. and others",
    collaboration = "LIGO Scientific, VIRGO",
    title = "{GWTC-2.1: Deep extended catalog of compact binary coalescences observed by LIGO and Virgo during the first half of the third observing run}",
    eprint = "2108.01045",
    archivePrefix = "arXiv",
    primaryClass = "gr-qc",
    reportNumber = "LIGO-P2100063",
    doi = "10.1103/PhysRevD.109.022001",
    journal = "Phys. Rev. D",
    volume = "109",
    number = "2",
    pages = "022001",
    year = "2024"
}

@article{gstlal,
    author = "Sachdev, Surabhi and others",
    title = "{The GstLAL Search Analysis Methods for Compact Binary Mergers in Advanced LIGO's Second and Advanced Virgo's First Observing Runs}",
    eprint = "1901.08580",
    archivePrefix = "arXiv",
    primaryClass = "gr-qc",
    month = "1",
    year = "2019"
}

@misc{gstlal_2,
      title={GstLAL: A software framework for gravitational wave discovery}, 
      author={Cannon, K and others},
      year={2020},
      eprint={2010.05082},
      archivePrefix={arXiv},
      primaryClass={astro-ph.IM},
      url={https://arxiv.org/abs/2010.05082}, 
}

@article{spiir,
  title = {SPIIR online coherent pipeline to search for gravitational waves from compact binary coalescences},
  author = {Chu, Qi and Kovalam, Manoj and Wen, Linqing and Slaven-Blair, Teresa and Bosveld, Joel and Chen, Yanbei and Clearwater, Patrick and Codoreanu, Alex and Du, Zhihui and Guo, Xiangyu and Guo, Xiaoyang and Kim, Kyungmin and Li, Tjonnie G. F. and Oloworaran, Victor and Panther, Fiona and Powell, Jade and Sengupta, Anand S. and Wette, Karl and Zhu, Xingjiang},
  journal = {Phys. Rev. D},
  volume = {105},
  issue = {2},
  pages = {024023},
  numpages = {17},
  year = {2022},
  month = {1},
  publisher = {American Physical Society},
  doi = {10.1103/PhysRevD.105.024023},
  url = {https://link.aps.org/doi/10.1103/PhysRevD.105.024023}
}

@article{ET1,
    author = "Punturo, M. and others",
    editor = "Ricci, Fulvio",
    title = "{The Einstein Telescope: A third-generation gravitational wave observatory}",
    doi = "10.1088/0264-9381/27/19/194002",
    journal = "Class. Quant. Grav.",
    volume = "27",
    pages = "194002",
    year = "2010"
}

@article{ET2,
    author = "Hild, S. and others",
    title = "{Sensitivity Studies for Third-Generation Gravitational Wave Observatories}",
    eprint = "1012.0908",
    archivePrefix = "arXiv",
    primaryClass = "gr-qc",
    doi = "10.1088/0264-9381/28/9/094013",
    journal = "Class. Quant. Grav.",
    volume = "28",
    pages = "094013",
    year = "2011"
}

@article{ET3,
    author = "Abac, Adrian and others",
    collaboration = "ET",
    title = "{The Science of the Einstein Telescope}",
    eprint = "2503.12263",
    archivePrefix = "arXiv",
    primaryClass = "gr-qc",
    reportNumber = "ET-0036C-25",
    month = "3",
    year = "2025"
}

@article{Reitze:2019iox,
    author = "Reitze, David and others",
    title = "{Cosmic Explorer: The U.S. Contribution to Gravitational-Wave Astronomy beyond LIGO}",
    eprint = "1907.04833",
    archivePrefix = "arXiv",
    primaryClass = "astro-ph.IM",
    reportNumber = "LIGO-P1900316",
    journal = "Bull. Am. Astron. Soc.",
    volume = "51",
    number = "7",
    pages = "035",
    year = "2019"
}

@article{sharpy,
    author = "Demasi, Gabriele and others",
    title = "{The Sequential Monte Carlo goes NUTS: Boosting Gravitational-Wave Inference}",
    eprint = "2601.02336",
    archivePrefix = "arXiv",
    primaryClass = "gr-qc",
    month = "1",
    year = "2026"
}

@article{Edwards:2023sak,
    author = "Edwards, Thomas D. P. and Wong, Kaze W. K. and Lam, Kelvin K. H. and Coogan, Adam and Foreman-Mackey, Daniel and Isi, Maximiliano and Zimmerman, Aaron",
    title = "{Differentiable and hardware-accelerated waveforms for gravitational wave data analysis}",
    eprint = "2302.05329",
    archivePrefix = "arXiv",
    primaryClass = "astro-ph.IM",
    doi = "10.1103/PhysRevD.110.064028",
    journal = "Phys. Rev. D",
    volume = "110",
    number = "6",
    pages = "064028",
    year = "2024"
}

@article{mbta,
   title={Low-latency analysis pipeline for compact binary coalescences in the advanced gravitational wave detector era},
   volume={33},
   ISSN={1361-6382},
   url={http://dx.doi.org/10.1088/0264-9381/33/17/175012},
   DOI={10.1088/0264-9381/33/17/175012},
   number={17},
   journal={Classical and Quantum Gravity},
   publisher={IOP Publishing},
   author={Adams, T and Buskulic, D and Germain, V and Guidi, G M and Marion, F and Montani, M and Mours, B and Piergiovanni, F and Wang, G},
   year={2016},
   month="8",
   pages={175012} }

@article{Harry:2016ijz,
    author = "Harry, Ian and Privitera, Stephen and Boh{\'e}, Alejandro and Buonanno, Alessandra",
    title = "{Searching for Gravitational Waves from Compact Binaries with Precessing Spins}",
    eprint = "1603.02444",
    archivePrefix = "arXiv",
    primaryClass = "gr-qc",
    doi = "10.1103/PhysRevD.94.024012",
    journal = "Phys. Rev. D",
    volume = "94",
    number = "2",
    pages = "024012",
    year = "2016"
}

@article{Harry:2017weg,
    author = "Harry, Ian and Calder{\'o}n Bustillo, Juan and Nitz, Alex",
    title = "{Searching for the full symphony of black hole binary mergers}",
    eprint = "1709.09181",
    archivePrefix = "arXiv",
    primaryClass = "gr-qc",
    reportNumber = "LIGO-DOCUMENT-P1700262, LIGO Document P1700262",
    doi = "10.1103/PhysRevD.97.023004",
    journal = "Phys. Rev. D",
    volume = "97",
    number = "2",
    pages = "023004",
    year = "2018"
}

@article{mlgw_bns,
   title={Combining effective-one-body accuracy and reduced-order-quadrature speed for binary neutron star merger parameter estimation with machine learning},
   volume={107},
   ISSN={2470-0029},
   url={http://dx.doi.org/10.1103/PhysRevD.107.084037},
   DOI={10.1103/physrevd.107.084037},
   number={8},
   journal={Physical Review D},
   publisher={American Physical Society (APS)},
   author={Tissino, Jacopo and Carullo, Gregorio and Breschi, Matteo and Gamba, Rossella and Schmidt, Stefano and Bernuzzi, Sebastiano},
   year={2023},
   month="4" }

@article{Cotesta:2018fcv,
    author = "Cotesta, Roberto and Buonanno, Alessandra and Boh{\'e}, Alejandro and Taracchini, Andrea and Hinder, Ian and Ossokine, Serguei",
    title = "{Enriching the Symphony of Gravitational Waves from Binary Black Holes by Tuning Higher Harmonics}",
    eprint = "1803.10701",
    archivePrefix = "arXiv",
    primaryClass = "gr-qc",
    doi = "10.1103/PhysRevD.98.084028",
    journal = "Phys. Rev. D",
    volume = "98",
    number = "8",
    pages = "084028",
    year = "2018"
}

@article{multibanding,
   title={Accelerating gravitational wave parameter estimation with multi-band template interpolation},
   volume={34},
   ISSN={1361-6382},
   url={http://dx.doi.org/10.1088/1361-6382/aa6d44},
   DOI={10.1088/1361-6382/aa6d44},
   number={11},
   journal={Classical and Quantum Gravity},
   publisher={IOP Publishing},
   author={Vinciguerra, Serena and Veitch, John and Mandel, Ilya},
   year={2017},
   month="5",
   pages={115006} }

@article{roq_1,
   title={Two-Step Greedy Algorithm for Reduced Order Quadratures},
   volume={57},
   ISSN={1573-7691},
   url={http://dx.doi.org/10.1007/s10915-013-9722-z},
   DOI={10.1007/s10915-013-9722-z},
   number={3},
   journal={Journal of Scientific Computing},
   publisher={Springer Science and Business Media LLC},
   author={Antil, Harbir and Field, Scott E. and Herrmann, Frank and Nochetto, Ricardo H. and Tiglio, Manuel},
   year={2013},
   month=may, pages={604–637} }

@article{roq_2,
   title={Gravitational wave parameter estimation with compressed likelihood evaluations},
   volume={87},
   ISSN={1550-2368},
   url={http://dx.doi.org/10.1103/PhysRevD.87.124005},
   DOI={10.1103/physrevd.87.124005},
   number={12},
   journal={Physical Review D},
   publisher={American Physical Society (APS)},
   author={Canizares, Priscilla and Field, Scott E. and Gair, Jonathan R. and Tiglio, Manuel},
   year={2013},
   month="6" }

@article{roq_3,
   title={Fast and accurate inference on gravitational waves from precessing compact binaries},
   volume={94},
   ISSN={2470-0029},
   url={http://dx.doi.org/10.1103/PhysRevD.94.044031},
   DOI={10.1103/physrevd.94.044031},
   number={4},
   journal={Physical Review D},
   publisher={American Physical Society (APS)},
   author={Smith, Rory and Field, Scott E. and Blackburn, Kent and Haster, Carl-Johan and Pürrer, Michael and Raymond, Vivien and Schmidt, Patricia},
   year={2016},
   month="8" }

@article{roq_4,
   title={Python-based reduced order quadrature building code for fast gravitational wave inference},
   volume={104},
   ISSN={2470-0029},
   url={http://dx.doi.org/10.1103/PhysRevD.104.063031},
   DOI={10.1103/physrevd.104.063031},
   number={6},
   journal={Physical Review D},
   publisher={American Physical Society (APS)},
   author={Qi, Hong and Raymond, Vivien},
   year={2021},
   month="9" }

@article{roq_5,
   title={Rapid parameter estimation of gravitational waves from binary neutron star coalescence using focused reduced order quadrature},
   volume={102},
   ISSN={2470-0029},
   url={http://dx.doi.org/10.1103/PhysRevD.102.104020},
   DOI={10.1103/physrevd.102.104020},
   number={10},
   journal={Physical Review D},
   publisher={American Physical Society (APS)},
   author={Morisaki, Soichiro and Raymond, Vivien},
   year={2020},
   month="11" }

@article{roq_6,
  title = {Rapid localization and inference on compact binary coalescences with the Advanced LIGO-Virgo-KAGRA gravitational-wave detector network},
  author = {Morisaki, Soichiro and Smith, Rory and Tsukada, Leo and Sachdev, Surabhi and Stevenson, Simon and Talbot, Colm and Zimmerman, Aaron},
  journal = {Phys. Rev. D},
  volume = {108},
  issue = {12},
  pages = {123040},
  numpages = {19},
  year = {2023},
  month = "12",
  publisher = {American Physical Society},
  doi = {10.1103/PhysRevD.108.123040},
  url = {https://link.aps.org/doi/10.1103/PhysRevD.108.123040}
}

@article{relative_binning,
    author = "Zackay, Barak and Dai, Liang and Venumadhav, Tejaswi",
    title = "{Relative Binning and Fast Likelihood Evaluation for Gravitational Wave Parameter Estimation}",
    eprint = "1806.08792",
    archivePrefix = "arXiv",
    primaryClass = "astro-ph.IM",
    month = "6",
    year = "2018"
}

@article{surrogate,
   title={Fast Prediction and Evaluation of Gravitational Waveforms Using Surrogate Models},
   volume={4},
   ISSN={2160-3308},
   url={http://dx.doi.org/10.1103/PhysRevX.4.031006},
   DOI={10.1103/physrevx.4.031006},
   number={3},
   journal={Physical Review X},
   publisher={American Physical Society (APS)},
   author={Field, Scott E. and Galley, Chad R. and Hesthaven, Jan S. and Kaye, Jason and Tiglio, Manuel},
   year={2014},
   month=jul }

@article{ROM_ANNs_1,
   title={Reduced-Order Modeling with Artificial Neurons for Gravitational-Wave Inference},
   volume={122},
   ISSN={1079-7114},
   url={http://dx.doi.org/10.1103/PhysRevLett.122.211101},
   DOI={10.1103/physrevlett.122.211101},
   number={21},
   journal={Physical Review Letters},
   publisher={American Physical Society (APS)},
   author={Chua, Alvin J.K. and Galley, Chad R. and Vallisneri, Michele},
   year={2019},
   month=may }

@article{NR_SUR_1,
   title={Fast and Accurate Prediction of Numerical Relativity Waveforms from Binary Black Hole Coalescences Using Surrogate Models},
   volume={115},
   ISSN={1079-7114},
   url={http://dx.doi.org/10.1103/PhysRevLett.115.121102},
   DOI={10.1103/physrevlett.115.121102},
   number={12},
   journal={Physical Review Letters},
   publisher={American Physical Society (APS)},
   author={Blackman, Jonathan and Field, Scott E. and Galley, Chad R. and Szilágyi, Béla and Scheel, Mark A. and Tiglio, Manuel and Hemberger, Daniel A.},
   year={2015},
   month=sep }

@article{NR_SUR_2,
   title={A Surrogate model of gravitational waveforms from numerical relativity simulations of precessing binary black hole mergers},
   volume={95},
   ISSN={2470-0029},
   url={http://dx.doi.org/10.1103/PhysRevD.95.104023},
   DOI={10.1103/physrevd.95.104023},
   number={10},
   journal={Physical Review D},
   publisher={American Physical Society (APS)},
   author={Blackman, Jonathan and Field, Scott E. and Scheel, Mark A. and Galley, Chad R. and Hemberger, Daniel A. and Schmidt, Patricia and Smith, Rory},
   year={2017},
   month=may }

@article{NR_SUR_3,
   title={Numerical relativity waveform surrogate model for generically precessing binary black hole mergers},
   volume={96},
   ISSN={2470-0029},
   url={http://dx.doi.org/10.1103/PhysRevD.96.024058},
   DOI={10.1103/physrevd.96.024058},
   number={2},
   journal={Physical Review D},
   publisher={American Physical Society (APS)},
   author={Blackman, Jonathan and Field, Scott E. and Scheel, Mark A. and Galley, Chad R. and Ott, Christian D. and Boyle, Michael and Kidder, Lawrence E. and Pfeiffer, Harald P. and Szilágyi, Béla},
   year={2017},
   month=jul }

@article{NR_SUR_4,
   title={Surrogate model of hybridized numerical relativity binary black hole waveforms},
   volume={99},
   ISSN={2470-0029},
   url={http://dx.doi.org/10.1103/PhysRevD.99.064045},
   DOI={10.1103/physrevd.99.064045},
   number={6},
   journal={Physical Review D},
   publisher={American Physical Society (APS)},
   author={Varma, Vijay and Field, Scott E. and Scheel, Mark A. and Blackman, Jonathan and Kidder, Lawrence E. and Pfeiffer, Harald P.},
   year={2019},
   month=mar }

@article{ROM_ANNs_2,
   title={Deep learning powered numerical relativity surrogate for binary black hole waveforms},
   volume={112},
   ISSN={2470-0029},
   url={http://dx.doi.org/10.1103/7bkx-hs53},
   DOI={10.1103/7bkx-hs53},
   number={4},
   journal={Physical Review D},
   publisher={American Physical Society (APS)},
   author={Freitas, Osvaldo Gramaxo and Theodoropoulos, Anastasios and Villanueva, Nino and Fernandes, Tiago and Nunes, Solange and Font, José A. and Onofre, Antonio and Torres-Forné, Alejandro and Martin-Guerrero, José D.},
   year={2025},
   month=aug }

@article{LIGOScientific:2017adf,
    author = "Abbott, B. P. and others",
    collaboration = "LIGO Scientific, Virgo, 1M2H, Dark Energy Camera GW-E, DES, DLT40, Las Cumbres Observatory, VINROUGE, MASTER",
    title = "{A gravitational-wave standard siren measurement of the Hubble constant}",
    eprint = "1710.05835",
    archivePrefix = "arXiv",
    primaryClass = "astro-ph.CO",
    reportNumber = "LIGO-P1700296, FERMILAB-PUB-17-472-A-AE",
    doi = "10.1038/nature24471",
    journal = "Nature",
    volume = "551",
    number = "7678",
    pages = "85--88",
    year = "2017"
}

@article{LIGOScientific:2019zcs,
    author = "Abbott, B. P. and others",
    collaboration = "LIGO Scientific, Virgo, VIRGO",
    title = "{A Gravitational-wave Measurement of the Hubble Constant Following the Second Observing Run of Advanced LIGO and Virgo}",
    eprint = "1908.06060",
    archivePrefix = "arXiv",
    primaryClass = "astro-ph.CO",
    reportNumber = "LIGO-P1900015",
    doi = "10.3847/1538-4357/abdcb7",
    journal = "Astrophys. J.",
    volume = "909",
    number = "2",
    pages = "218",
    year = "2021"
}

@article{Burns:2019tqz,
    author = "Burns, Eric and others",
    title = "{A Summary of Multimessenger Science with Neutron Star Mergers}",
    eprint = "1903.03582",
    archivePrefix = "arXiv",
    primaryClass = "astro-ph.HE",
    month = "3",
    year = "2019"
}

@article{LIGOScientific:2018dkp,
    author = "Abbott, B. P. and others",
    collaboration = "LIGO Scientific, Virgo",
    title = "{Tests of General Relativity with GW170817}",
    eprint = "1811.00364",
    archivePrefix = "arXiv",
    primaryClass = "gr-qc",
    reportNumber = "LIGO-P1800059",
    doi = "10.1103/PhysRevLett.123.011102",
    journal = "Phys. Rev. Lett.",
    volume = "123",
    number = "1",
    pages = "011102",
    year = "2019"
}

@article{Baker:2017hug,
    author = "Baker, T. and Bellini, E. and Ferreira, P. G. and Lagos, M. and Noller, J. and Sawicki, I.",
    title = "{Strong constraints on cosmological gravity from GW170817 and GRB 170817A}",
    eprint = "1710.06394",
    archivePrefix = "arXiv",
    primaryClass = "astro-ph.CO",
    doi = "10.1103/PhysRevLett.119.251301",
    journal = "Phys. Rev. Lett.",
    volume = "119",
    number = "25",
    pages = "251301",
    year = "2017"
}

@misc{chollet2015keras,
  title={Keras},
  author={Chollet, Fran\c{c}ois and others},
  year={2015},
  publisher={GitHub},
  journal = {GitHub repository},
  howpublished = {\url{https://github.com/fchollet/keras}}
}

@article{degeneracy_1,
    author = "Frattale Mascioli, Adriano and Crescimbeni, Francesco and Pacilio, Costantino and Pani, Paolo and Pannarale, Francesco",
    title = "{Taming systematics in distance and inclination measurements with gravitational waves: Role of the detector network and higher-order modes}",
    eprint = "2504.12473",
    archivePrefix = "arXiv",
    primaryClass = "gr-qc",
    doi = "10.1103/23cf-j34y",
    journal = "Phys. Rev. D",
    volume = "112",
    number = "6",
    pages = "062003",
    year = "2025"
}

@article{degeneracy_2,
    author = "Usman, Samantha A. and Mills, Joseph C. and Fairhurst, Stephen",
    title = "{Constraining the Inclinations of Binary Mergers from Gravitational-wave Observations}",
    eprint = "1809.10727",
    archivePrefix = "arXiv",
    primaryClass = "gr-qc",
    doi = "10.3847/1538-4357/ab0b3e",
    journal = "Astrophys. J.",
    volume = "877",
    number = "2",
    pages = "82",
    year = "2019"
}

@article{degeneracy_3,
    author = "London, Lionel and Khan, Sebastian and Fauchon-Jones, Edward and Garc{\'\i}a, Cecilio and Hannam, Mark and Husa, Sascha and Jim{\'e}nez-Forteza, Xisco and Kalaghatgi, Chinmay and Ohme, Frank and Pannarale, Francesco",
    title = "{First higher-multipole model of gravitational waves from spinning and coalescing black-hole binaries}",
    eprint = "1708.00404",
    archivePrefix = "arXiv",
    primaryClass = "gr-qc",
    doi = "10.1103/PhysRevLett.120.161102",
    journal = "Phys. Rev. Lett.",
    volume = "120",
    number = "16",
    pages = "161102",
    year = "2018"
}

@article{degeneracy_4,
   title={Dark Sirens to Resolve the Hubble–Lemaître Tension},
   volume={905},
   ISSN={2041-8213},
   url={http://dx.doi.org/10.3847/2041-8213/abcaf5},
   DOI={10.3847/2041-8213/abcaf5},
   number={2},
   journal={The Astrophysical Journal Letters},
   publisher={American Astronomical Society},
   author={Borhanian, Ssohrab and Dhani, Arnab and Gupta, Anuradha and Arun, K. G. and Sathyaprakash, B. S.},
   year={2020},
   month=dec, pages={L28} }

@article{degeneracy_5,
   title={Identifying when precession can be measured in gravitational waveforms},
   volume={103},
   ISSN={2470-0029},
   url={http://dx.doi.org/10.1103/PhysRevD.103.124023},
   DOI={10.1103/physrevd.103.124023},
   number={12},
   journal={Physical Review D},
   publisher={American Physical Society (APS)},
   author={Green, Rhys and Hoy, Charlie and Fairhurst, Stephen and Hannam, Mark and Pannarale, Francesco and Thomas, Cory},
   year={2021},
   month=jun }

@article{degeneracy_6,
  title = {Measuring the Hubble Constant with Neutron Star Black Hole Mergers},
  author = {Vitale, Salvatore and Chen, Hsin-Yu},
  journal = {Phys. Rev. Lett.},
  volume = {121},
  issue = {2},
  pages = {021303},
  numpages = {6},
  year = {2018},
  month = "7",
  publisher = {American Physical Society},
  doi = {10.1103/PhysRevLett.121.021303},
  url = {https://link.aps.org/doi/10.1103/PhysRevLett.121.021303}
}

@article{degeneracy_7,
   title={Breaking bad degeneracies with Love relations: Improving gravitational-wave measurements through universal relations},
   volume={107},
   ISSN={2470-0029},
   url={http://dx.doi.org/10.1103/PhysRevD.107.043010},
   DOI={10.1103/physrevd.107.043010},
   number={4},
   journal={Physical Review D},
   publisher={American Physical Society (APS)},
   author={Xie, Yiqi and Chatterjee, Deep and Holder, Gilbert and Holz, Daniel E. and Perkins, Scott and Yagi, Kent and Yunes, Nicolás},
   year={2023},
   month=feb }

@article{gerosa_2018,
  title = {Spin orientations of merging black holes formed from the evolution of stellar binaries},
  author = {Gerosa, Davide and Berti, Emanuele and O'Shaughnessy, Richard and Belczynski, Krzysztof and Kesden, Michael and Wysocki, Daniel and Gladysz, Wojciech},
  journal = {Phys. Rev. D},
  volume = {98},
  issue = {8},
  pages = {084036},
  numpages = {21},
  year = {2018},
  month = "10",
  publisher = {American Physical Society},
  doi = {10.1103/PhysRevD.98.084036},
  url = {https://link.aps.org/doi/10.1103/PhysRevD.98.084036}
}

@article{Samsing_2018,
   title={Eccentric black hole mergers forming in globular clusters},
   volume={97},
   ISSN={2470-0029},
   url={http://dx.doi.org/10.1103/PhysRevD.97.103014},
   DOI={10.1103/physrevd.97.103014},
   number={10},
   journal={Physical Review D},
   publisher={American Physical Society (APS)},
   author={Samsing, Johan},
   year={2018},
   month=may }

@article{Mapelli:2021taw,
    author = "Mapelli, Michela",
    title = "{Formation Channels of Single and Binary Stellar-Mass Black Holes}",
    eprint = "2106.00699",
    archivePrefix = "arXiv",
    primaryClass = "astro-ph.HE",
    doi = "10.1007/978-981-15-4702-7_16-1",
    year = "2021"
}

@article{population_2,
    author = "Abac, A. G. and others",
    collaboration = "LIGO Scientific, VIRGO, KAGRA",
    title = "{GWTC-4.0: Population Properties of Merging Compact Binaries}",
    eprint = "2508.18083",
    archivePrefix = "arXiv",
    primaryClass = "astro-ph.HE",
    reportNumber = "LIGO-P2400004",
    month = "8",
    year = "2025"
}

@article{population_1,
  title = {Population of Merging Compact Binaries Inferred Using Gravitational Waves through GWTC-3},
  author = {Abbott, R. and others},
  collaboration = {LIGO Scientific Collaboration, Virgo Collaboration, and KAGRA Collaboration},
  journal = {Phys. Rev. X},
  volume = {13},
  issue = {1},
  pages = {011048},
  numpages = {75},
  year = {2023},
  month = "3",
  publisher = {American Physical Society},
  doi = {10.1103/PhysRevX.13.011048},
  url = {https://link.aps.org/doi/10.1103/PhysRevX.13.011048}
}

@article{Owen_1999,
   title="{Matched filtering of gravitational waves from inspiraling compact binaries: Computational cost and template placement}",
   volume={60},
   ISSN={1089-4918},
   url={http://dx.doi.org/10.1103/PhysRevD.60.022002},
   DOI={10.1103/physrevd.60.022002},
   number={2},
   journal={Physical Review D},
   publisher={American Physical Society (APS)},
   author={Owen, Benjamin J. and Sathyaprakash, B. S.},
   year={1999},
   month="6" }

@article{Owen_1996,
  title = "{Search templates for gravitational waves from inspiraling binaries: Choice of template spacing}",
  author = {Owen, Benjamin J.},
  journal = {Phys. Rev. D},
  volume = {53},
  issue = {12},
  pages = {6749--6761},
  numpages = {0},
  year = {1996},
  month = {6},
  publisher = {American Physical Society},
  doi = {10.1103/PhysRevD.53.6749},
  url = {https://link.aps.org/doi/10.1103/PhysRevD.53.6749}
}

@article{Blondel2024TheEO,
  title={The Elements of Differentiable Programming},
  author={Mathieu Blondel and Vincent Roulet},
  journal={ArXiv},
  year={2024},
  volume={abs/2403.14606},
  url={https://api.semanticscholar.org/CorpusID:268553523}
}

@article{2006:Skilling,
author = {Skilling, John},
year = {2006},
month = {12},
pages = {833-860},
title = {Nested sampling for general Bayesian computation. Bayesian Anal. 1(4), 833-860},
volume = {1},
journal = {Bayesian Analysis},
doi = {10.1214/06-BA127}
}

@article{Estelles:2021gvs,
    author = "Estell{\'e}s, H{\'e}ctor and Colleoni, Marta and Garc{\'\i}a-Quir{\'o}s, Cecilio and Husa, Sascha and Keitel, David and Mateu-Lucena, Maite and Planas, Maria de Lluc and Ramos-Buades, Antoni",
    title = "{New twists in compact binary waveform modeling: A fast time-domain model for precession}",
    eprint = "2105.05872",
    archivePrefix = "arXiv",
    primaryClass = "gr-qc",
    reportNumber = "LIGO-P2100136",
    doi = "10.1103/PhysRevD.105.084040",
    journal = "Phys. Rev. D",
    volume = "105",
    number = "8",
    pages = "084040",
    year = "2022"
}

@article{DUANE1987216,
title = {Hybrid Monte Carlo},
journal = {Physics Letters B},
volume = {195},
number = {2},
pages = {216-222},
year = {1987},
issn = {0370-2693},
doi = {https://doi.org/10.1016/0370-2693(87)91197-X},
url = {https://www.sciencedirect.com/science/article/pii/037026938791197X},
author = {Simon Duane and A.D. Kennedy and Brian J. Pendleton and Duncan Roweth}
}

@article{hmc_betancourt,
author = {Betancourt, Michael},
year = {2017},
month = {1},
pages = {},
title = {A Conceptual Introduction to Hamiltonian Monte Carlo},
doi = {10.48550/arXiv.1701.02434}
}

@article{Abbott_2017,
   title={Gravitational Waves and Gamma-Rays from a Binary Neutron Star Merger: GW170817 and GRB 170817A},
   volume={848},
   ISSN={2041-8213},
   url={http://dx.doi.org/10.3847/2041-8213/aa920c},
   DOI={10.3847/2041-8213/aa920c},
   number={2},
   journal={The Astrophysical Journal Letters},
   publisher={American Astronomical Society},
   author={Abbott, B. P. and others},
   year={2017},
   month=oct, pages={L13} }

@software{alex_nitz_2024_10473621,
  author       = {Alex Nitz and
                  Ian Harry and
                  Duncan Brown and
                  Christopher M. Biwer and
                  Josh Willis and
                  Tito Dal Canton and
                  Collin Capano and
                  Thomas Dent and
                  Larne Pekowsky and
                  Gareth S Cabourn Davies and
                  Soumi De and
                  Miriam Cabero and
                  Shichao Wu and
                  Andrew R. Williamson and
                  Bernd Machenschalk and
                  Duncan Macleod and
                  Francesco Pannarale and
                  Prayush Kumar and
                  Steven Reyes and
                  dfinstad and
                  Sumit Kumar and
                  Márton Tápai and
                  Leo Singer and
                  Praveen Kumar and
                  veronica-villa and
                  maxtrevor and
                  Bhooshan Uday Varsha Gadre and
                  Sebastian Khan and
                  Stephen Fairhurst and
                  Arthur Tolley},
  title        = {gwastro/pycbc: v2.3.3 release of PyCBC},
  month        = jan,
  year         = 2024,
  publisher    = {Zenodo},
  version      = {v2.3.3},
  doi          = {10.5281/zenodo.10473621},
  url          = {https://doi.org/10.5281/zenodo.10473621},
}

@inproceedings{
yallup2025nested,
title={Nested Slice Sampling},
author={David Yallup and Namu Kroupa and Will Handley},
booktitle={Frontiers in Probabilistic Inference: Learning meets Sampling},
year={2025},
url={https://openreview.net/forum?id=ekbkMSuPo4}
}

@misc{cabezas2024,
      title={BlackJAX: Composable Bayesian inference in JAX}, 
      author={Alberto Cabezas and Adrien Corenflos and Junpeng Lao and Rémi Louf and Antoine Carnec and Kaustubh Chaudhari and Reuben Cohn-Gordon and Jeremie Coullon and Wei Deng and Sam Duffield and Gerardo Durán-Martín and Marcin Elantkowski and Dan Foreman-Mackey and Michele Gregori and Carlos Iguaran and Ravin Kumar and Martin Lysy and Kevin Murphy and Juan Camilo Orduz and Karm Patel and Xi Wang and Rob Zinkov},
      year={2024},
      eprint={2402.10797},
      archivePrefix={arXiv},
      primaryClass={cs.MS},
      url={https://arxiv.org/abs/2402.10797}, 
}

@article{Prathaban:2025qgg,
    author = "Prathaban, Metha and Yallup, David and Alvey, James and Yang, Ming and Templeton, Will and Handley, Will",
    title = "{Gravitational-wave inference at GPU speed: A bilby-like nested sampling kernel within blackjax-ns}",
    eprint = "2509.04336",
    archivePrefix = "arXiv",
    primaryClass = "gr-qc",
    month = "9",
    year = "2025"
}

@article{Mihaylov:2023bkc,
    author = {Mihaylov, Deyan P. and Ossokine, Serguei and Buonanno, Alessandra and Estelles, Hector and Pompili, Lorenzo and P{\"u}rrer, Michael and Ramos-Buades, Antoni},
    title = "{pySEOBNR: a software package for the next generation of effective-one-body multipolar waveform models}",
    eprint = "2303.18203",
    archivePrefix = "arXiv",
    primaryClass = "gr-qc",
    month = "3",
    year = "2023"
}

@article{LIGOScientific:2018hze,
    author = "Abbott, B. P. and others",
    collaboration = "LIGO Scientific, Virgo",
    title = "{Properties of the binary neutron star merger GW170817}",
    eprint = "1805.11579",
    archivePrefix = "arXiv",
    primaryClass = "gr-qc",
    doi = "10.1103/PhysRevX.9.011001",
    journal = "Phys. Rev. X",
    volume = "9",
    number = "1",
    pages = "011001",
    year = "2019"
}

@misc{lalsuite,
    author         = "{LIGO Scientific Collaboration}
                      and {Virgo Collaboration}
                      and {KAGRA Collaboration}",
    title          = "{LVK} {A}lgorithm {L}ibrary - {LALS}uite",
    howpublished   = "Free software (GPL)",
    doi            = "10.7935/GT1W-FZ16",
    year           = "2018"
}

@misc{XLA,
  title={OpenXLA},
  author={{Arm Limited} and {Google LLC} and {NVIDIA Corporation}},
  publisher={GitHub},
  journal = {GitHub repository},
  howpublished = {\url{https://github.com/openxla/xla}}
}

@ARTICLE{matplotlib,
  author={Hunter, John D.},
  journal={Computing in Science \& Engineering}, 
  title={Matplotlib: A 2D Graphics Environment}, 
  year={2007},
  volume={9},
  number={3},
  pages={90-95},
  doi={10.1109/MCSE.2007.55}}

@ARTICLE{numpy,
       author = {{Harris}, Charles R. and {Millman}, K. Jarrod and {van der Walt}, St{\'e}fan J. and {Gommers}, Ralf and {Virtanen}, Pauli and {Cournapeau}, David and {Wieser}, Eric and {Taylor}, Julian and {Berg}, Sebastian and {Smith}, Nathaniel J. and {Kern}, Robert and {Picus}, Matti and {Hoyer}, Stephan and {van Kerkwijk}, Marten H. and {Brett}, Matthew and {Haldane}, Allan and {del R{\'\i}o}, Jaime Fern{\'a}ndez and {Wiebe}, Mark and {Peterson}, Pearu and {G{\'e}rard-Marchant}, Pierre and {Sheppard}, Kevin and {Reddy}, Tyler and {Weckesser}, Warren and {Abbasi}, Hameer and {Gohlke}, Christoph and {Oliphant}, Travis E.},
        title = "{Array programming with NumPy}",
      journal = {Nature},
         year = 2020,
        month = sep,
       volume = {585},
       number = {7825},
        pages = {357-362},
          doi = {10.1038/s41586-020-2649-2},
archivePrefix = {arXiv},
       eprint = {2006.10256},
 primaryClass = {cs.MS},
       adsurl = {https://ui.adsabs.harvard.edu/abs/2020Natur.585..357H}
}

@ARTICLE{scipy,
       author = {{Virtanen}, Pauli and {Gommers}, Ralf and {Oliphant}, Travis E. and {Haberland}, Matt and {Reddy}, Tyler and {Cournapeau}, David and {Burovski}, Evgeni and {Peterson}, Pearu and {Weckesser}, Warren and {Bright}, Jonathan and {van der Walt}, St{\'e}fan J. and {Brett}, Matthew and {Wilson}, Joshua and {Millman}, K. Jarrod and {Mayorov}, Nikolay and {Nelson}, Andrew R.~J. and {Jones}, Eric and {Kern}, Robert and {Larson}, Eric and {Carey}, C.~J. and {Polat}, {\.I}lhan and {Feng}, Yu and {Moore}, Eric W. and {VanderPlas}, Jake and {Laxalde}, Denis and {Perktold}, Josef and {Cimrman}, Robert and {Henriksen}, Ian and {Quintero}, E.~A. and {Harris}, Charles R. and {Archibald}, Anne M. and {Ribeiro}, Ant{\^o}nio H. and {Pedregosa}, Fabian and {van Mulbregt}, Paul and {SciPy 1.  0 Contributors}},
        title = "{SciPy 1.0: fundamental algorithms for scientific computing in Python}",
      journal = {Nature Medicine},
         year = 2020,
        month = feb,
       volume = {17},
        pages = {261-272},
          doi = {10.1038/s41592-019-0686-2},
archivePrefix = {arXiv},
       eprint = {1907.10121},
 primaryClass = {cs.MS},
       adsurl = {https://ui.adsabs.harvard.edu/abs/2020NaMet..17..261V}
}

@article{LIGOScientific:2026ctl,
    author = "Abac, A. G. and others",
    collaboration = "LIGO Scientific, VIRGO, KAGRA",
    title = "{GWTC-5.0: Population Properties of Merging Compact Binaries}",
    eprint = "2605.27226",
    archivePrefix = "arXiv",
    primaryClass = "astro-ph.HE",
    reportNumber = "LIGO-P2600045",
    month = "5",
    year = "2026"
}

@article{LIGOScientific:2026uyd,
    collaboration = "LIGO Scientific, VIRGO, KAGRA",
    title = "{GWTC-5.0: Constraints on the Cosmic Expansion Rate and Modified Gravitational-wave Propagation}",
    eprint = "2605.27227",
    archivePrefix = "arXiv",
    primaryClass = "astro-ph.CO",
    reportNumber = "LIGO-P2600018",
    month = "5",
    year = "2026"
}

@ARTICLE{scikit,
       author = {{Pedregosa}, Fabian and {Varoquaux}, Ga{\"e}l and {Gramfort}, Alexandre and {Michel}, Vincent and {Thirion}, Bertrand and {Grisel}, Olivier and {Blondel}, Mathieu and {M{\"u}ller}, Andreas and {Nothman}, Joel and {Louppe}, Gilles and {Prettenhofer}, Peter and {Weiss}, Ron and {Dubourg}, Vincent and {Vanderplas}, Jake and {Passos}, Alexandre and {Cournapeau}, David and {Brucher}, Matthieu and {Perrot}, Matthieu and {Duchesnay}, {\'E}douard},
        title = "{Scikit-learn: Machine Learning in Python}",
      journal = {Journal of Machine Learning Research},
         year = 2011,
        month = oct,
       volume = {12},
        pages = {2825-2830},
          doi = {10.48550/arXiv.1201.0490},
archivePrefix = {arXiv},
       eprint = {1201.0490},
 primaryClass = {cs.LG},
       adsurl = {https://ui.adsabs.harvard.edu/abs/2011JMLR...12.2825P}
}

@article{Coulter:2017wya,
    author = "Coulter, D. A. and others",
    title = "{Swope Supernova Survey 2017a (SSS17a), the Optical Counterpart to a Gravitational Wave Source}",
    eprint = "1710.05452",
    archivePrefix = "arXiv",
    primaryClass = "astro-ph.HE",
    doi = "10.1126/science.aap9811",
    journal = "Science",
    volume = "358",
    pages = "1556",
    year = "2017"
}

@article{Purrer:2026wgp,
    author = {P{\"u}rrer, Michael and Girish, Ashwin and Thomas, Lucy M. and Field, Scott E. and Varma, Vijay},
    title = "{Fast, accurate, and differentiable: a neural-network surrogate for NRSur7dq4 precessing binary black hole waveforms}",
    eprint = "2607.24960",
    archivePrefix = "arXiv",
    primaryClass = "gr-qc",
    month = "7",
    year = "2026"
}
\end{document}